%% file: main.tex
\pdfoutput=1

\documentclass[12pt,a4paper]{article}

\usepackage{ifthen} 
\newboolean{pdflatex}
\setboolean{pdflatex}{true} 

\newboolean{articletitles}
\setboolean{articletitles}{true} 

\newboolean{uprightparticles}
\setboolean{uprightparticles}{false} 

\def\paperauthors{LHCb collaboration} 
\def\paperasciititle{Prompt and nonprompt psi(2S) production in pPb collisions at sqrt(s)=8.16 TeV} 
\def\papertitle{Prompt and nonprompt \psitwos production in $p$Pb collisions at $\sqsnn=8.16\tev$} 
\def\paperkeywords{{High Energy Physics}, {LHCb}} 
\def\papercopyright{\the\year\ CERN for the benefit of the LHCb collaboration} 
\def\paperlicence{CC BY 4.0 licence}
\def\paperlicenceurl{https://creativecommons.org/licenses/by/4.0/}

\input{preamble}

\begin{document}

\renewcommand{\thefootnote}{\fnsymbol{footnote}}
\setcounter{footnote}{1}

\input{title-LHCb-PAPER}

\renewcommand{\thefootnote}{\arabic{footnote}}
\setcounter{footnote}{0}

\cleardoublepage


\pagestyle{plain} 
\setcounter{page}{1}
\pagenumbering{arabic}

\input{intro}
\input{detector}
\input{selection}

\input{systematicunc}

\input{results}
\input{acknowledgements}

\input{appendix}

\addcontentsline{toc}{section}{References}
\bibliographystyle{LHCb}
\bibliography{main,LHCb-PAPER,LHCb-CONF,LHCb-DP,LHCb-TDR,LHCb-Phys}

\newpage
\input{Authorship_LHCb-PAPER-2023-024}

\end{document}

%% file: preamble.tex
\usepackage[top=1in, bottom=1.25in, left=1in, right=1in]{geometry}

\usepackage{microtype}
\usepackage{lineno}  
\usepackage{xspace} 
\usepackage{caption} 

\usepackage{graphicx}  
\usepackage{color}
\usepackage{colortbl}
\graphicspath{{./figs/}} 
\usepackage{booktabs}
\usepackage{amsmath} 
\usepackage{amssymb}
\usepackage{amsfonts}
\usepackage{upgreek} 
\usepackage{amsbsy}
\newcommand*\patchAmsMathEnvironmentForLineno[1]{%
\expandafter\let\csname old#1\expandafter\endcsname\csname #1\endcsname
\expandafter\let\csname oldend#1\expandafter\endcsname\csname
end#1\endcsname
 \renewenvironment{#1}%
   {\linenomath\csname old#1\endcsname}%
   {\csname oldend#1\endcsname\endlinenomath}%
}
\newcommand*\patchBothAmsMathEnvironmentsForLineno[1]{%
  \patchAmsMathEnvironmentForLineno{#1}%
  \patchAmsMathEnvironmentForLineno{#1*}%
}
\AtBeginDocument{%
\patchBothAmsMathEnvironmentsForLineno{equation}%
\patchBothAmsMathEnvironmentsForLineno{align}%
\patchBothAmsMathEnvironmentsForLineno{flalign}%
\patchBothAmsMathEnvironmentsForLineno{alignat}%
\patchBothAmsMathEnvironmentsForLineno{gather}%
\patchBothAmsMathEnvironmentsForLineno{multline}%
\patchBothAmsMathEnvironmentsForLineno{eqnarray}%
}

\usepackage{hyperxmp}

\usepackage[pdftex,
            pdfauthor={\paperauthors},
            pdftitle={\paperasciititle},
            pdfkeywords={\paperkeywords},
            pdfcopyright={Copyright (C) \papercopyright},
            pdflicenseurl={\paperlicenceurl}]{hyperref}

\usepackage[bottom,flushmargin,hang,multiple]{footmisc}

\usepackage[all]{hypcap} 

\input{lhcb-symbols-def} 

\usepackage{cite} 
\usepackage{mciteplus}

%% file: lhcb-symbols-def.tex
\usepackage{xspace} 
\usepackage{upgreek}

\def\lhcb   {\mbox{LHCb}\xspace}

\def\MagUp {\mbox{\em Mag\kern -0.05em Up}\xspace}

\ifthenelse{\boolean{uprightparticles}}%
{

 \def\Pmu         {\ensuremath{\upmu}\xspace}

 \def\Ppsi        {\ensuremath{\uppsi}\xspace}

 \def\PDelta      {\ensuremath{\Delta}\xspace}                 
 \def\PXi         {\ensuremath{\Xi}\xspace}                 
 \def\PLambda     {\ensuremath{\Lambda}\xspace}                 
 \def\PSigma      {\ensuremath{\Sigma}\xspace}                 
 \def\POmega      {\ensuremath{\Omega}\xspace}                 
 \def\PUpsilon    {\ensuremath{\Upsilon}\xspace}
 \let\oldPi\Pi
 \def\PPi         {\ensuremath{\oldPi}\xspace}

 \def\PB      {\ensuremath{\mathrm{B}}\xspace}                 
                  
 \def\PD      {\ensuremath{\mathrm{D}}\xspace}

 \def\PJ      {\ensuremath{\mathrm{J}}\xspace}                 
 \def\PK      {\ensuremath{\mathrm{K}}\xspace}

 \def\Pb      {\ensuremath{\mathrm{b}}\xspace}                 
 \def\Pc      {\ensuremath{\mathrm{c}}\xspace}                 
                  
 \def\Pe      {\ensuremath{\mathrm{e}}\xspace}

 \def\Pi      {\ensuremath{\mathrm{i}}\xspace}

 \def\Ps      {\ensuremath{\mathrm{s}}\xspace}

 \def\thebaroffset{0.0em}
}
{

 \def\Pmu         {\ensuremath{\mu}\xspace}

 \def\Ppsi        {\ensuremath{\psi}\xspace}                 
                  
 \mathchardef\PDelta="7101
 \mathchardef\PXi="7104
 \mathchardef\PLambda="7103
 \mathchardef\PSigma="7106
 \mathchardef\POmega="710A
 \mathchardef\PUpsilon="7107
 \mathchardef\PPi="7105
                  
 \def\PB      {\ensuremath{B}\xspace}                 
                  
 \def\PD      {\ensuremath{D}\xspace}

 \def\PJ      {\ensuremath{J}\xspace}                 
 \def\PK      {\ensuremath{K}\xspace}

 \def\Pb      {\ensuremath{b}\xspace}                 
 \def\Pc      {\ensuremath{c}\xspace}                 
                  
 \def\Pe      {\ensuremath{e}\xspace}

 \def\Pi      {\ensuremath{i}\xspace}

 \def\Ps      {\ensuremath{s}\xspace}

 \def\thebaroffset{0.18em}
}
\newcommand{\offsetoverline}[2][\thebaroffset]{\kern #1\overline{\kern -#1 #2}}%

\makeatletter
\ifcase \@ptsize \relax
  \newcommand{\miniscule}{\@setfontsize\miniscule{4}{5}}
\or
  \newcommand{\miniscule}{\@setfontsize\miniscule{5}{6}}
\or
  \newcommand{\miniscule}{\@setfontsize\miniscule{5}{6}}
\fi
\makeatother

\DeclareRobustCommand{\optbar}[1]{\shortstack{{\miniscule (\rule[.5ex]{1.25em}{.18mm})}
  \\ [-.7ex] $#1$}}

\def\ep         {{\ensuremath{\Pe^+}}\xspace}

\def\mup        {{\ensuremath{\Pmu^+}}\xspace}

\def\squark    {{\ensuremath{\Ps}}\xspace}

\def\cquark    {{\ensuremath{\Pc}}\xspace}

\def\bquark    {{\ensuremath{\Pb}}\xspace}

\def\KorKbar {\kern \thebaroffset\optbar{\kern -\thebaroffset \PK}{}\xspace}

\def\D       {{\ensuremath{\PD}}\xspace}

\def\DorDbar {\kern \thebaroffset\optbar{\kern -\thebaroffset \PD}\xspace}

\def\Dp      {{\ensuremath{\D^+}}\xspace}
\def\Dm      {{\ensuremath{\D^-}}\xspace}

\def\DpDm    {\ensuremath{\Dp {\kern -0.16em \Dm}}\xspace}

\def\B       {{\ensuremath{\PB}}\xspace}

\def\BorBbar {\kern \thebaroffset\optbar{\kern -\thebaroffset \PB}\xspace}

\def\Bd      {{\ensuremath{\B^0}}\xspace}

\def\BdorBdbar {\kern \thebaroffset\optbar{\kern -\thebaroffset \Bd}\xspace}

\def\Bs      {{\ensuremath{\B^0_\squark}}\xspace}

\def\BsorBsbar {\kern \thebaroffset\optbar{\kern -\thebaroffset \Bs}\xspace}

\def\jpsi     {{\ensuremath{{\PJ\mskip -3mu/\mskip -2mu\Ppsi}}}\xspace}
\def\psitwos  {{\ensuremath{\Ppsi{(2S)}}}\xspace}

\def\Y#1S{\ensuremath{\PUpsilon{(#1S)}}\xspace}

\def\LorLbar     {\kern \thebaroffset\optbar{\kern -\thebaroffset \PLambda}\xspace}

\def\to                 {\ensuremath{\rightarrow}\xspace}

\def\AT#1     {\ensuremath{A_{\mathrm{T}}^{#1}}\xspace}           

\def\C#1      {\ensuremath{\mathcal{C}_{#1}}\xspace}                       
\def\Cp#1     {\ensuremath{\mathcal{C}_{#1}^{'}}\xspace}                    
\def\Ceff#1   {\ensuremath{\mathcal{C}_{#1}^{\mathrm{(eff)}}}\xspace}        
\def\Cpeff#1  {\ensuremath{\mathcal{C}_{#1}^{'\mathrm{(eff)}}}\xspace}       
\def\Ope#1    {\ensuremath{\mathcal{O}_{#1}}\xspace}                       
\def\Opep#1   {\ensuremath{\mathcal{O}_{#1}^{'}}\xspace}                    

\newcommand{\nospaceunit}[1]{\ensuremath{\text{#1}}}       
\newcommand{\aunit}[1]{\ensuremath{\text{\,#1}}}       

\newcommand{\tev}{\aunit{Te\kern -0.1em V}\xspace}
\newcommand{\gev}{\aunit{Ge\kern -0.1em V}\xspace}
\newcommand{\mev}{\aunit{Me\kern -0.1em V}\xspace}
\newcommand{\kev}{\aunit{ke\kern -0.1em V}\xspace}
\newcommand{\ev}{\aunit{e\kern -0.1em V}\xspace}
 
\newcommand{\mevc}{\ensuremath{\aunit{Me\kern -0.1em V\!/}c}\xspace}
\newcommand{\gevc}{\ensuremath{\aunit{Ge\kern -0.1em V\!/}c}\xspace}
\newcommand{\mevcc}{\ensuremath{\aunit{Me\kern -0.1em V\!/}c^2}\xspace}
\newcommand{\gevcc}{\ensuremath{\aunit{Ge\kern -0.1em V\!/}c^2}\xspace}

\def\mum  {\ensuremath{\,\upmu\nospaceunit{m}}\xspace}

\def\nb {\aunit{nb}\xspace}
\def\invnb {\ensuremath{\nb^{-1}}\xspace}

\newcommand{\chisq}{\ensuremath{\chi^2}\xspace}

\newcommand{\chisqip}{\ensuremath{\chi^2_{\text{IP}}}\xspace}

\def\gsim{{~\raise.15em\hbox{$>$}\kern-.85em
          \lower.35em\hbox{$\sim$}~}\xspace}
\def\lsim{{~\raise.15em\hbox{$<$}\kern-.85em
          \lower.35em\hbox{$\sim$}~}\xspace}

\def\sqsnn {\ensuremath{\protect\sqrt{s_{\scriptscriptstyle\text{NN}}}}\xspace}
\def\pt         {\ensuremath{p_{\mathrm{T}}}\xspace}

\def\evtgen     {\mbox{\textsc{EvtGen}}\xspace}

\def\geant      {\mbox{\textsc{Geant4}}\xspace}

\def\photos     {\mbox{\textsc{Photos}}\xspace}

\def\tell1  {TELL1\xspace}
\def\ukl1   {UKL1\xspace}

\newcommand{\ie}{\mbox{\itshape i.e.}\xspace}

\newcommand{\lhcborcid}[1]{\href{https://orcid.org/#1}{\hspace*{0.1em}\raisebox{-0.45ex}{\includegraphics[width=1em]{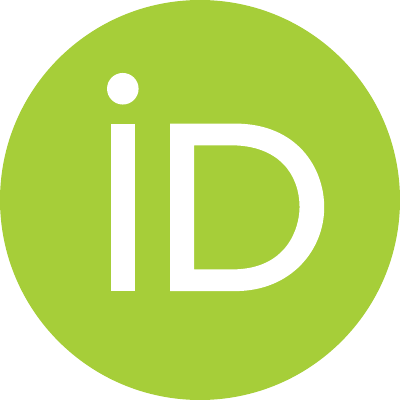}}}}


%% file: title-LHCb-PAPER.tex

\begin{titlepage}
\pagenumbering{roman}

\vspace*{-1.5cm}
\centerline{\large EUROPEAN ORGANIZATION FOR NUCLEAR RESEARCH (CERN)}
\vspace*{1.5cm}
\noindent
\begin{tabular*}{\linewidth}{lc@{\extracolsep{\fill}}r@{\extracolsep{0pt}}}
\ifthenelse{\boolean{pdflatex}}
{\vspace*{-1.5cm}\mbox{\!\!\!\includegraphics[width=.14\textwidth]{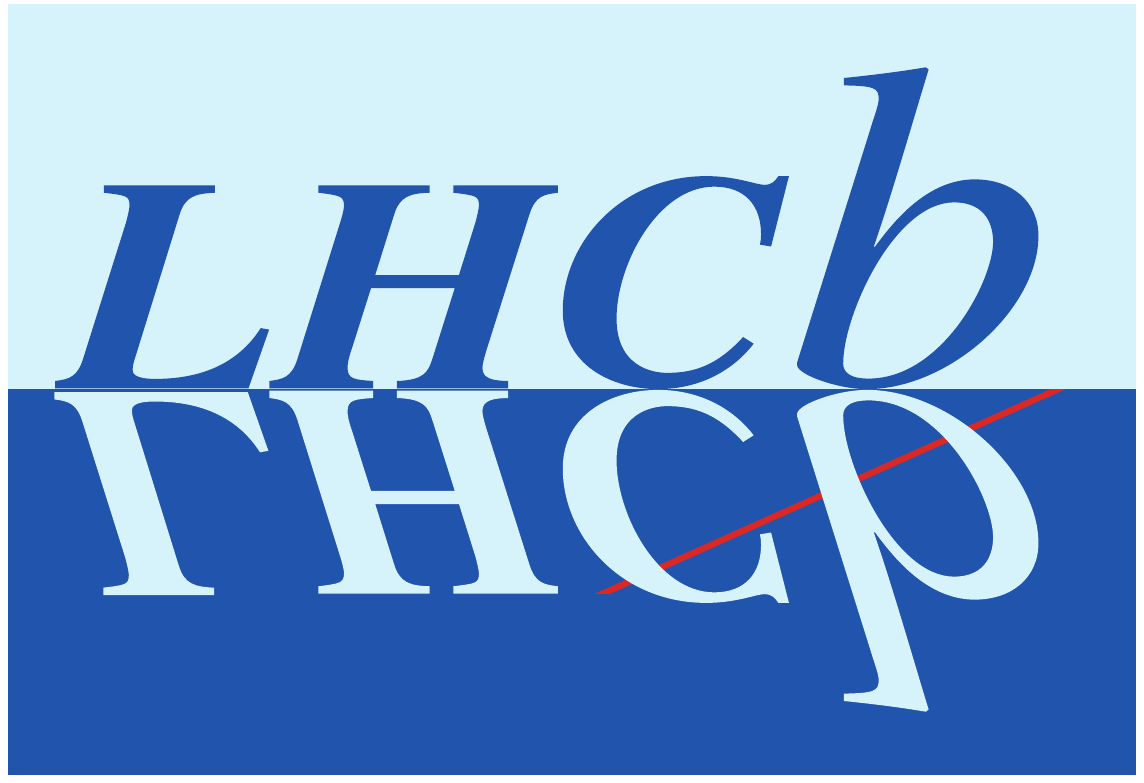}} & &}%
{\vspace*{-1.2cm}\mbox{\!\!\!\includegraphics[width=.12\textwidth]{figs/lhcb-logo.eps}} & &}%
\\
 & & CERN-EP-2023-293 \\
 & & LHCb-PAPER-2023-024 \\
 & & 22 April 2024 \\
 & & \\
\end{tabular*}

\vspace*{3.5cm}

{\normalfont\bfseries\boldmath\huge
\begin{center}
      \papertitle
\end{center}
}

\vspace*{2.0cm}

\begin{center}
\paperauthors\footnote{Authors are listed at the end of this paper.}
\end{center}

\vspace{\fill}

\begin{abstract}
  \noindent
  The production of \psitwos mesons in proton-lead collisions at a centre-of-mass energy 
  per nucleon pair of $\sqsnn=8.16\,{\rm TeV}$ is studied with the \lhcb detector using data corresponding to an integrated luminosity of 34\invnb. 
  The prompt and nonprompt \psitwos production cross-sections and the ratio of the \psitwos to \jpsi cross-section
  are measured as a function of the meson transverse momentum and rapidity in the nucleon-nucleon centre-of-mass frame, together
  with forward-to-backward ratios and nuclear modification factors. The production of prompt \psitwos is observed 
  to be more suppressed compared to $pp$ collisions than the prompt \jpsi production, while the nonprompt productions have similar suppression 
  factors. 
\end{abstract}

\vspace*{2.0cm}

\begin{center}
  Published in JHEP 04 (2024) 111
\end{center}

\vspace{\fill}

{\footnotesize 
\centerline{\copyright~\papercopyright. \href{\paperlicenceurl}{\paperlicence}.}}
\vspace*{2mm}

\end{titlepage}


\newpage
\setcounter{page}{2}
\mbox{~}

%% file: intro.tex
\section{Introduction}
\label{sec:intro}

The study of quark-gluon plasma (QGP) requires a broad range of observables.
Amongst many potential probes likely to be affected by the presence of a colour-deconfined medium, quarkonium states such as \jpsi and \psitwos were 
first proposed by Matsui and Satz in 1986~\cite{Matsui:1986dk}.
The charm and anticharm quark pairs produced in hard scattering processes are expected to be dissociated by colour screening in the presence of QGP, 
which would lead to a suppression of \jpsi and \psitwos production in heavy-ion collisions relative to production in proton-proton collisions.
The theoretical understanding of the quarkonium bound-state dynamics has improved significantly over the past 30 years~\cite{Mocsy:2013syh}, 
with the development of lattice quantum chromodynamics (QCD) and effective field theory approaches, and experimental measurements at the SPS, RHIC and LHC accelerators reveal patterns indicative of deconfinement~\cite{Andronic:2015wma}. More specifically, a low transverse-momentum contribution to \jpsi production was observed at the LHC~\cite{Abelev:2012rv,Abelev:2013ila,Adam:2015rba,Adam:2015isa,Adam:2016rdg},
which has been interpreted as a signature of charmonium production originating from recombination of unbound charm
quarks generated either during the lifetime of
the deconfined medium~\cite{Thews:2000rj} or at the phase boundary~\cite{BraunMunzinger:2000px}.

In this context, the measurement of \psitwos meson production in heavy-nucleus collisions, and its comparison with \jpsi production,
 plays a crucial role in the interpretation of charmonium measurements.
The \psitwos state has a larger spatial extension and a smaller binding energy than the \jpsi state~\cite{Satz:2005hx}.
In addition, its production does not contain feed-down contributions
from decays of $\chi_c$ states, in contrast to the \jpsi case.
Uncertainties in the predicted overall charm cross-section
cancel out in the ratio of \psitwos to \jpsi production in nucleus-nucleus collisions.
The charmonium production ratio constitutes an observable that can
constrain or discriminate between theoretical models~\cite{Abelevetal:2014cna}.
First measurements of the \psitwos to \jpsi production ratio in nucleus-nucleus collisions at the LHC
have been performed by the CMS~\cite{CMS:2014vjg,CMS:2017uuv,CMS:2016wgo}, ALICE~\cite{Adam:2015isa,ALICE:2022jeh} and ATLAS~\cite{ATLAS:2018hqe} collaborations, at center of mass per nucleon pair energies of \sqsnn~=~2.76\tev and 5.20\tev.

Effects that are not related to deconfinement,
known as cold nuclear matter (CNM) effects,
play an important role in the interpretation of the available 
data from ultra-relativistic nucleus-nucleus collisions. An accurate quantification of the CNM effects is necessary to distinguish them from deconfinement-related phenomena.
The size of the CNM effects can be measured in proton--nucleus or deuteron--nucleus collisions,
which have been studied at various collision energies~\cite{Andronic:2015wma}.
At the LHC collision energies, the main modifications of
charmonium production compared to $pp$ collisions are related to the modification of the gluon flux.
This is treated using collinear factorisation with nuclear parton distribution
functions (nPDFs)~\cite{deFlorian:2011fp,Kovarik:2015cma,Eskola:2021nhw,AbdulKhalek:2022fyi,Duwentaster:2022kpv}
or using the colour glass condensate (CGC) approach to describe
the saturation regime of QCD~\cite{Gelis:2010nm,Fujii:2006ab}.
Furthermore, small-angle gluon radiation arising from the interference
between initial and final-state radiation, called coherent energy loss,
was proposed as one of the dominant causes of nuclear modification in
quarkonium production in proton--nucleus collisions~\cite{Arleo:2012rs}.
Calculations based on these
approaches~\cite{Vogt:2004dh,Vogt:2010aa,Lansberg:2016deg, Ma:2015sia,Ducloue:2015gfa, Arleo:2012rs}
are able to describe \jpsi production measurements at the
LHC~\cite{Abelev:2013yxa,LHCb-PAPER-2013-052,Adam:2015iga,Adam:2015jsa,Aad:2015ddl,Sirunyan:2017mzd,LHCb-PAPER-2017-014}
reasonably well within their respective uncertainties.
In these computations, the slightly different kinematics induced by the mass difference between the \jpsi and the \psitwos states and the feed-down originating from $\chi_c$ decays contributing to the \jpsi production are considered negligible compared to the uncertainties in the models themselves. The modification of quarkonium production in
these models affects the initial state only but not the 
\jpsi or \psitwos final states, that is to say identical 
modification of the production of these two mesons is predicted. 
In contrast to these phenomenological expectations, indications of a relative suppression
of \psitwos meson production with respect to the production of the \jpsi meson
were observed by the PHENIX collaboration at RHIC~\cite{Adare:2013ezl,Adare:2016psx}
in proton--nucleus or deuteron--nucleus collisions at \sqsnn=200\gev and by the ALICE~\cite{Abelev:2014zpa,Adam:2016ohd,ALICE:2020vjy,ALICE:2020tsj}, CMS~\cite{CMS:2018gbb} and LHCb~\cite{LHCb-PAPER-2015-058} collaborations at the LHC in 
proton--nucleus collisions at \mbox{\sqsnn=5.02\tev} and 8.16\tev.

This study aims to improve the understanding of the relative production of \psitwos and \jpsi  
 in proton--nucleus compared to $pp$ collisions.
 The similar previous measurement by the LHCb collaboration~\cite{LHCb-PAPER-2015-058} did not have sufficient 
 precision to draw conclusions on the relative suppression between the two states. 
The analysis separates prompt and nonprompt production, where the former consists of mesons produced directly in the collisions or in decays of higher-mass promptly-produced charmonium states, and the latter consists of mesons produced in decays of $b$-hadrons.
Comparisons with models including factorisation breaking via hadronic or partonic interactions influencing the fate of the $c\bar{c}$ pair
after its creation~\cite{Ferreiro:2014bia,Ma:2017rsu} are provided.

%% file: detector.tex
\section{Detector, data sample and observables}
\label{sec:Detector}

The \lhcb detector~\cite{Alves:2008zz,LHCb-DP-2014-002} is a single-arm forward
spectrometer covering the \mbox{pseudorapidity} range $2<\eta <5$,
designed for the study of particles containing \bquark or \cquark
quarks. The detector includes a high-precision tracking system
consisting of a silicon-strip vertex detector surrounding the
interaction region~\cite{LHCb-DP-2014-001}, a large-area silicon-strip detector located
upstream of a dipole magnet with a bending power of about
$4{\mathrm{\,Tm}}$, and three stations of silicon-strip detectors and straw
drift tubes~\cite{LHCb-DP-2013-003} placed downstream of the magnet.
The tracking system provides a measurement of the momentum of charged particles with
a relative uncertainty that varies from 0.5\% at low momentum to 1.0\% at 200\gevc.
The minimum distance of a track to a primary vertex (PV), the impact parameter,
is measured with a resolution of $(15+29/\pt)\mum$,
where \pt is the transverse momentum in the LHCb frame, in\,\gevc.
Different types of charged hadrons are distinguished using information
from two ring-imaging Cherenkov detectors~\cite{LHCb-DP-2012-003}.
Photons, electrons and hadrons are identified by a calorimeter system consisting of
scintillating-pad and preshower detectors, an electromagnetic
calorimeter and a hadronic calorimeter. Muons are identified by a
system composed of alternating layers of iron and multiwire
proportional chambers~\cite{LHCb-DP-2012-002}.

This analysis is based on data acquired during the 2016 LHC heavy-ion run with collisions of protons and
$^{208}$Pb ions at a centre-of-mass energy per nucleon pair of \mbox{$\sqrt{s_{\rm NN}}=8.16\,{\rm TeV}$}. During the run, the direction of the ion and 
proton beams were exchanged in the accelerator and data recorded with these two
different configurations. 
Since the energy per nucleon in the proton beam is larger than in the lead beam,
the nucleon-nucleon centre-of-mass system has a rapidity in the laboratory frame of 0.465 ($-0.465$),
when the proton (lead) beam travels from the vertex detector towards the muon chambers.
Consequently, the LHCb detector covers two different acceptance regions,
\begin{itemize}
\item $1.5<y^*<4.0$ when the proton beam travels from the vertex detector towards the muon chambers, denoted $p$Pb, and
\item $-5.0<y^*<-2.5$ when the proton beam travels from the muon chambers towards the vertex detector, denoted Pb$p$,
\end{itemize}
where $y^*$ is the rapidity in the centre-of-mass frame of the colliding nucleons with respect to the proton beam direction.
The data samples correspond to an integrated luminosity of $13.6\pm 0.3\invnb$ of $p$Pb collisions and $20.8\pm0.5\invnb$ of Pb$p$ collisions~\cite{LHCb-PAPER-2014-047}. 
The instantaneous luminosity while these samples were recorded ranged between 0.5 and $1.0\times 10^{29}\,{\rm cm}^{-2}{\rm s}^{-1}$, where the average number of collisions per bunch crossing is less than 0.1.

\section{Definition of the observables}

The differential prompt and nonprompt production cross-sections for \psitwos mesons are measured in the range $0<\pt<14\gevc$ and $1.5<y^*<4.0$ for $p$Pb collisions and $-5.0<y^*<-2.5$ for Pb$p$ collisions, in bins of \pt and $y^*$. 
The \psitwos to \jpsi cross-section ratio 
\begin{equation}\label{eq:ratio}
  \frac{\text{d}^2 \sigma_{\psitwos} /\text{d} \pt \text{d}y^*}{\text{d}^2 \sigma_{\jpsi} / \text{d}\pt\text{d}y^*      } =
  \frac{N(\psitwos)}{N(\jpsi)} \times \frac{\epsilon^{\rm tot}(\jpsi)}{\epsilon^{\rm tot}(\psitwos)}\times
  \frac{{\cal B}\left(\jpsi\to\mup\mu^-\right)}{{\cal B}\left(\psitwos\to\ep e^-\right)}
\end{equation}
is obtained from the ratio of measured \psitwos to \jpsi yields in the same dataset, 
$N(\psitwos)$ and 
$N(\jpsi)$, reconstructed in the $\mup\mu^-$ final state, corrected for acceptance and detection efficiencies, $\epsilon^{\rm tot}(\psitwos)$ and $\epsilon^{\rm tot}(\jpsi)$, and for the branching fractions
into $\mu^+\mu^-$, \mbox{${\cal B}\left(\psitwos\to\ep e^-\right)=(7.93\pm0.17)\times 10^{-3}$} and 
\mbox{${\cal B}\left(\jpsi\to\mup\mu^-\right)= (5.961 \pm 0.033)\times 10^{-2}$}~\cite{PDG2022}.
The branching fraction of the $\psitwos$ decay into two electrons is more precisely known than for the decay into two muons, and is therefore used with the assumption of lepton universality, neglecting the mass difference between muons and electrons. No 
systematic uncertainty is assigned for this assumption. 

The absolute \psitwos production cross-section is then derived from this ratio
multiplied by the \jpsi production cross-section measured with less restrictive selection criteria on the muon particle identification, published in Ref.~\cite{LHCb-PAPER-2017-014},
\begin{equation}
  \frac{\text{d}^2 \sigma_{\psitwos}}{\text{d}\pt \text{d}y^*} = \frac{\text{d}^2 \sigma_{\psitwos} /\text{d} \pt \text{d}y^*}{\text{d}^2 \sigma_{\jpsi} / \text{d}\pt\text{d}y^*} \cdot \left[\frac{\text{d}^2 \sigma_{\jpsi}}{\text{d}\pt \text{d}y^*}\right]_{\rm pub}.
\end{equation}
Nuclear effects are quantified using the nuclear modification factor, $R_{p{\rm Pb}}$, 
\begin{equation}\label{eq:rpa}
R_{p{\rm Pb}} (\pt,y^*) \equiv \frac{1}{A} \frac{{\rm d}^2 \sigma_{p{\rm Pb}}(\pt,y^*)/{\rm d}\pt{\rm d}y^*}{{\rm d}^2\sigma_{pp}(\pt,y^*)/{\rm d}\pt{\rm d}y^*},
\end{equation}
where
$A=208$ is the mass number of the Pb ion,
${\rm d}^2 \sigma_{p{\rm Pb}}(\pt,y^*)/{\rm d}\pt{\rm d}y^*$ is the production cross-section in $p$Pb or Pb$p$ collisions and
${\rm d}^2 \sigma_{pp}(\pt,y^*)/{\rm d}\pt{\rm d}y^*$ the reference production cross-section in $pp$ collisions at the same 
centre-of-mass energy. In the absence of nuclear effects, the nuclear modification factor is equal to unity.

Similarly to the cross-section, the \psitwos nuclear modification factor is derived from corrected yield ratios and the published \jpsi modification factor~\cite{LHCb-PAPER-2018-049},
\begin{equation}
R^{\psitwos}_{p{\rm Pb}} (\pt,y^*) =
 \frac{\left[ \frac{{\rm d}^2 \sigma_{\psitwos} / {\rm d} \pt {\rm d}y^*}{{\rm d}^2 \sigma_{\jpsi} / {\rm d}\pt {\rm d}y^*} \right]_{p{\rm Pb}}}{
 \left[ \frac{{\rm d}^2 \sigma_{\psitwos} / {\rm d} \pt {\rm d}y^*}{{\rm d}^2 \sigma_{\jpsi} / {\rm d}\pt {\rm d}y^*} \right]_{pp}} \cdot
   \left[R^{\jpsi}_{p{\rm Pb}} (\pt,y^*)\right]_{\rm pub}.
\end{equation}
The nuclear modification factor in Pb$p$ collisions, $R^{\psitwos}_{{\rm Pb}p} (\pt,y^*)$ is
computed analogously. 
The reference \psitwos to \jpsi cross-section ratio in $pp$ collisions 
is obtained from measurements at 7\tev by the LHCb 
collaboration~\cite{LHCb-PAPER-2018-049} assuming that the 
ratio between \psitwos and \jpsi production in a given \pt bin is 
the same at 7\tev and at 8.16\tev and is independent of the charmonium-state rapidity in the LHCb acceptance. 
This assumption is consistent with the results of the present analysis and with previous measurements~\cite{LHCb-PAPER-2018-049}. No systematic uncertainty is assigned for this assumption. 

The forward-to-backward ratios compare production in 
$p$Pb and Pb$p$ collisions in the common acceptance range
$2.5<|y^*|<4.0$, without the need to know the reference 
cross-section in $pp$ collisions. It is defined as
\begin{equation}
R_{\rm FB}(\pt,y^*) = \frac{\left[\frac{{\rm d}\sigma(\pt,y^*)}{{\rm d}\pt}\right]_{p{\rm Pb}}}{\left[\frac{{\rm d}\sigma(\pt,-y^*)}{{\rm d}\pt}\right]_{{\rm Pb}p}}. 
\end{equation}

%% file: selection.tex

\section{Event selection and cross-section determination}
\label{sec:selection}

\subsection{Selection}

An online event selection is performed by a trigger system consisting
of a hardware stage, which selects events containing at least one muon with \pt larger than 500\mevc,
followed by a software stage. In the first stage of the software trigger, two muon
tracks with $\pt>500\mevc$ are required to form a \jpsi or \psitwos candidate with invariant mass 
$M_{\mu^+\mu^-}>2.5 \gevcc$.  In the second stage, \jpsi and \psitwos candidates with an
invariant mass within 120\mevcc of the known value of the \jpsi or \psitwos mass~\cite{PDG2022} are selected. 

At the analysis stage,  each event is required to have at least one PV 
reconstructed from at least four tracks measured in the vertex detector. For events with multiple PVs, the PV that has the smallest \chisqip 
with respect to the \jpsi or the \psitwos candidate is chosen. Here, \chisqip is defined as the difference between the vertex-fit \chisq calculated with the \jpsi or \psitwos meson candidate included in or excluded from the PV fit. Each identified muon track is required to be in the pseudorapidity range $2<\eta<5$, have a good-quality track fit, and have $\pt>750\mevc$ and $\pt>1000\mevc$ 
for \jpsi and \psitwos candidates, respectively. The stricter selection for \psitwos is due to a larger combinatorial background. The two
muon tracks of the \jpsi or \psitwos candidate must form a good-quality vertex. 

\subsection{Determination of signal yields}

The reconstructed vertex of the \jpsi or \psitwos mesons originating from $b$-hadron decays tends to be separated from the PV. 
This property is used to distinguish between prompt and nonprompt \jpsi or \psitwos mesons by exploiting the pseudoproper time defined as 
\begin{equation}
t_{z} \equiv \frac{(z-z_{\rm PV})\times M}{ p_{z}},
\end{equation}
where $z$ and $z_{\rm PV}$ are the coordinates along the beam axis of the \jpsi or \psitwos decay vertex and  PV
positions,  $p_z$ is the $z$ component of the \jpsi or \psitwos momentum and $M$ the known \jpsi or \psitwos mass.  
The yields of \jpsi and \psitwos signals, for the prompt and nonprompt categories,
are determined from a two-dimensional unbinned
maximum-likelihood fit to their invariant-mass and pseudoproper-time distributions, performed independently for each $(\pt, y^*)$ bin.

\begin{figure}[!b]
\includegraphics[width=0.495\textwidth]{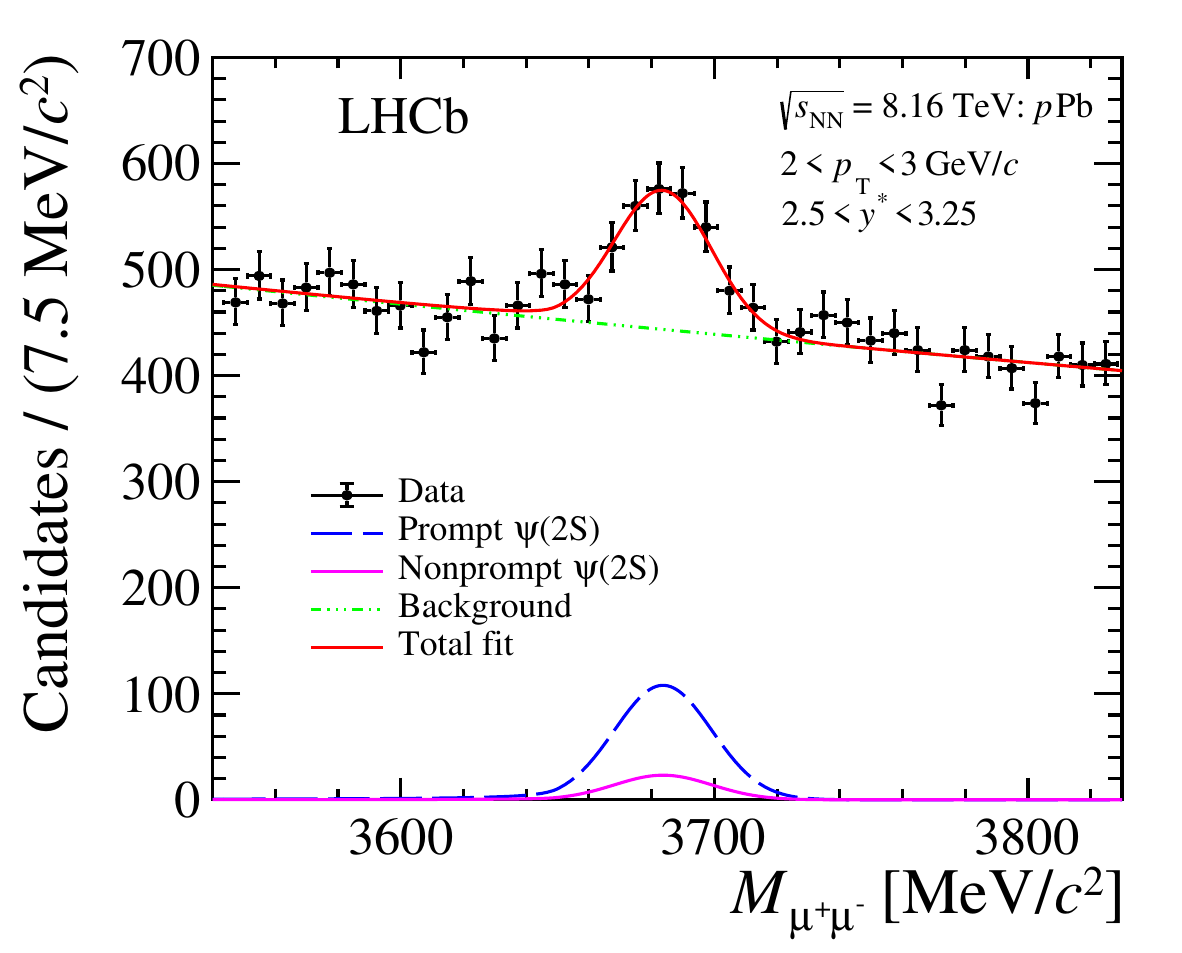}
\includegraphics[width=0.495\textwidth]{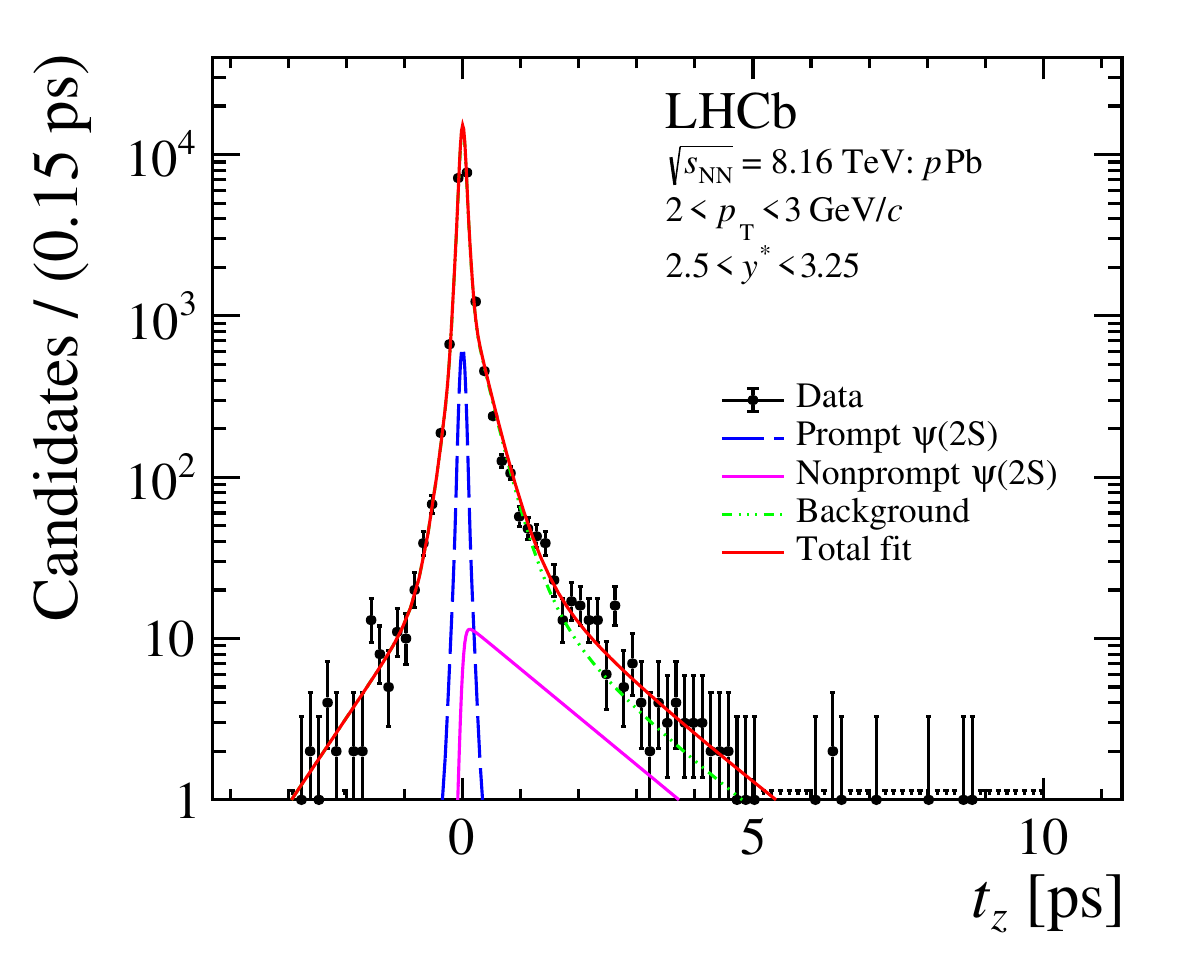}
\includegraphics[width=0.495\textwidth]{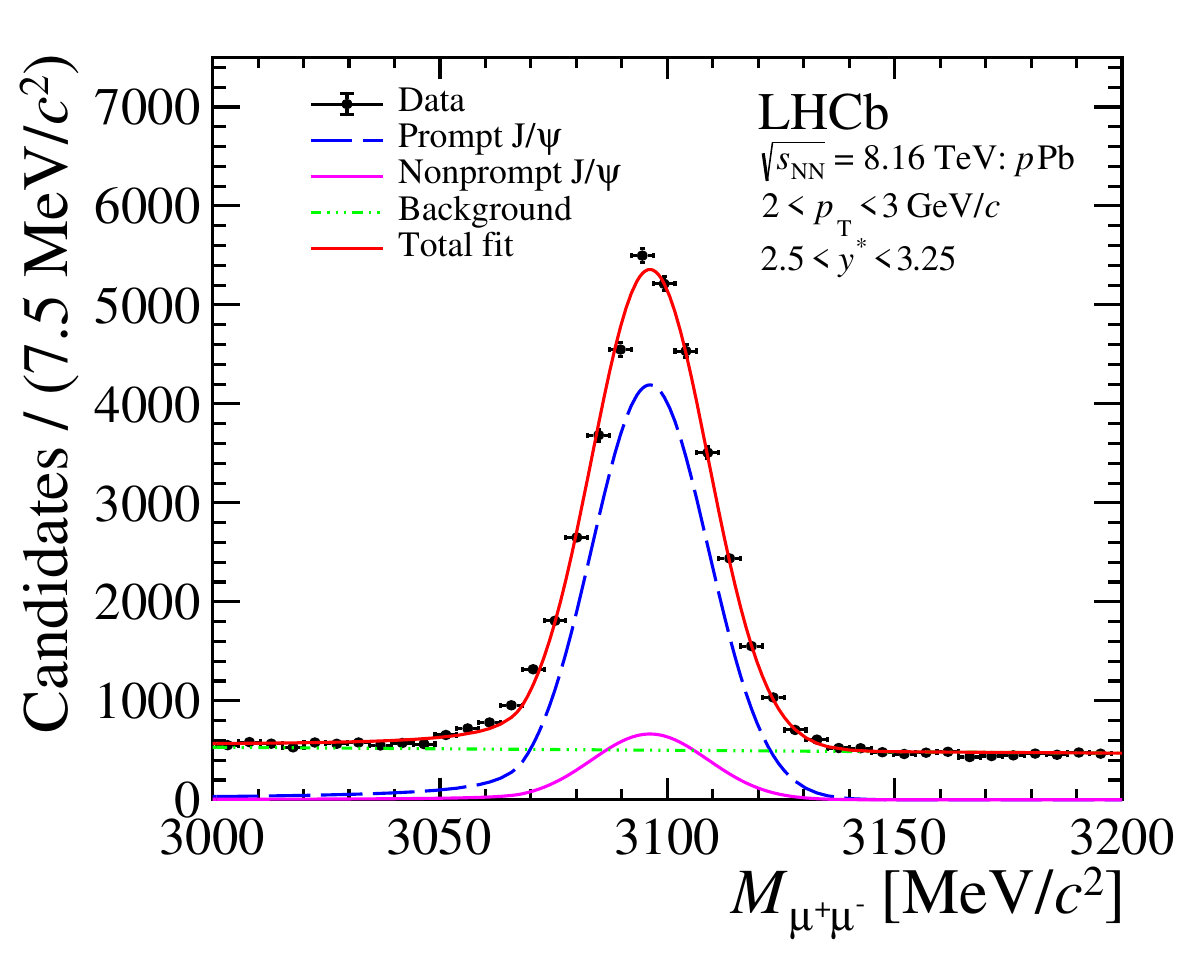}
\includegraphics[width=0.495\textwidth]{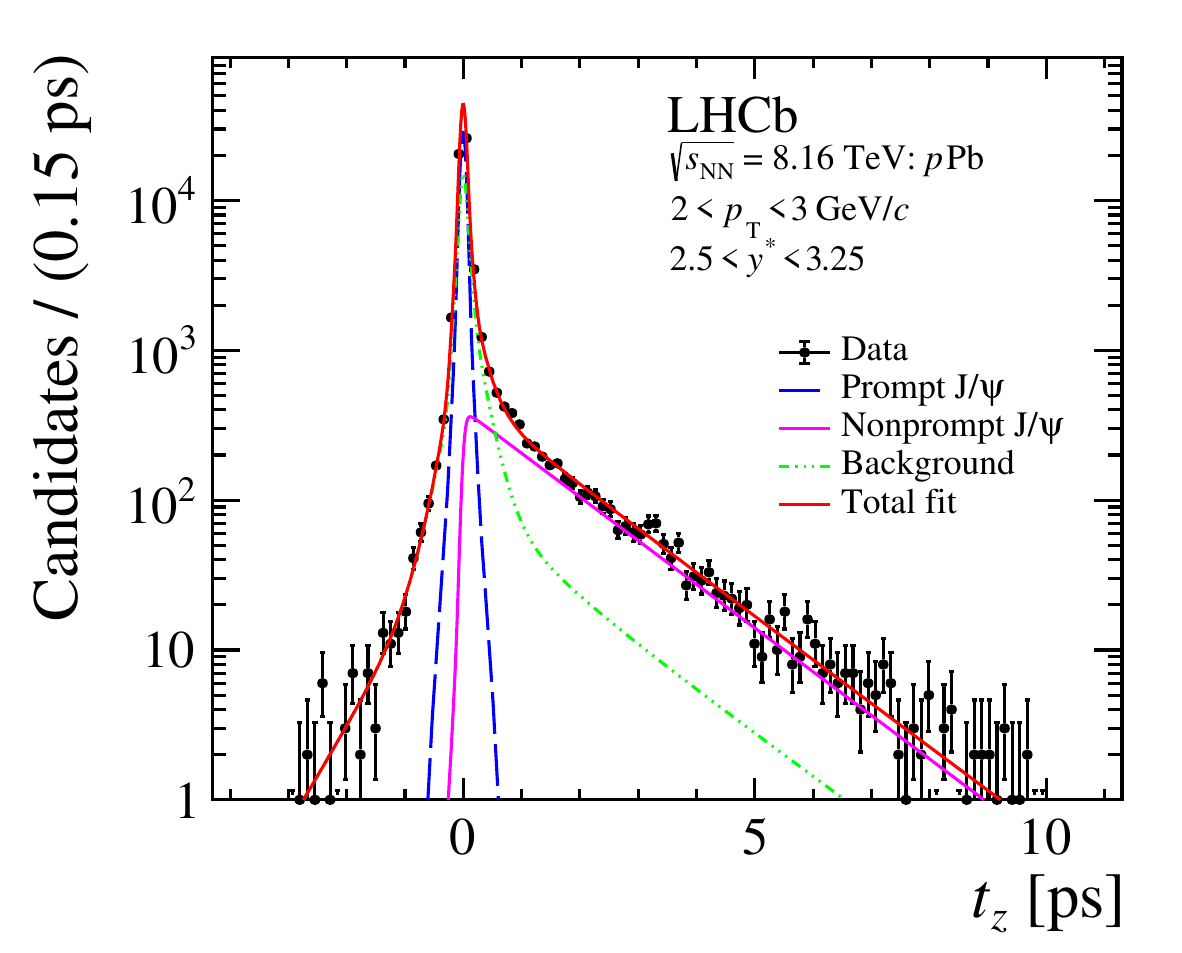}
\caption{Distributions of (left) invariant mass and (right) pseudoproper time for (upper) \psitwos and (lower)  \jpsi candidates in the bin $2<\pt<3\gevc$ and
 $2.5<y^*<3.25$. The data are overlaid with the fit results.}
\label{fig:masstzfit}
\end{figure}

The invariant-mass distribution consists of two components: the \jpsi or \psitwos signal and the 
background, which is only of combinatorial origin. In the fit function, the signal is described by a Crystal Ball function~\cite{Skwarnicki:1986xj}, 
and the combinatorial background by an exponential function. The $t_z$ distribution of prompt \jpsi or \psitwos is described by a Dirac  $\delta$-function $\delta(t_z)$, and that of nonprompt mesons by an exponential function. Both functions are convolved with a triple-Gaussian function to take into account the pseudoproper-time resolution. The background $t_z$ distribution is described by an empirical function derived from the shape 
observed in the region $3200<M_{\mu^+\mu^-}<3250\mevcc$. This background is composed of muons from semileptonic decays of $b$ and $c$ hadrons and
from decays of pions and kaons in the detector. The distribution is parameterised as a sum of a Dirac $\delta$-function and of five exponential functions, 
three for positive $t_z$ values and two for negative $t_z$ values, convolved with the sum of two 
Gaussian functions.
An example of the \psitwos and \jpsi invariant mass and the pseudoproper-time distributions for one $(\pt,y^*)$ bin is shown in Fig.~\ref{fig:masstzfit} 
for the $p$Pb sample, where the one-dimensional projections of the fit result are drawn on the distributions. 

\subsection{Efficiencies}

The total efficiency, $\epsilon_{\rm tot}$, is the product of the geometrical acceptance and the efficiencies for 
charged-track reconstruction,
 particle identification, and candidate and trigger selections. Samples of simulated events are used to evaluate 
these efficiencies except for the charged-track reconstruction and particle identification efficiencies, which are determined from calibration data samples. 
Inelastic $p$Pb and Pb$p$ collisions are simulated by the \textsc{Epos} event generator, which is tuned to describe LHC data~\cite{Pierog:2013ria}. The $\psitwos\to\mu^+\mu^-$ and $\jpsi\to\mu^+\mu^-$ signal candidates are generated separately 
with the {\sc Pythia}8  generator~\cite{Sjostrand:2007gs} in $pp$ collisions with beams having momenta equal to the momenta per nucleon 
of the $p$ and Pb beams. These events are merged with the \textsc{Epos} collisions to build the samples out of which the
efficiencies are computed.  The decays of hadrons are simulated by \evtgen~\cite{Lange:2001uf}, in which final-state electromagnetic
radiation is modelled with \photos~\cite{Golonka:2005pn}. The interaction of the particles with the detector, and the detector response, are 
implemented using the \geant toolkit~\cite{Allison:2006ve, *Agostinelli:2002hh} as described in Ref.~\cite{LHCb-PROC-2011-006}. 

The charged-track reconstruction efficiency is first evaluated in simulation and is corrected per track using calibration samples. 
For this purpose, \jpsi candidates are formed with one fully reconstructed ``tag'' track and one ``probe'' track, reconstructed partially 
with a subset of the tracking sub-detectors and both tracks identified as muons~\cite{LHCB-DP-2013-002}, in data and in simulation. 
The tag-and-probe correction factors are extracted from $p$Pb and Pb$p$ data and \jpsi simulation samples, taking into account their dependence on the particle multiplicity. They are applied to the 
\psitwos and \jpsi simulation in order to obtain the track reconstruction efficiencies in the \psitwos and \jpsi \pt and $y^*$ bins of the analysis.

The muon identification efficiency is determined for each track in data with a tag-and-probe method taking into account the efficiency variation as 
a function of track momentum, pseudorapidity and detector occupancy. 
Calibration samples of \jpsi mesons are selected by applying a tight identification criterion on one of the muons  and no identification requirements on the 
second muon~\cite{PIDCALIB}.
However, the calibration samples collected in $p$Pb and Pb$p$ collisions are limited in size. The efficiency is thus evaluated 
using the calibration samples collected in $pp$ collisions, taking into account the differences in the detector occupancy between $pp$, $p$Pb and 
Pb$p$ collisions, which affects the muon identification performance. While the muon identification efficiency is observed to be robust against the variation of detector occupancies, the probability of hadron misidentification presents a stronger dependence on hit or track multiplicity. The \psitwos and \jpsi simulation samples are assigned per-candidate weights according to the
efficiencies determined per track in data. These weighted samples are used to compute the muon identification efficiency in bins of  \pt and $y^*$ of the \psitwos or 
\jpsi meson.

\begin{figure}[t]
\includegraphics[width=0.49\textwidth]{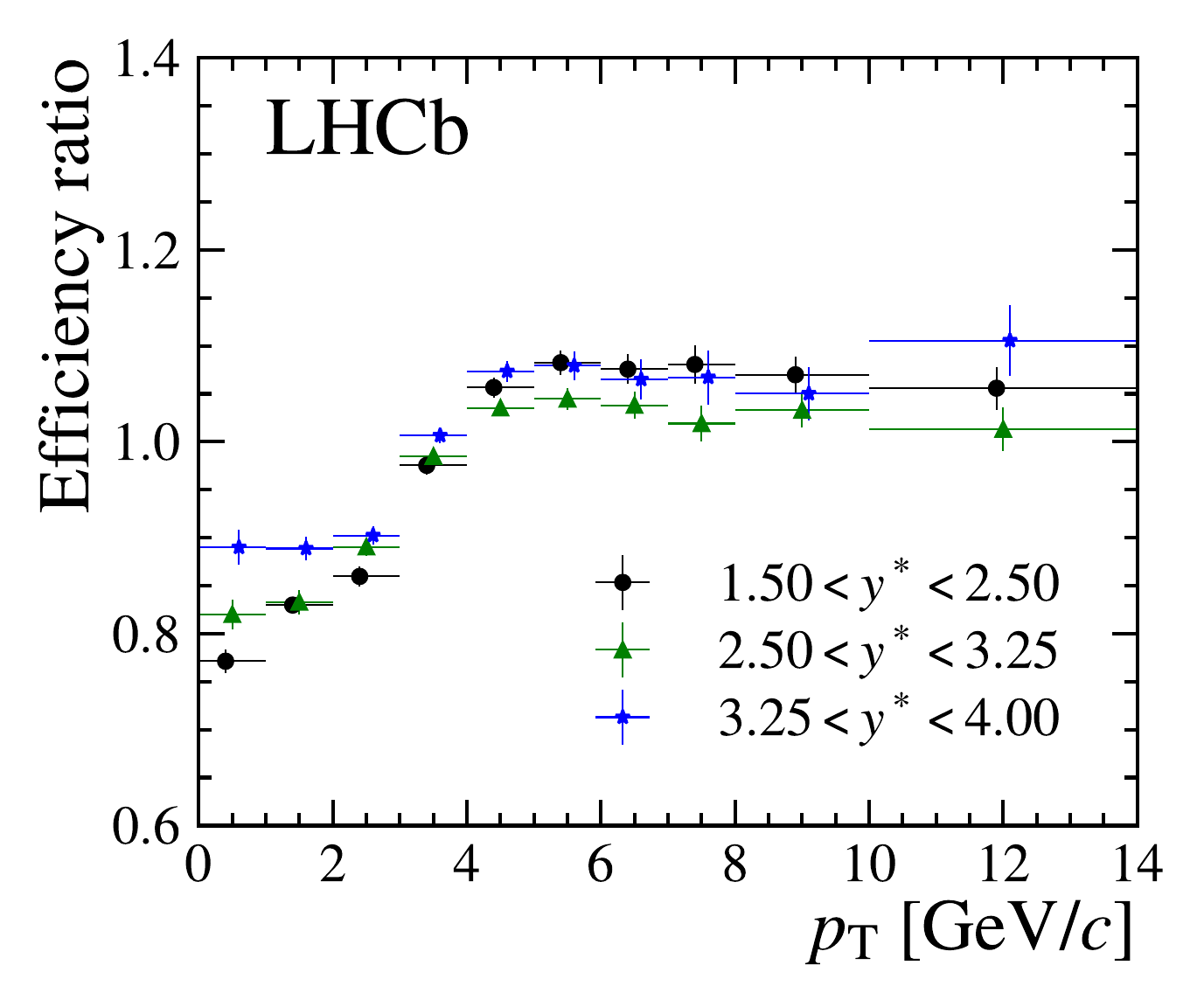}
\includegraphics[width=0.49\textwidth]{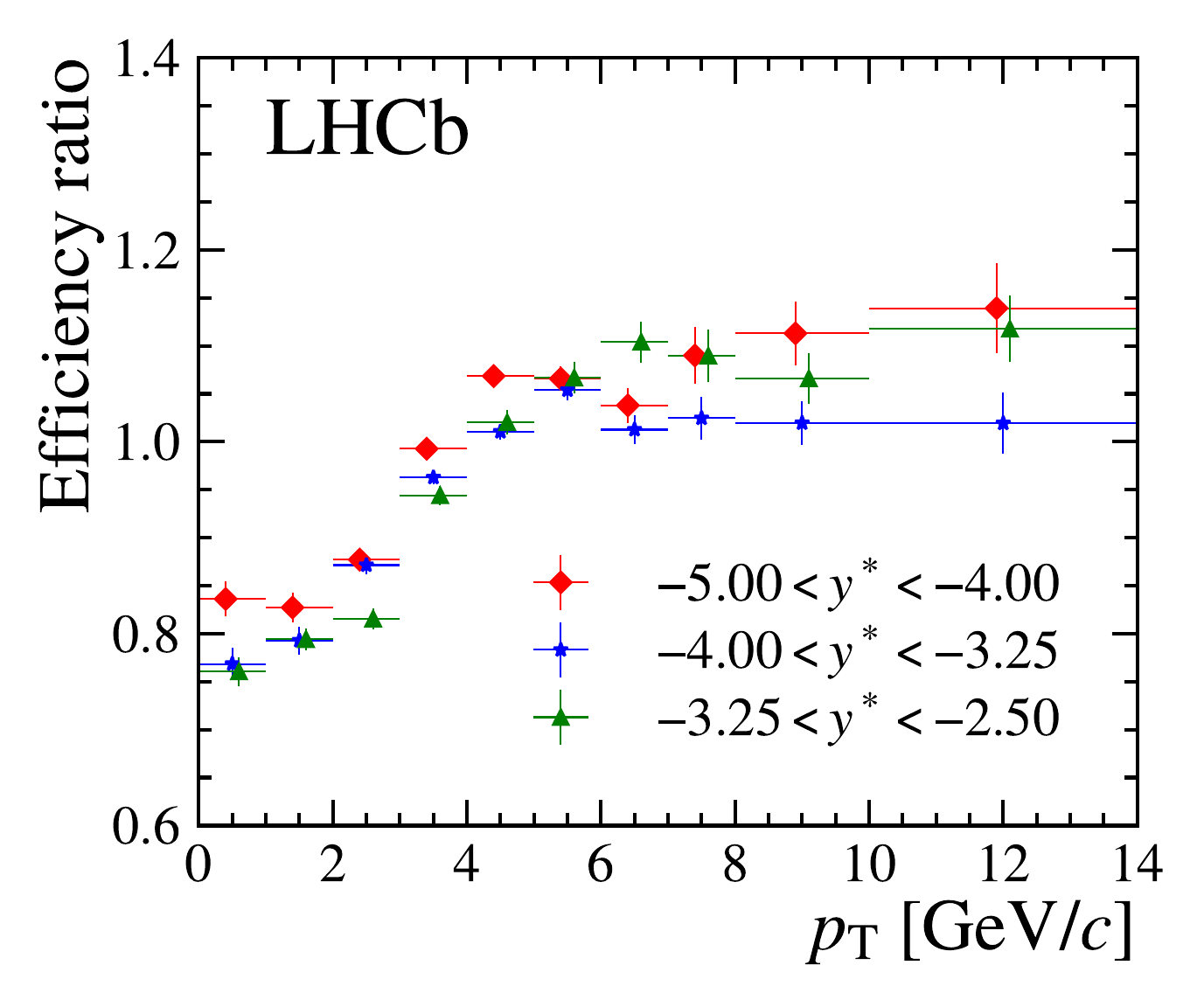}
\caption{Ratio of \jpsi to \psitwos total efficiencies as a function of the quarkonium \pt in different $y^*$ intervals for (left) $p$Pb and
(right) Pb$p$ collisions.
\label{fig:efftot}
}
\end{figure}

The total efficiencies are found to be consistent for the prompt and nonprompt quarkonia. 
The ratio of the \jpsi and \psitwos efficiencies is shown in Fig.~\ref{fig:efftot} in each analysis bin, for $p$Pb and Pb$p$ collisions. 
The uncertainties include both statistical uncertainties and the systematic uncertainties described in the following section. 

%% file: systematicunc.tex
\section{Systematic uncertainties}

The systematic uncertainties on the measurement of the cross-section ratio for prompt and nonprompt \psitwos to \jpsi production
are summarised in Table~\ref{tab:syst}.
The efficiencies depend on the \jpsi and \psitwos polarisation at production. Measurements by the ALICE and the LHCb collaborations
 in $pp$ collisions~\cite{ALICE:2018crw,LHCb-PAPER-2013-008,LHCb-PAPER-2013-067} indicate a polarisation close to zero in the kinematic 
 region covered in the analysis, with a negligible effect on the total detection efficiency. It is assumed that the \psitwos and \jpsi mesons are also produced with no polarisation in the $p$Pb and Pb$p$ collisions. No systematic uncertainty is assigned to this assumption.

\renewcommand{\arraystretch}{1.1}
\begin{table}[!b]
\caption{\small Summary of relative systematic uncertainties in $p$Pb and Pb$p$ collisions on the ratio of \psitwos to \jpsi cross-sections for prompt 
  and nonprompt production. Uncertainties that are computed bin-by-bin are expressed as ranges from the minimum
  to maximum values. All uncertainties are 
  assumed to be fully correlated between bins, apart
  from the uncertainty arising from the simulation
  sample size. Values are expressed as percentage.}
\label{tab:syst}
\centering
\begin{tabular}{@{}lrr@{}}\toprule
Source & $p$Pb & Pb$p$  \\ \midrule
Signal extraction & 2.2  & 2.2  \\
Particle identification & 0--1.7 &  0--2.0 \\
Efficiency extrapolation at high multiplicity & 5.0 &  5.0 \\
Tracking  & 0--0.1 & 0--0.1  \\
Hardware trigger & 0--1.1 & 0--1.1  \\
Particle multiplicity & 5.0 &  5.0 \\
Simulation sample size & 0.3--3.2 & 0.4--4.0 \\
$\frac{{\cal B}(\jpsi\to\mu^+\mu^-)}{{\cal B} (\psitwos\to\mu^+\mu^-)}$ (assuming lepton universality) & 2.2 & 2.2  \\
 \bottomrule
\end{tabular}
\end{table}

The \psitwos and \jpsi meson yields are affected by the choice of the modelling of the signal mass shape in the fit. An uncertainty is evaluated using an alternative fit model where the signal mass shape is described by the sum of a Crystal Ball function 
and a Gaussian function. The relative difference of the signal yields between the two fits is 2.2\%, 
which is assigned as a systematic uncertainty that is fully correlated between bins. The uncertainty associated with the choice of the shape of the $t_z$ distribution is negligible.
 
The uncertainty on the muon identification efficiency has several contributions. The size of the calibration sample affects the statistical precision 
of the efficiency determined with the tag-and-probe method described in the previous section. The impact of the binning in muon momentum, 
pseudorapidity and detector occupancy used in that method is estimated by varying the binning scheme. Finally, an uncertainty due to the 
method used to determine the number of signal candidates in the calibration samples is also considered. The total systematic uncertainty due to these 
three sources varies between 0 and 2\% and is assumed
to be fully correlated between bins. The particle identification efficiency
is determined from data control samples obtained from $pp$ collisions~\cite{PIDCALIB}. The particle multiplicity in $p$Pb collisions
is larger than in $pp$ collisions and the efficiencies
in the high multiplicity bins are extrapolated from lower 
multiplicity values. The uncertainty associated with this 
procedure is estimated by comparing the \jpsi double-differential 
absolute cross-section obtained in this analysis with that published in Ref.~\cite{LHCb-PAPER-2017-014}, which was obtained with looser 
selection requirements. They agree within 5\%, 
which is taken as the uncertainty related to the particle identification efficiency 
extrapolation. 
The uncertainty related to charged track reconstruction, which largely cancels in the efficiency ratio, varies from 0 to 0.1\% and is correlated between bins.

The hardware trigger efficiencies obtained from the simulation are validated by 
comparing them with the efficiencies measured in control samples in data~\cite{LHCb-DP-2012-004}.  
A systematic uncertainty is evaluated by comparing the results in simulation and in data. 
This uncertainty varies between 0 and 1.1\% and is assumed to be correlated between bins. 

The reconstruction and PID efficiencies depend on the particle multiplicity. The observed particle multiplicity distributions are compatible between
events with \psitwos and \jpsi mesons, however  the uncertainties are large. Since these distributions could be different~\cite{ALICE:2022gpu}, efficiencies are recomputed 
varying the multiplicity distributions between \psitwos and \jpsi 
events within their statistical uncertainties. Variations of
5\% are observed and this is assigned as the systematic uncertainty
related to possible different multiplicity distributions in events
with \jpsi or \psitwos candidates.  
The finite size of the simulation sample used for the efficiency determination introduces a systematic uncertainty, which varies between 0.3\% and 4.0\%.
The uncertainty on the ratio of branching fractions is 2.2\%.

The systematic uncertainties are assumed to be uncorrelated between the cross-section ratios and the \jpsi cross-section measurement in Ref.~\cite{LHCb-PAPER-2018-049} when computing the 
\psitwos cross-sections. They are also assumed to uncorrelated with the ratio
of reference cross-sections in $pp$ collisions at 8.16\tev used to 
compute the modification factors and their ratios.

%% file: results.tex
\section{Results}

\subsection{\texorpdfstring{Ratios of $\boldsymbol{\psitwos}$ to $\boldsymbol{\jpsi}$ cross-sections}{Ratios of psi(2S) to J/psi cross-sections}}

\begin{figure}[t]
\begin{center}
\includegraphics[width=7.85cm]{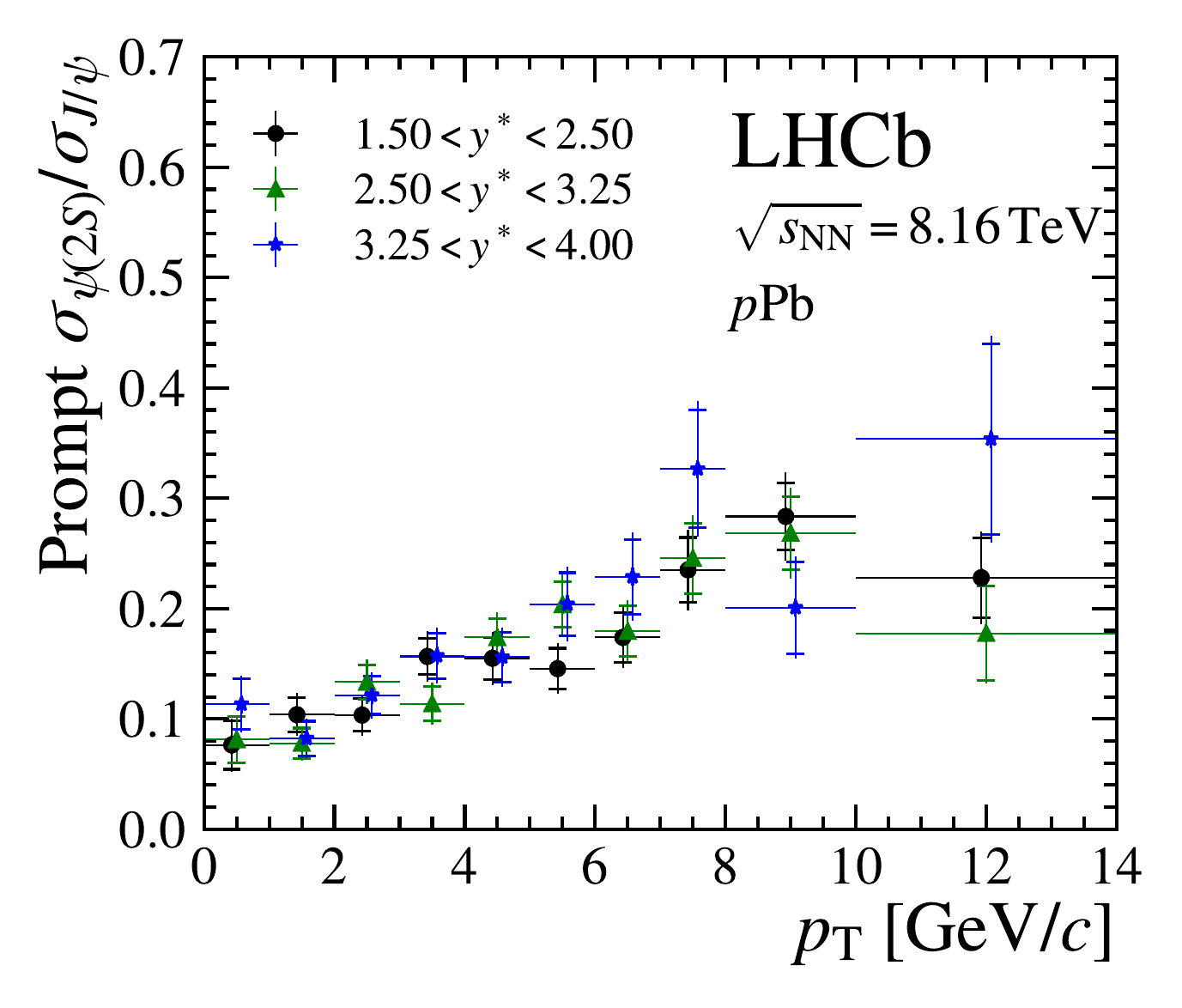}
\includegraphics[width=7.85cm]{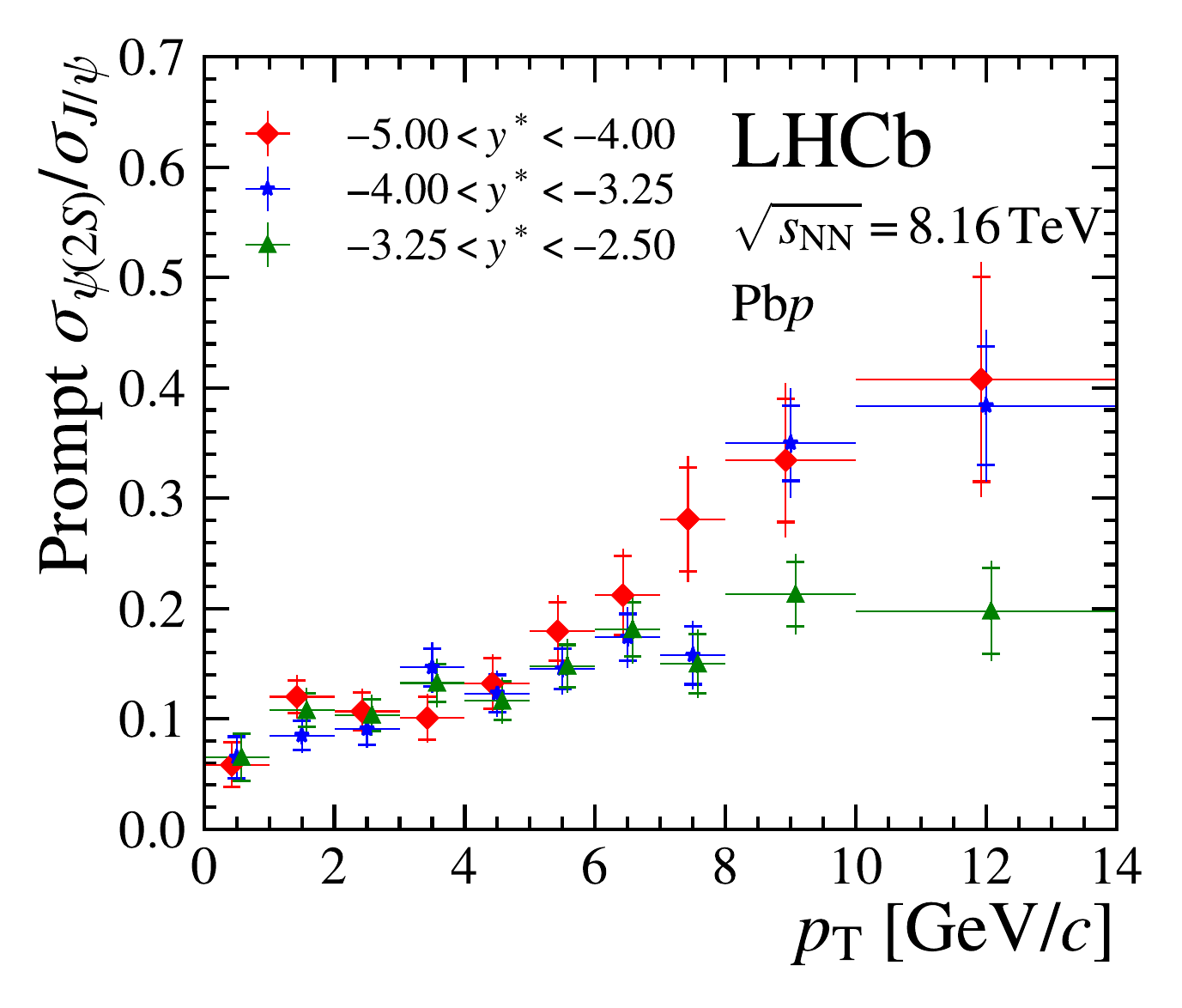}
\caption{\small Ratio of prompt \psitwos to prompt \jpsi production cross-section in (left) $p$Pb and (right) Pb$p$ collisions,
as a function of \pt for the different rapidity intervals.
Horizontal error bars are the bin widths, vertical error bars represent the statistical and total uncertainties.}
\label{fig:promptcrossratio}
\end{center}
\end{figure}

\begin{figure}[t]
\begin{center}
\includegraphics[width=7.85cm]{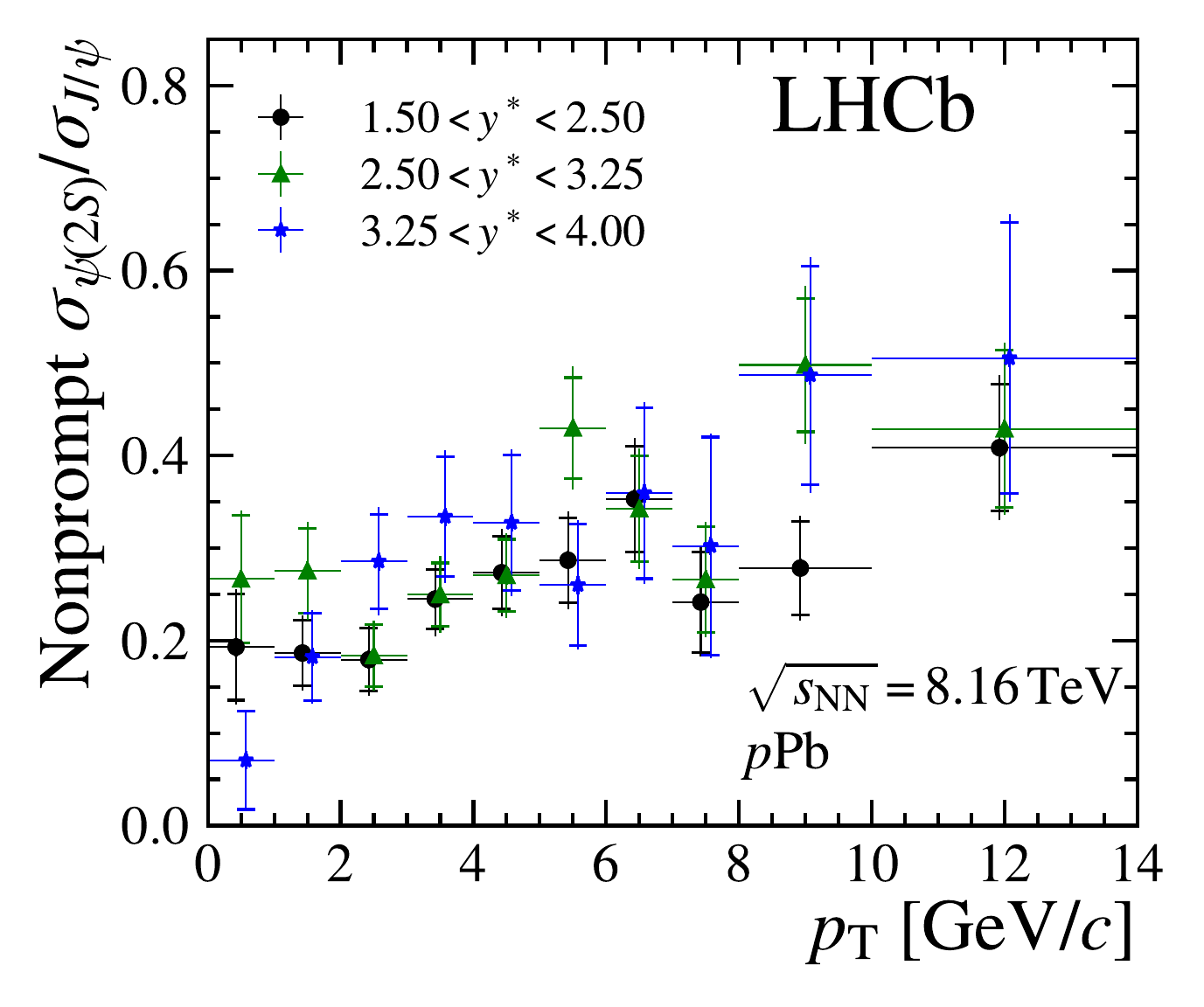}
\includegraphics[width=7.85cm]{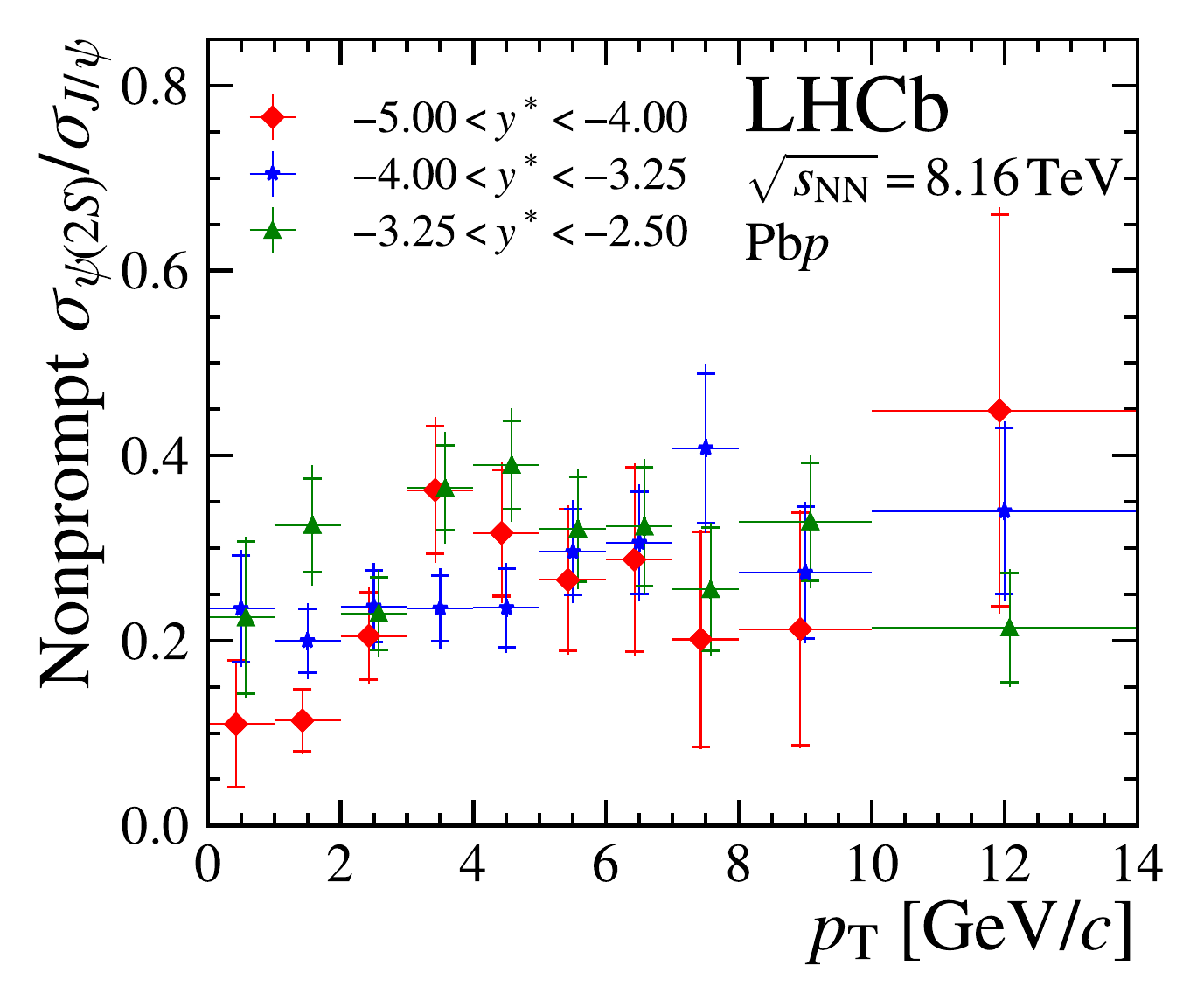}
\caption{\small Ratio of nonprompt \psitwos to nonprompt \jpsi  production cross-section in (left) $p$Pb 
and (right) Pb$p$ collisions, as a function of \pt for the different rapidity intervals.
Horizontal error bars are the bin widths, vertical error bars represent the statistical and total uncertainties.}
\label{fig:nonpromptcrossratio}
\end{center}
\end{figure}

The prompt production cross-section ratios, $\frac{\sigma_{\psitwos}}{\sigma_{\jpsi}}$, are shown in Fig.~\ref{fig:promptcrossratio} as a function of \pt in the three rapidity intervals of the analysis of $p$Pb and Pb$p$ collisions. All numerical values corresponding to the figures are available in the appendices A -- E. 
The production cross-section ratios for the nonprompt production
are shown in Fig.~\ref{fig:nonpromptcrossratio}.

\begin{figure}[htbp]
\begin{center}
\includegraphics[width=7.85cm]{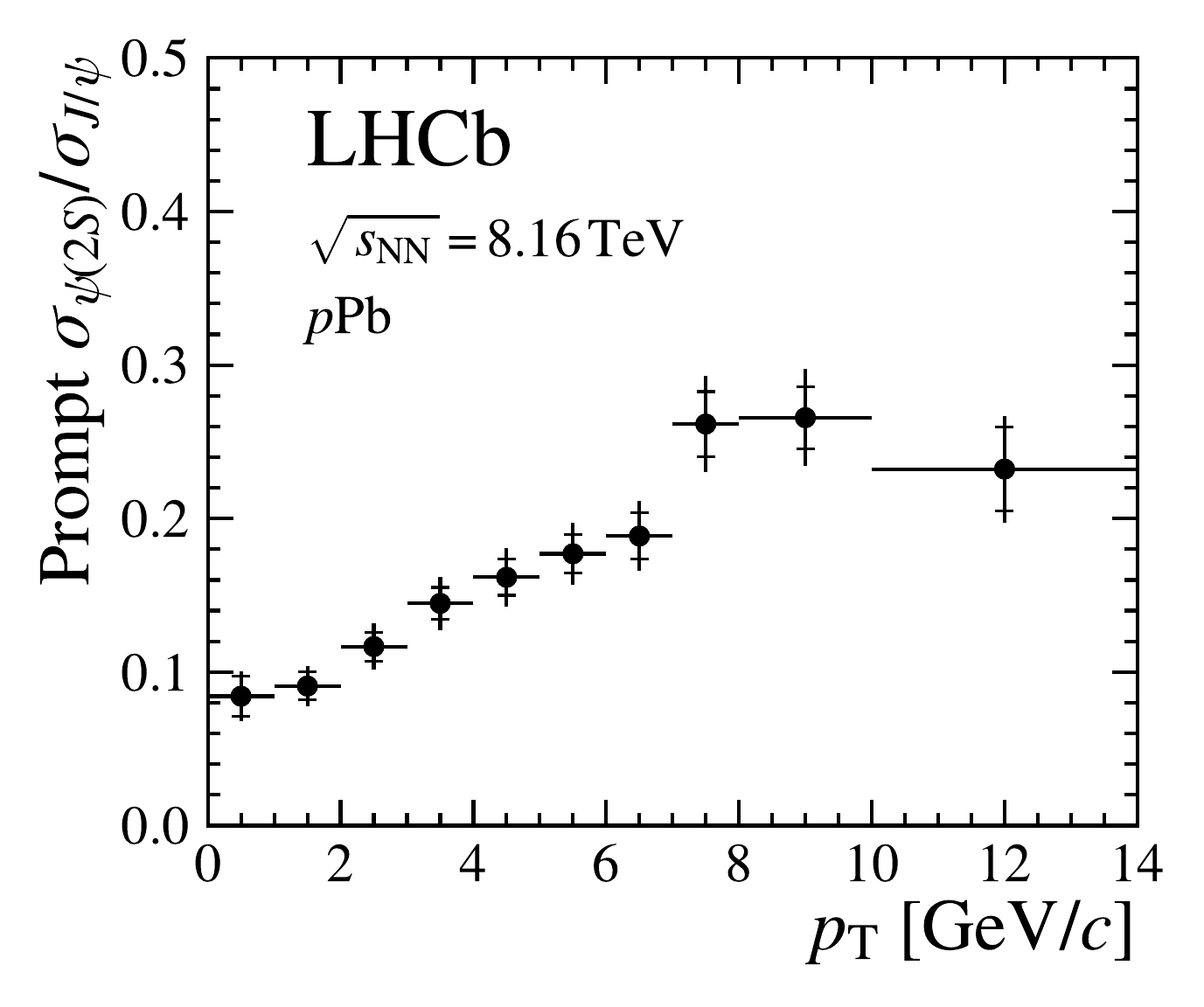}
\includegraphics[width=7.85cm]{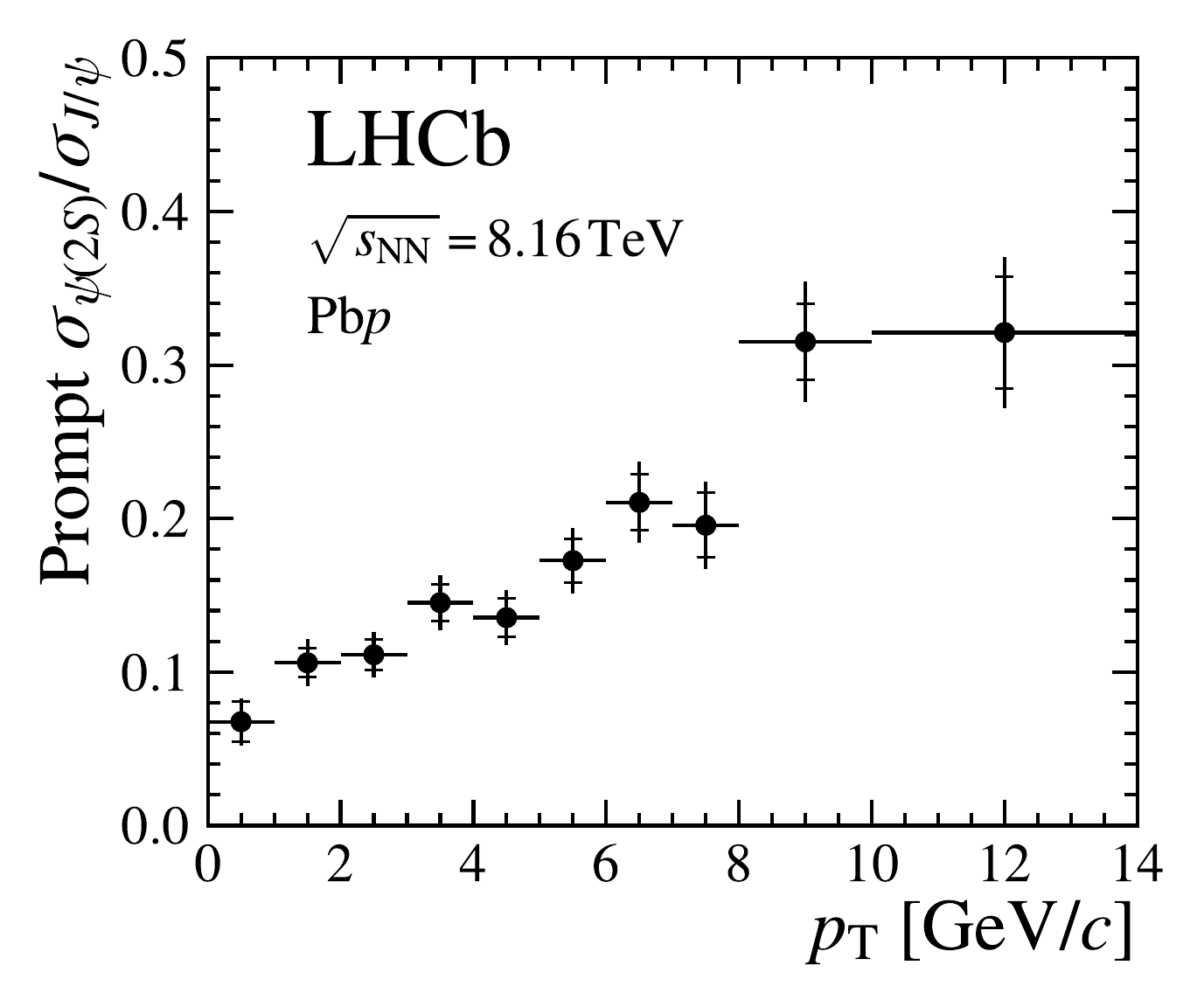}
\caption{\small Ratio of prompt \psitwos to prompt \jpsi production cross-sections in (left) $p$Pb collisions, integrated over $1.5<y^*<4.0$, and (right)
Pb$p$ collisions, integrated over $-5.0<y^*<-2.5$,
as a function of \pt. Horizontal error bars are the bin widths, vertical error bars represent the statistical and total uncertainties.}
\label{fig:promptcrossratio_pt}
\end{center}
\end{figure}

\begin{figure}[htbp]
\begin{center}
\includegraphics[width=7.85cm]{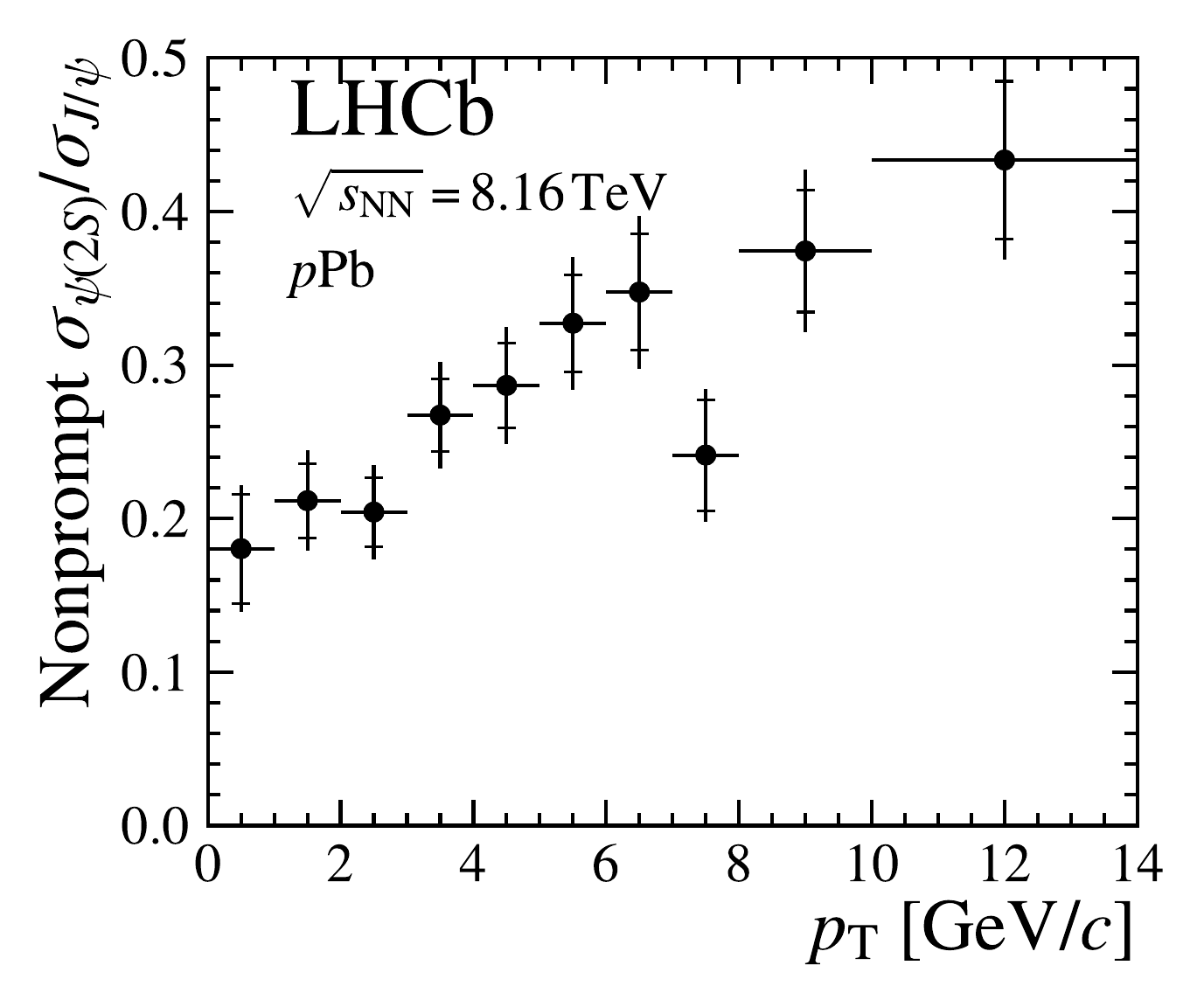}
\includegraphics[width=7.85cm]{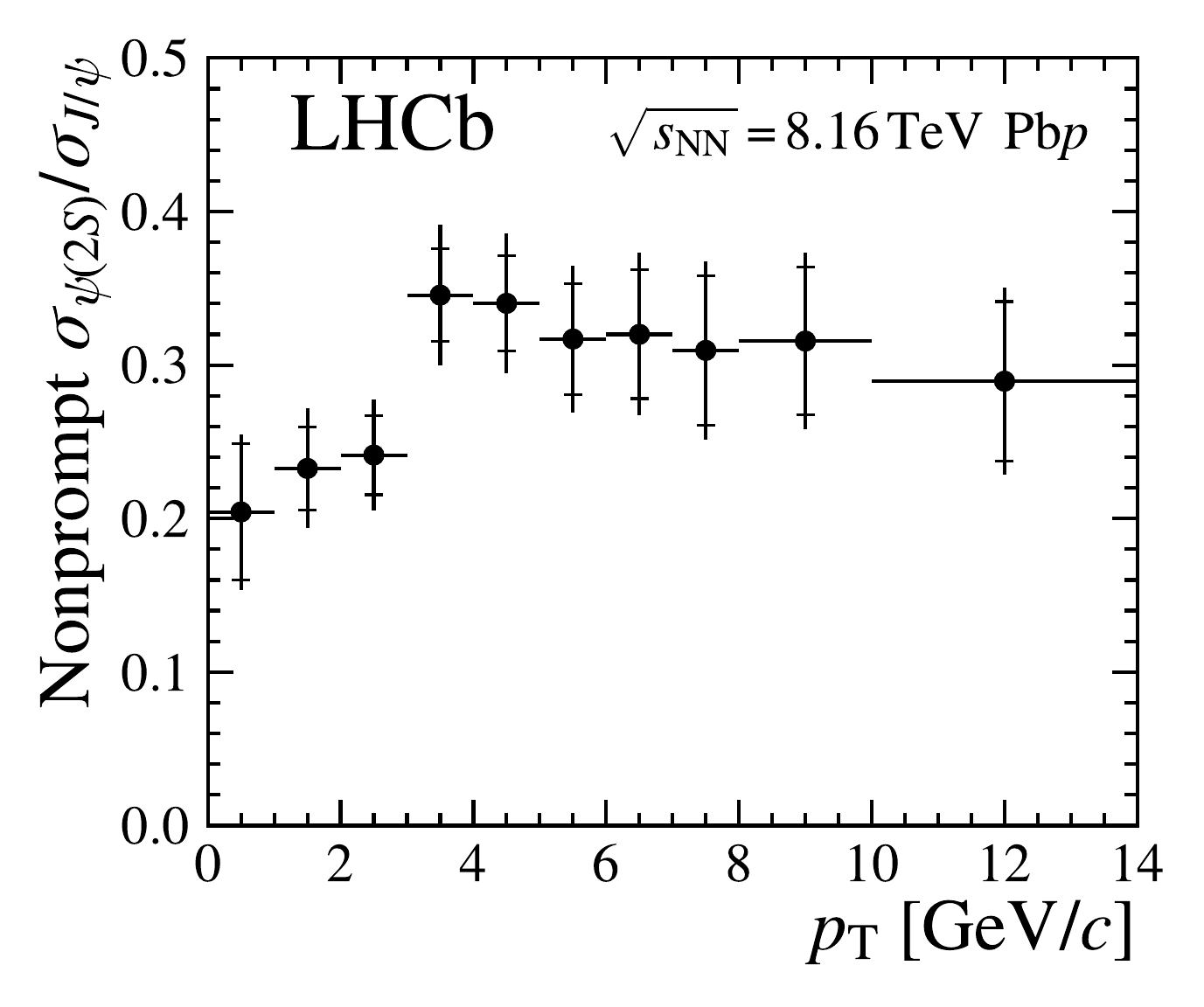}
\caption{\small Ratio of nonprompt \psitwos to nonprompt \jpsi production cross-section
 in (left) $p$Pb collisions, integrated over $1.5<y^*<4.0$,
and (right) Pb$p$ collisions, integrated over $-5.0<y^*<-2.5$, as a function of \pt. Horizontal error bars are the bin widths,
vertical error bars represent the statistical and total uncertainties.}
\label{fig:nonpromptcrossratio_pt}
\end{center}
\end{figure}

\begin{figure}[htbp]
\begin{center}
\includegraphics[width=7.85cm]{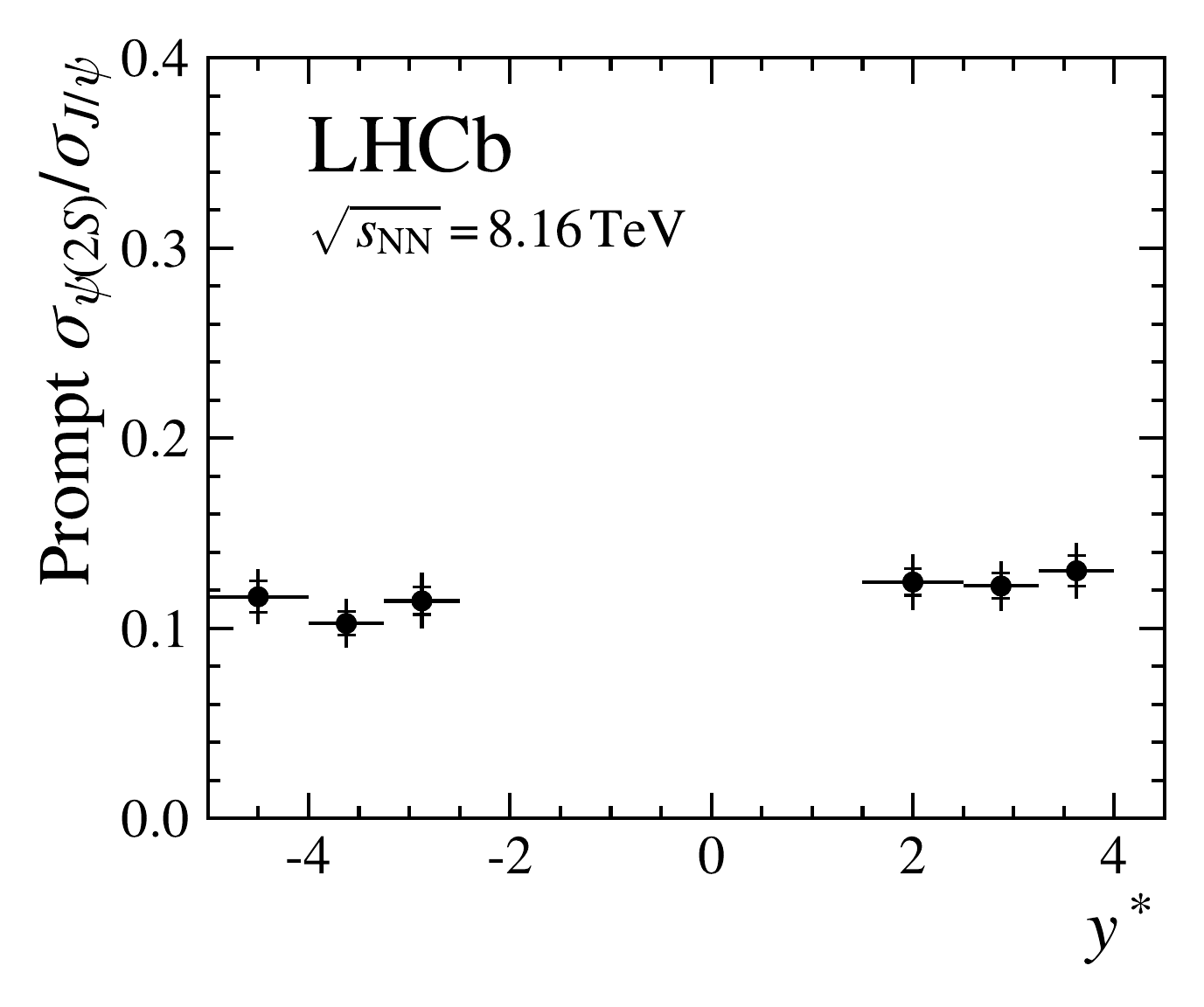}
\includegraphics[width=7.85cm]{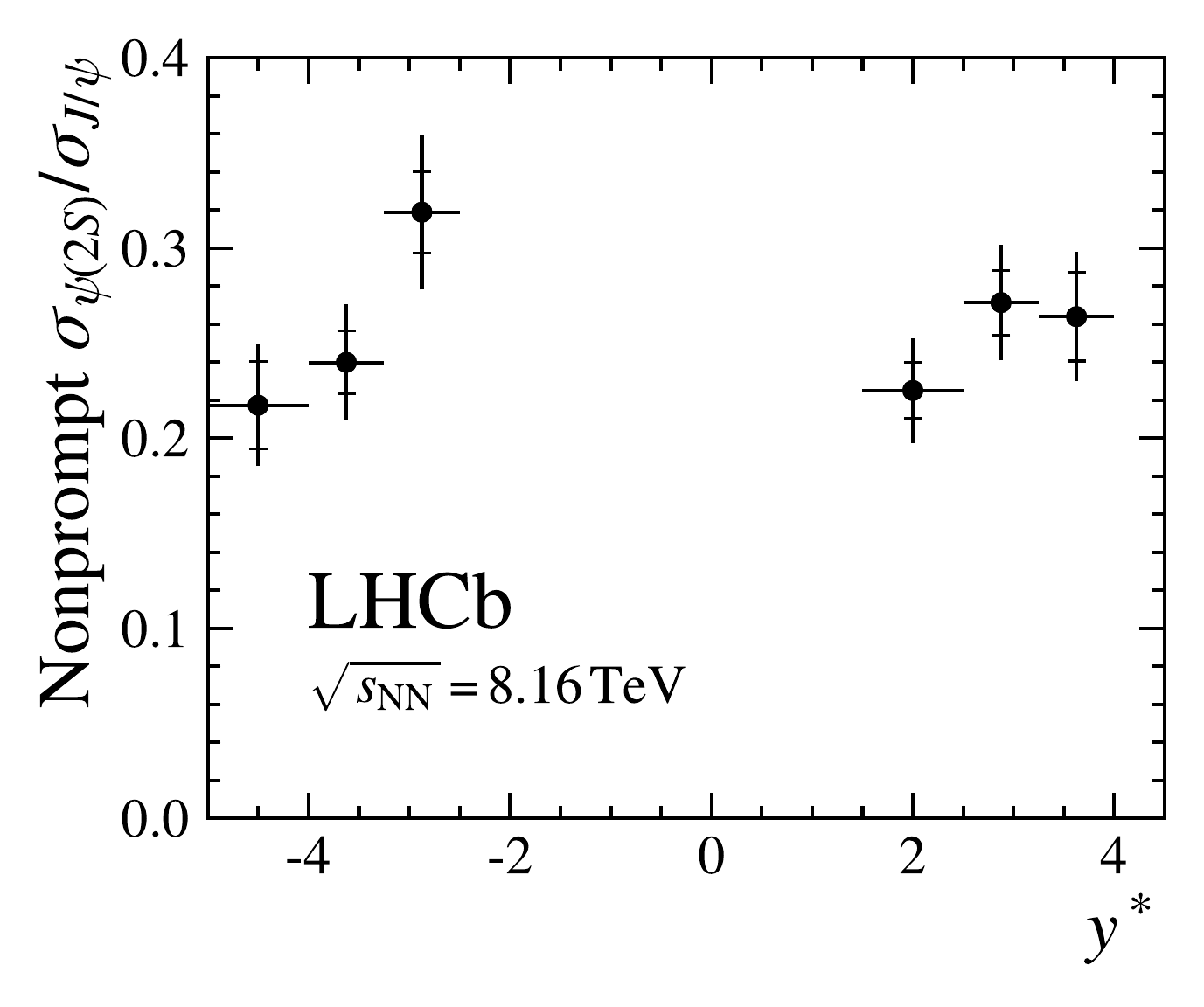}
\caption{\small Ratio of (left) prompt and (right) nonprompt \psitwos to \jpsi production cross-section, as a
function of rapidity, integrated over \pt.
Horizontal error bars are the bin widths, vertical error bars represent the statistical and total uncertainties.}
\label{fig:crossratio_y}
\end{center}
\end{figure}

The prompt production cross-section ratios as a function of \pt, integrated in the range $1.5<y^*<4.0$ for $p$Pb collisions  and 
$-5.0<y^*<-2.5$ for Pb$p$ collisions, are shown in Fig.~\ref{fig:promptcrossratio_pt}.
The corresponding values for nonprompt production are 
shown in Fig.~\ref{fig:nonpromptcrossratio_pt}.
The production cross-section ratios are also extracted as a function of rapidity integrated in the range
$0<\pt<14\gevc$. They are shown in Fig.~\ref{fig:crossratio_y}.

\subsection{\texorpdfstring{Cross-sections for $\boldsymbol{\psitwos}$ production}{Cross-sections for psi(2S) production}}

\begin{figure}[t]
\begin{center}
\includegraphics[width=7.85cm]{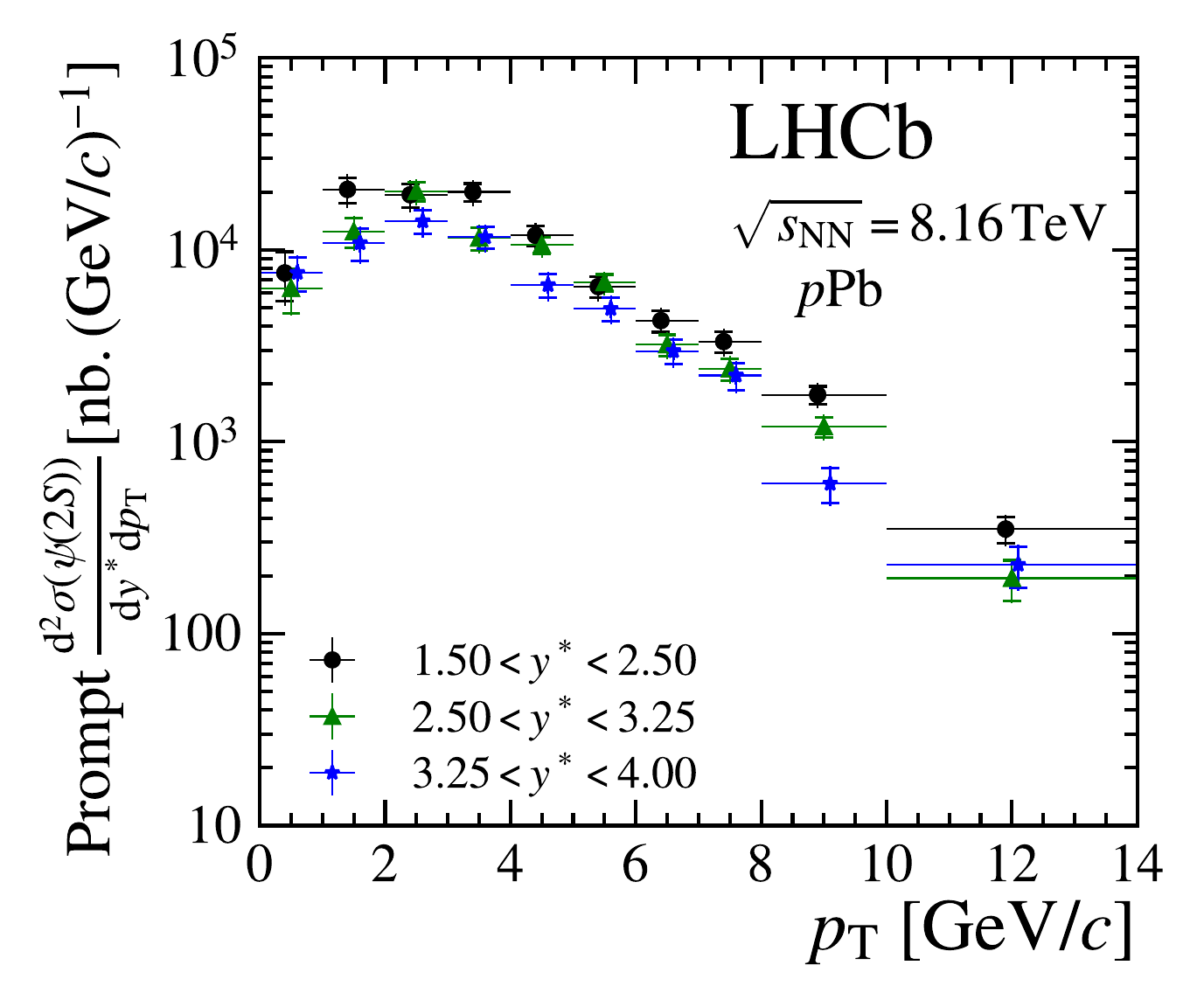}
\includegraphics[width=7.85cm]{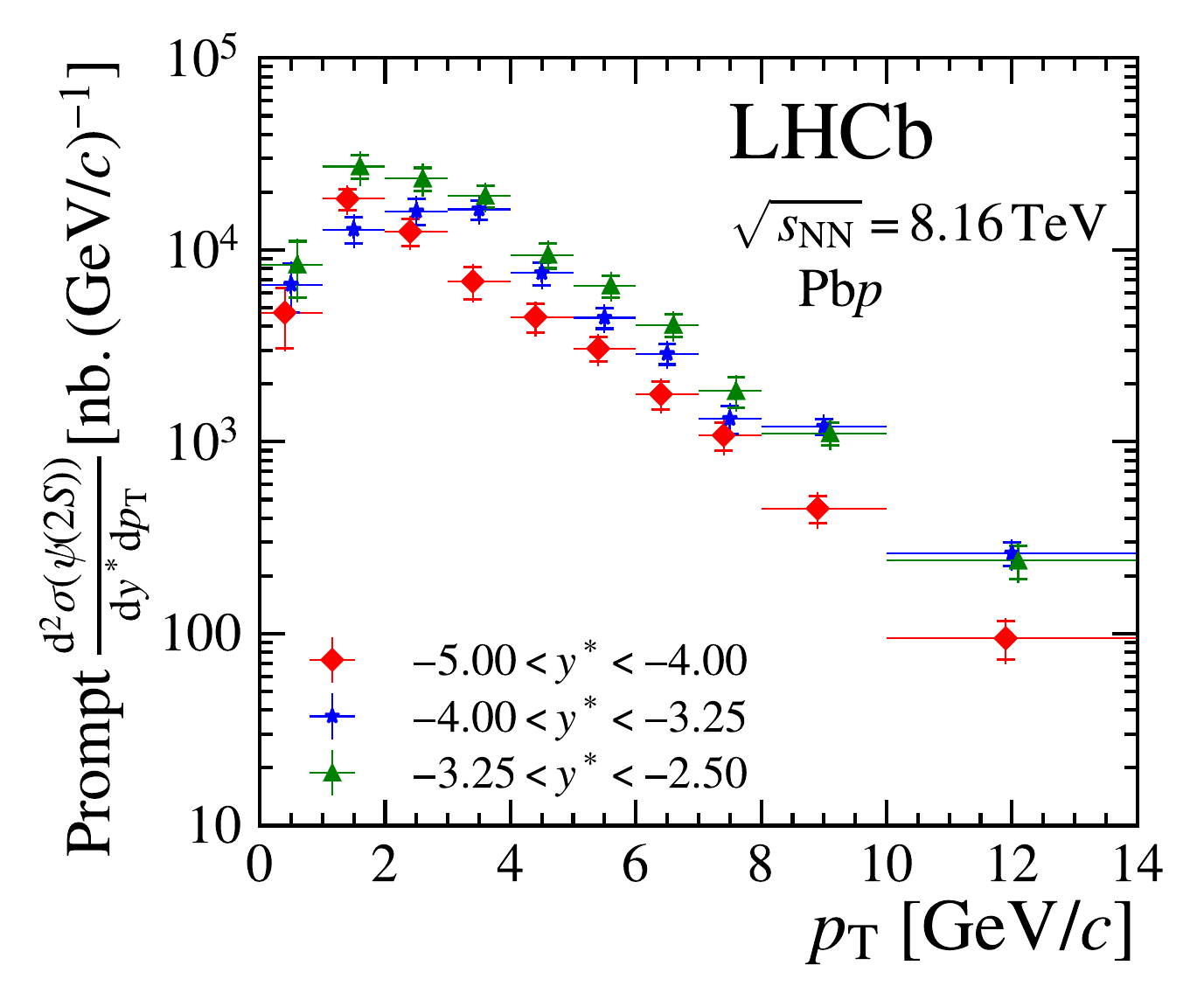}
\caption{\small Absolute prompt \psitwos production cross-section in (left)  $p$Pb and (right) Pb$p$  collisions,
as a function of \pt for the different rapidity intervals. Horizontal error bars are the bin widths, vertical error bars represent the statistical and total
uncertainties.}
\label{fig:promptcrosssection_psi2s}
\end{center}
\end{figure}

\begin{figure}[ht]
\begin{center}
\includegraphics[width=7.85cm]{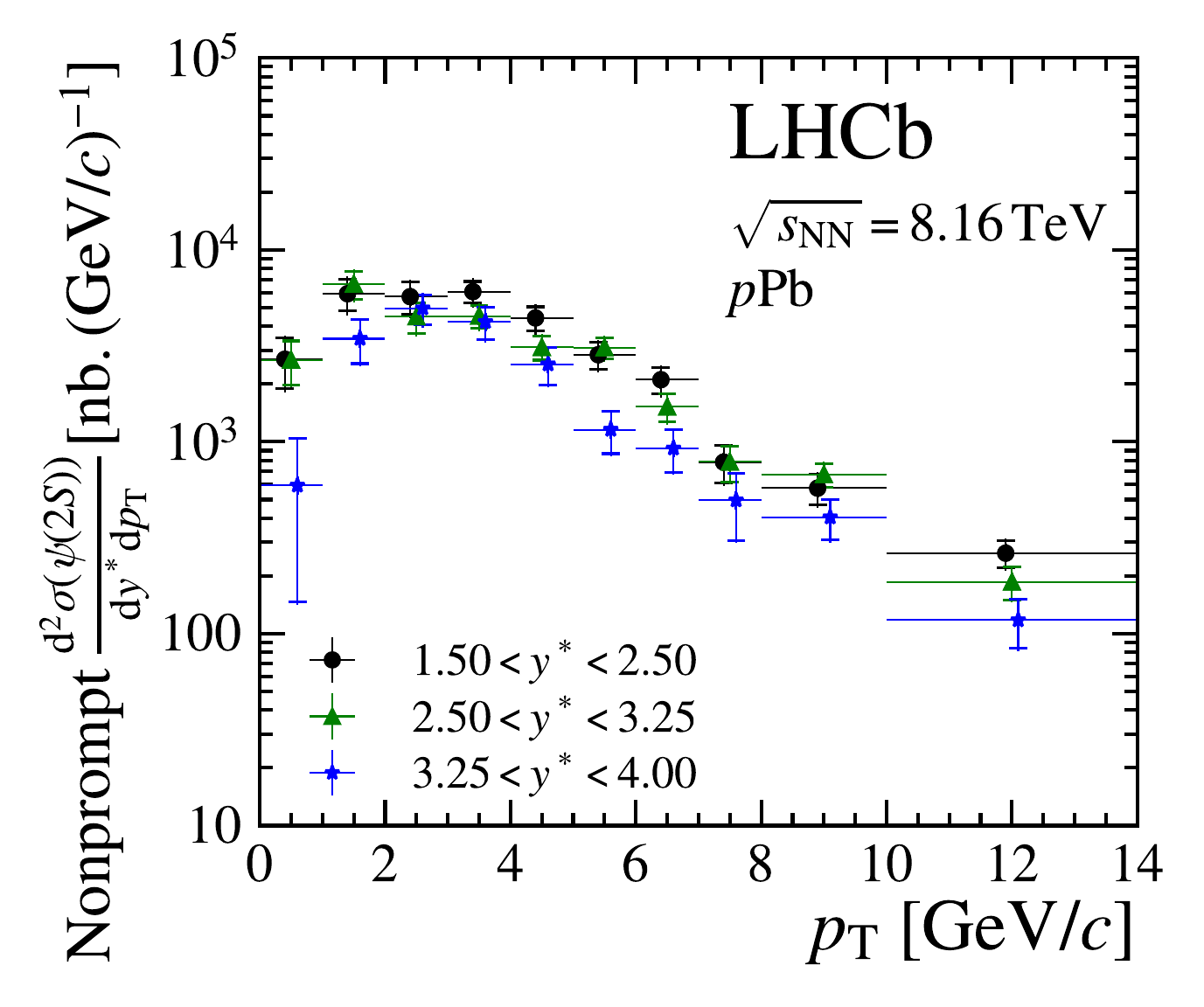}
\includegraphics[width=7.85cm]{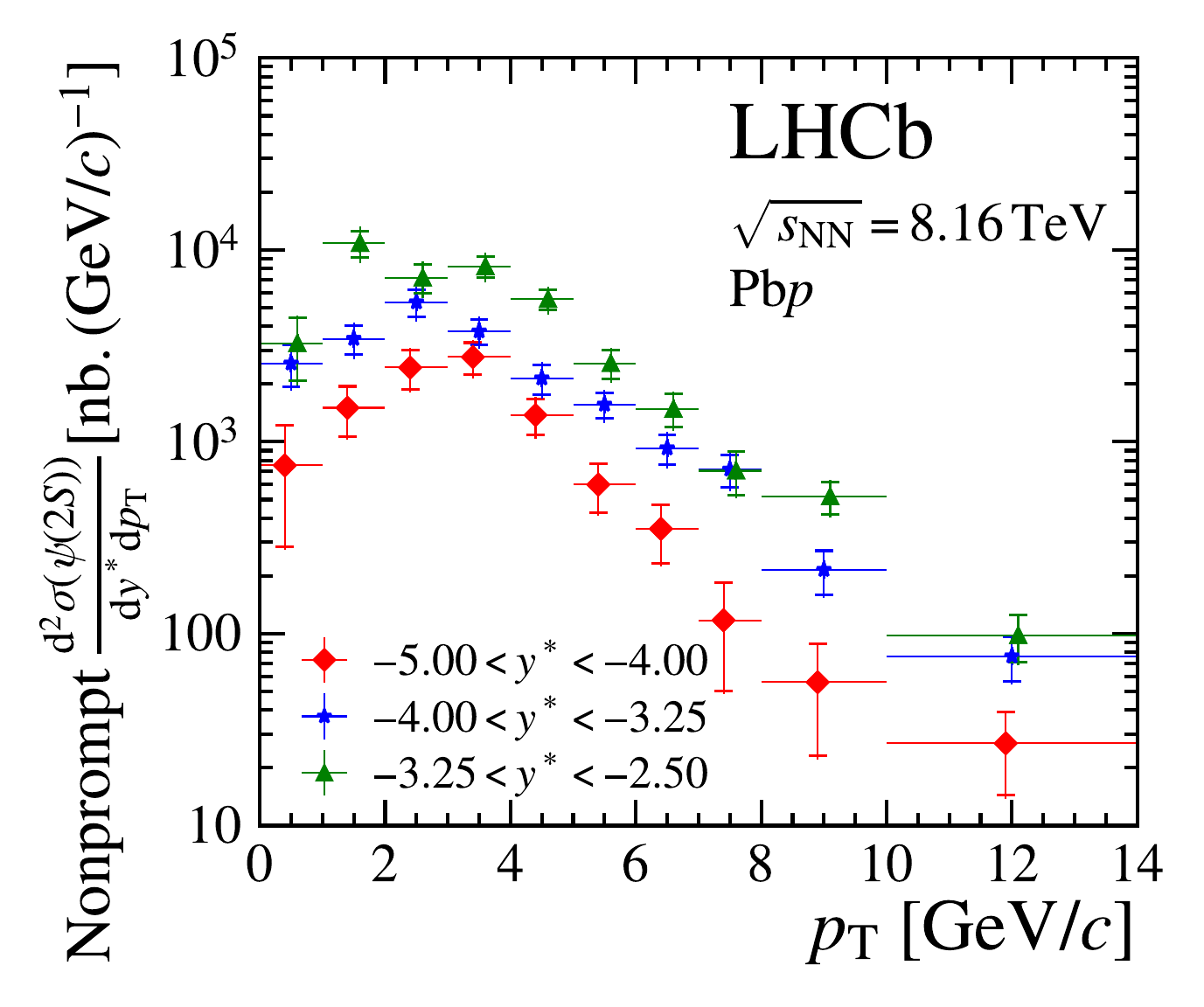}
\caption{\small Absolute nonprompt \psitwos production cross-section in (left) $p$Pb  and (right) Pb$p$ collisions,
as a function of \pt for the different rapidity intervals. Horizontal error bars are the bin widths, vertical error bars represent the statistical and total
uncertainties.}
\label{fig:nonpromptcrosssection_psi2s}
\end{center}
\end{figure}

\begin{figure}[ht]
\begin{center}
\includegraphics[width=7.85cm]{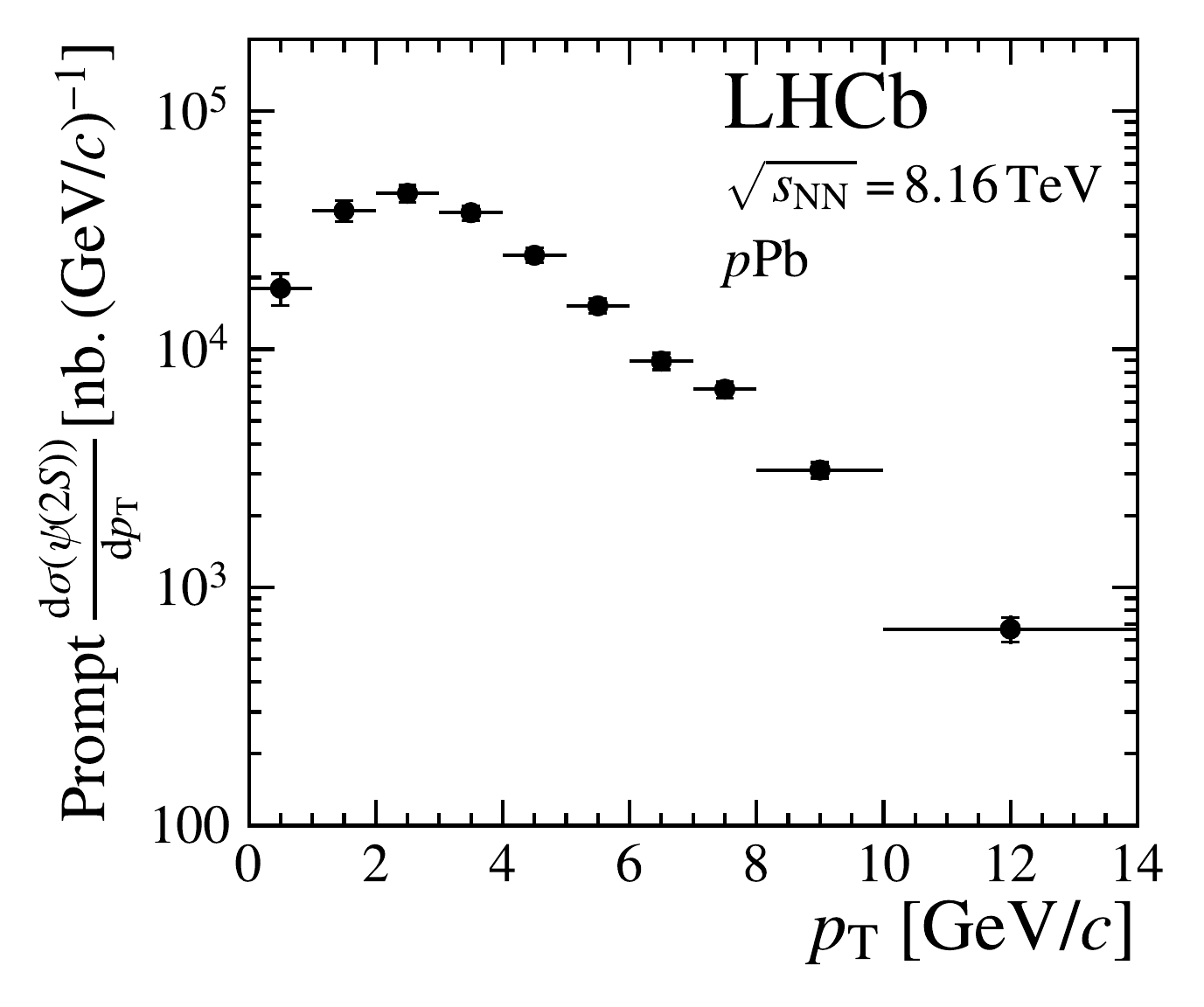}
\includegraphics[width=7.85cm]{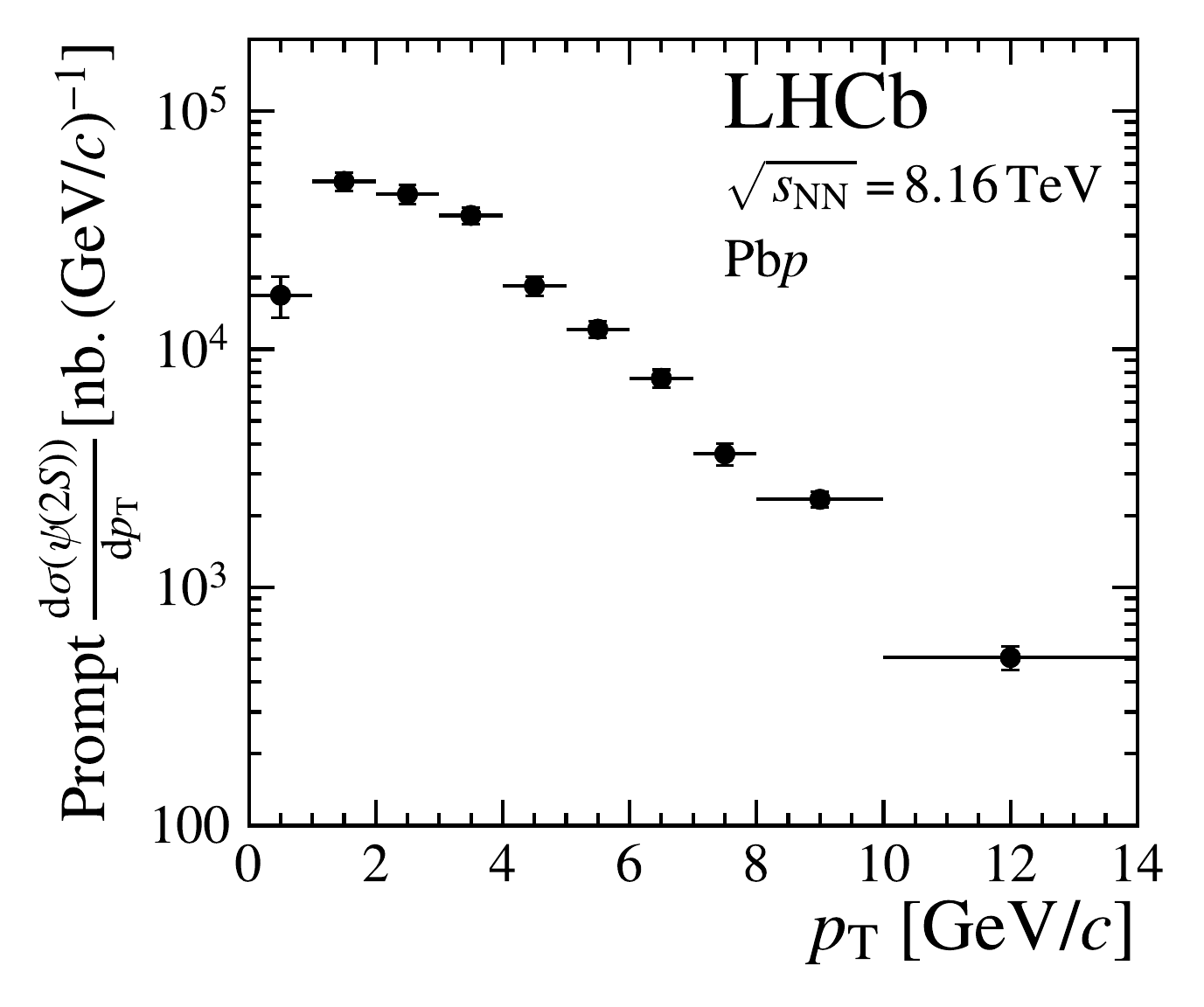}
\includegraphics[width=7.85cm]{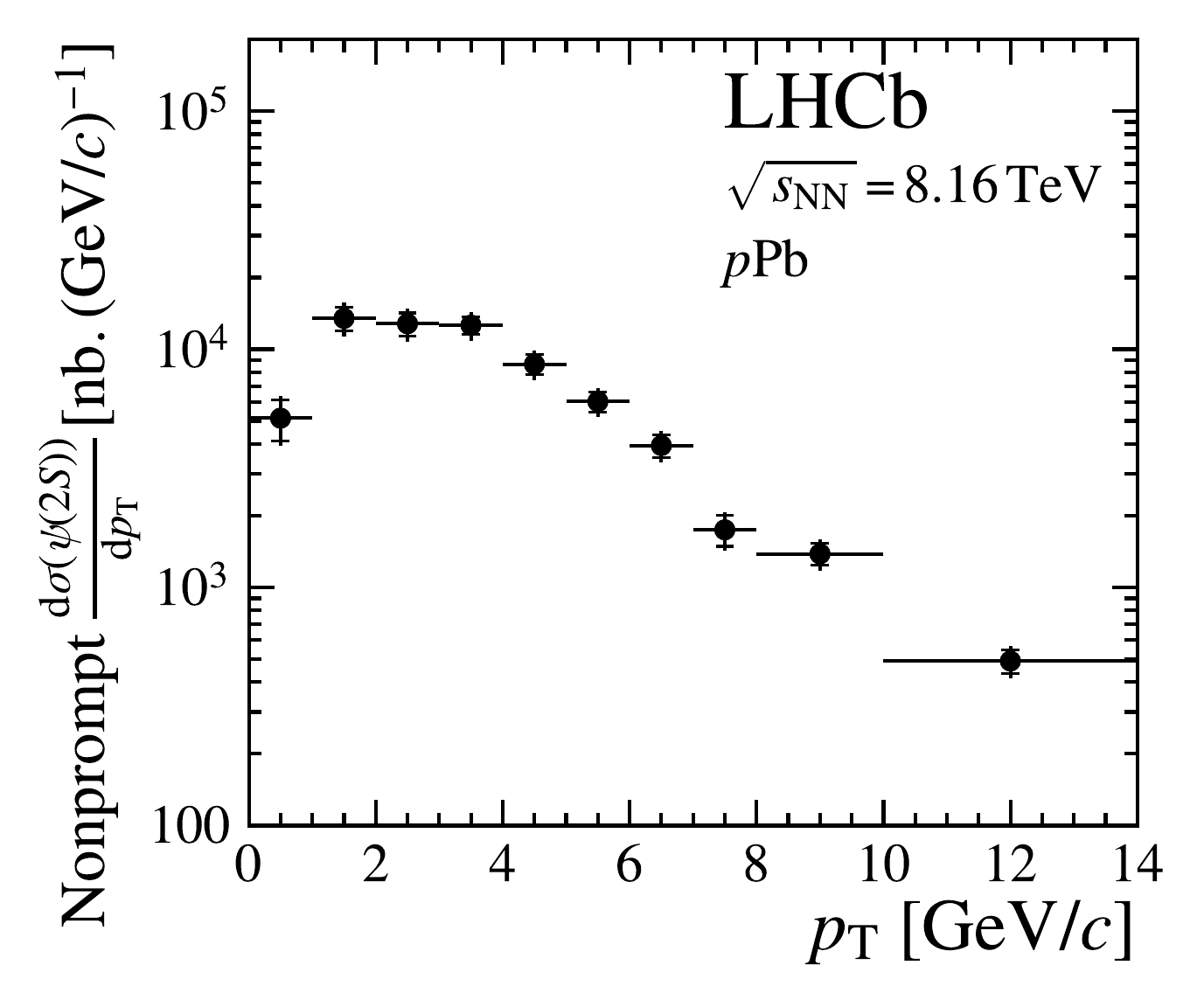}
\includegraphics[width=7.85cm]{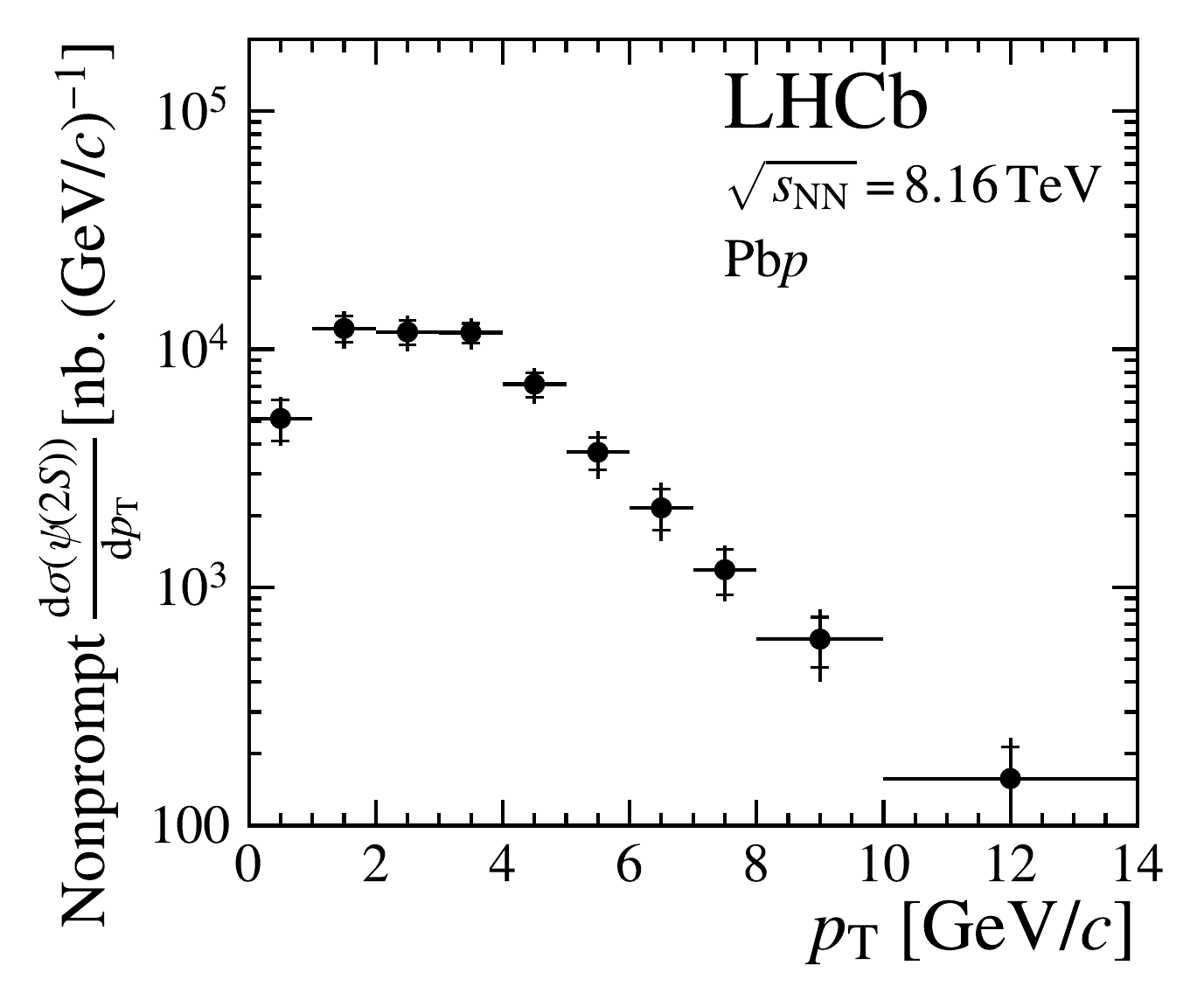}
\caption{\small Absolute production cross-section
as a function of \pt for (top left) prompt \psitwos in $p$Pb collisions, integrated over $1.5<y^*<4.0$, (top right) prompt \psitwos in Pb$p$ collisions, integrated over $-5.0<y^*<-2.5$, 
(bottom left)
nonprompt \psitwos in $p$Pb collisions, integrated over $1.5<y^*<4.0$,  (bottom right) nonprompt \psitwos in Pb$p$ collisions, integrated over $-5.0<y^*<-2.5$. Horizontal error bars are the bin widths,
vertical error bars represent the statistical and total uncertainties.}
\label{fig:integrated_pt_psi2s}
\end{center}
\end{figure}

The absolute \psitwos cross-sections are obtained multiplying the ratios shown in the previous section by the 
measured values of the \jpsi cross-sections~\cite{LHCb-PAPER-2017-014},
 in the same \pt and $y^*$ bins. 
The prompt \psitwos production cross-section
is shown in Fig.~\ref{fig:promptcrosssection_psi2s} and the nonprompt cross-section in Fig.~\ref{fig:nonpromptcrosssection_psi2s}, 
as a function of \pt for different rapidity intervals, for $p$Pb and Pb$p$ collisions. The cross-sections, integrated over $y^*$ in the LHCb acceptance and as a function of \pt are shown in Fig.~\ref{fig:integrated_pt_psi2s}.

\subsection{\texorpdfstring{Nuclear modification factors for $\boldsymbol{\psitwos}$ production}{Nuclear modification factors for psi(2S) production}}

\begin{figure}[htpb]
\begin{center}
\includegraphics[width=7.85cm]{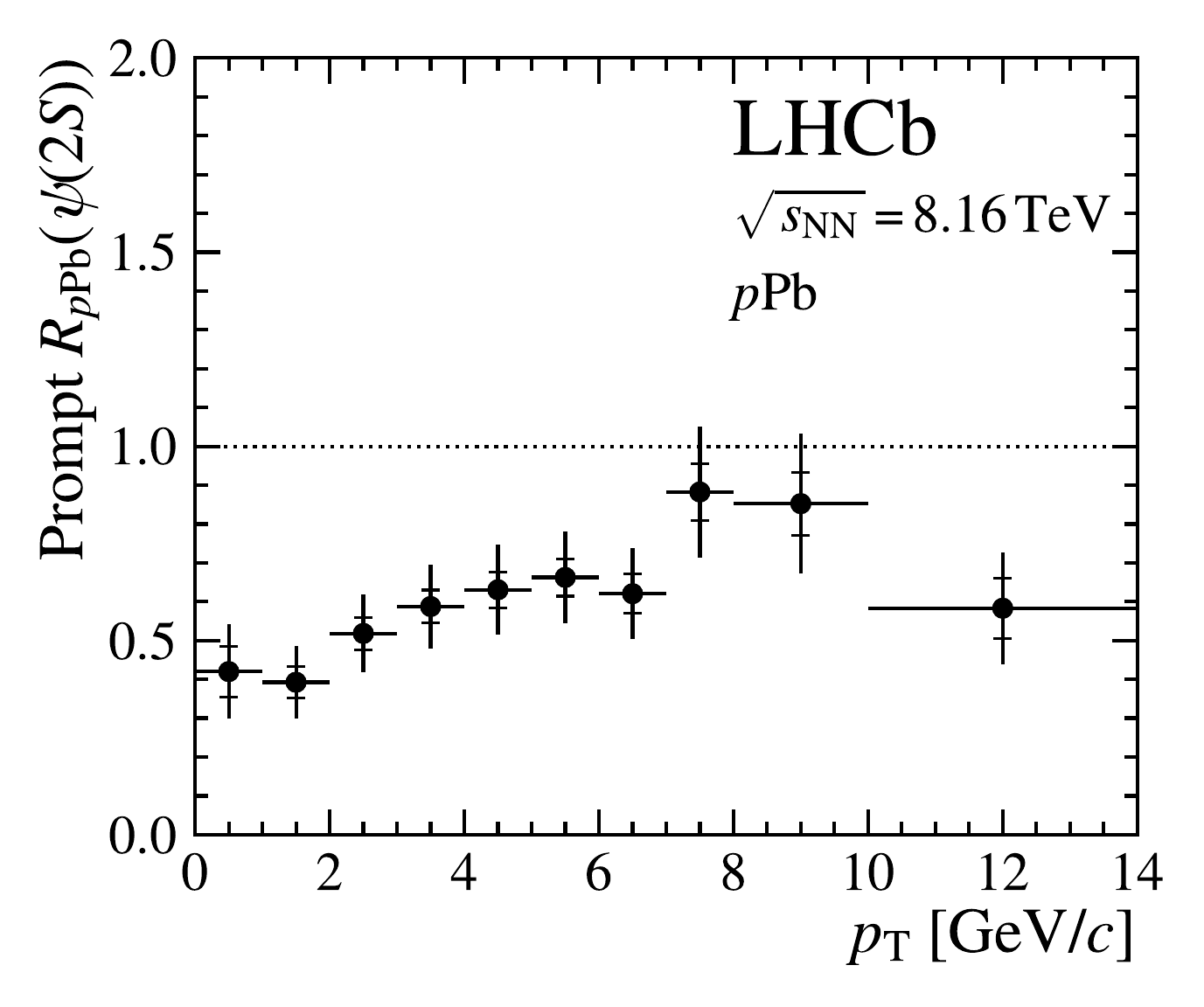}
\includegraphics[width=7.85cm]{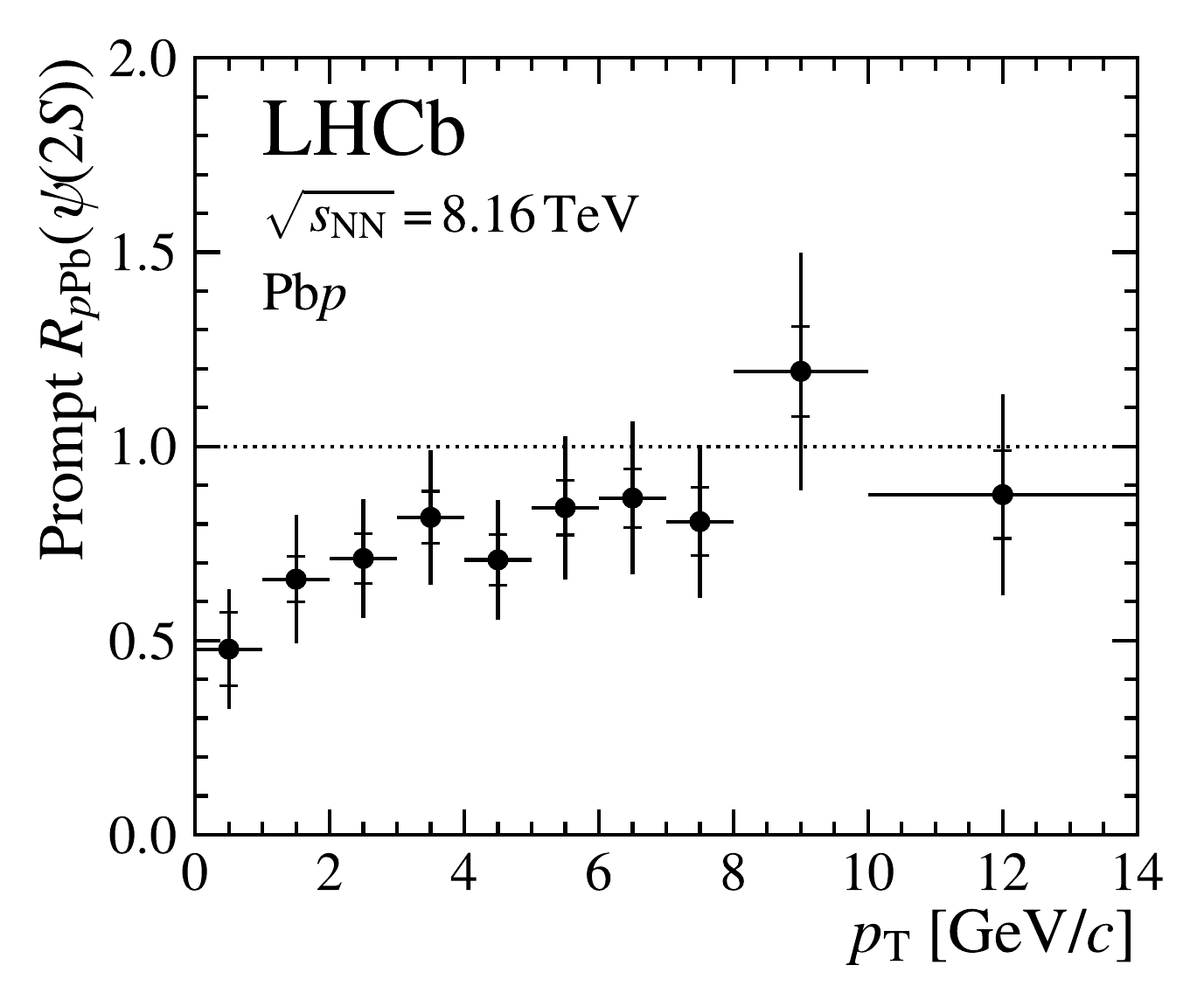}
\includegraphics[width=7.85cm]{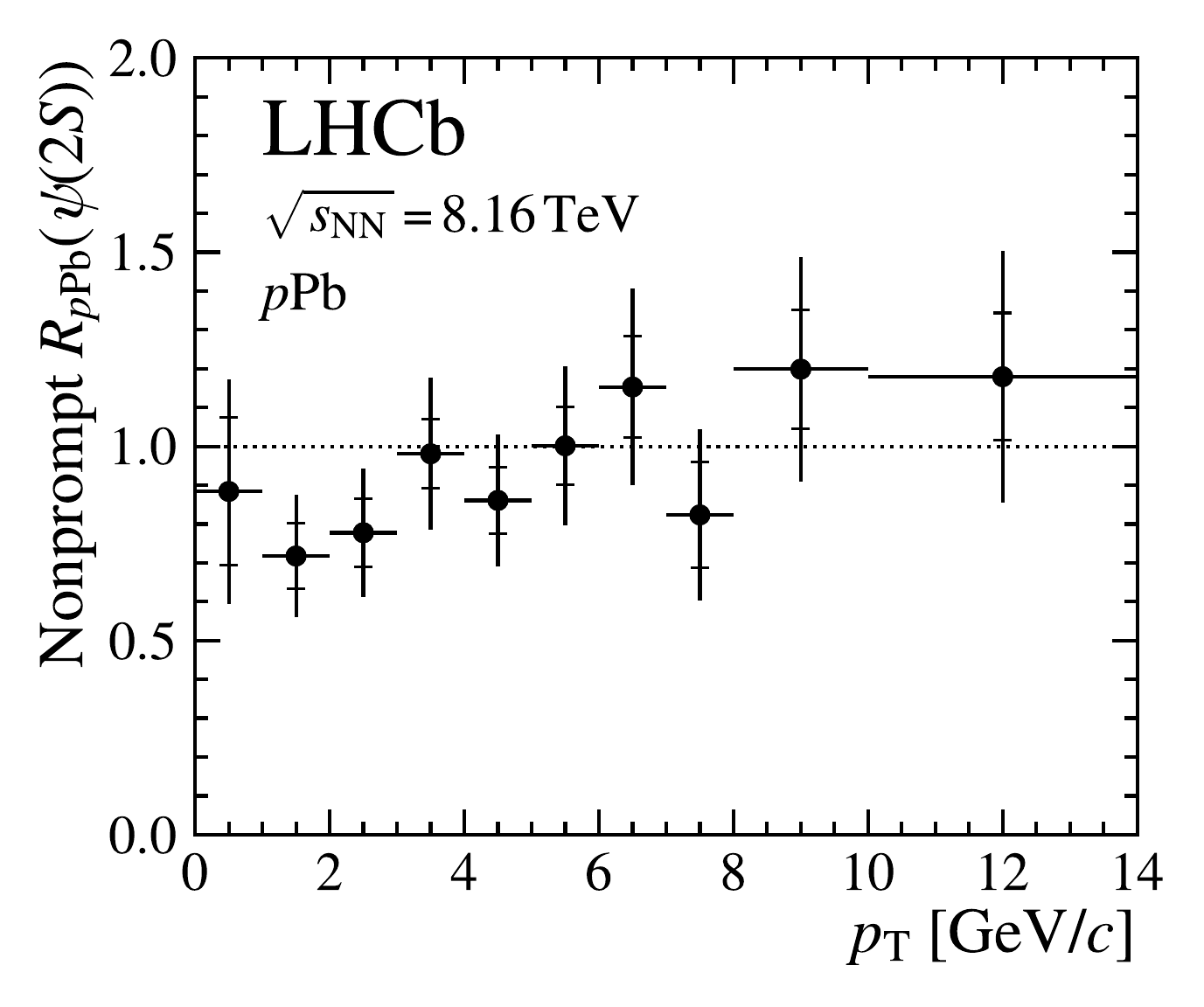}
\includegraphics[width=7.85cm]{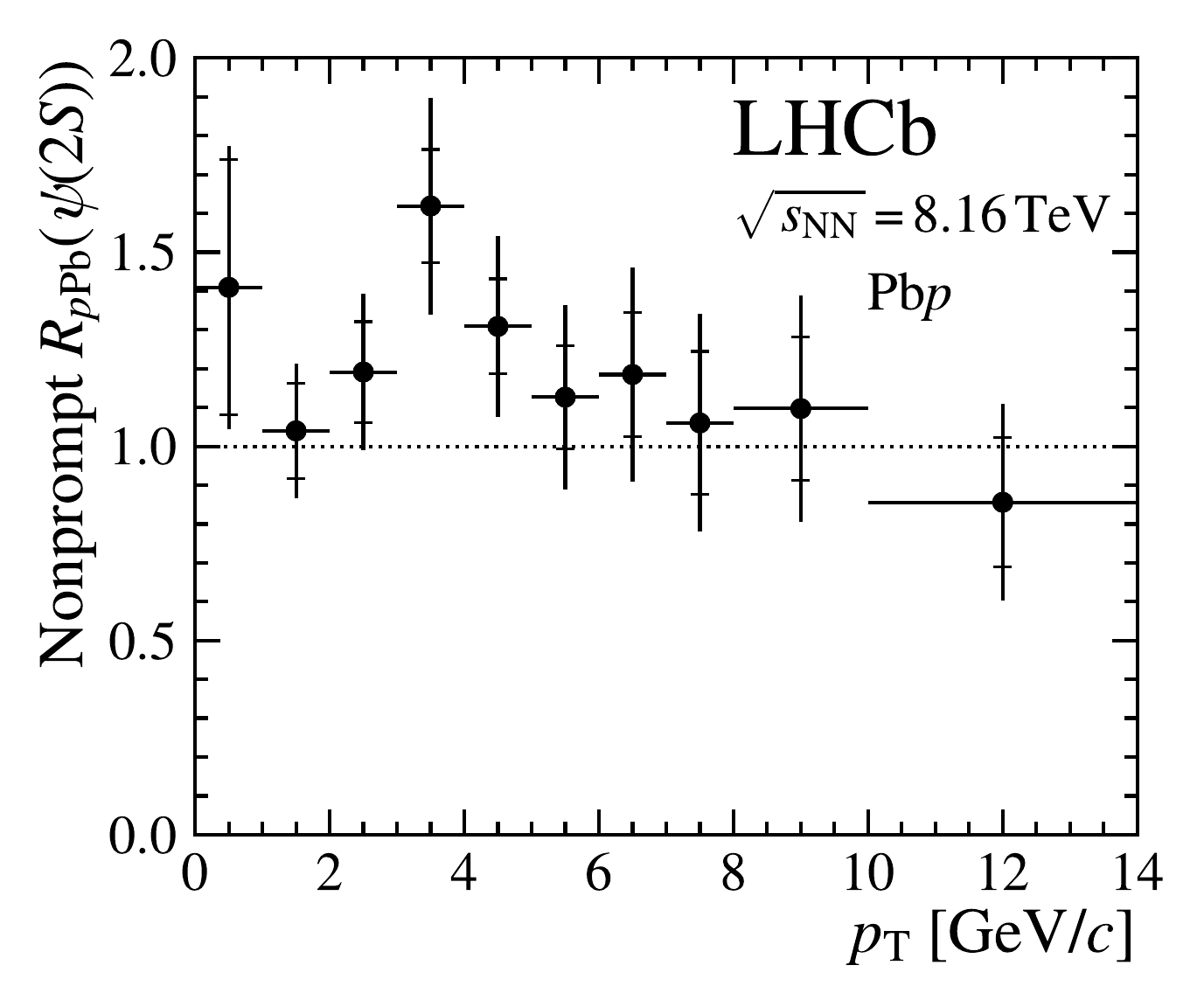}
\caption{\small Nuclear modification factor as a function of \pt, for (top left) prompt \psitwos in $p$Pb collisions, integrated over $1.5<y^{*}<4.0$, (top right)
prompt \psitwos in Pb$p$ collisions, integrated over $-5.0<y^{*}<-2.5$, (bottom left) nonprompt \psitwos in $p$Pb collisions, integrated over $1.5<y^{*}<4.0$,
 and (bottom right) nonprompt \psitwos in Pb$p$ collisions,  integrated over $-5.0<y^{*}<-2.5$. Horizontal error bars are the bin widths,
vertical error bars represent the statistical and total uncertainties.
 }
 \label{fig:rpa_psi2s_pt}
 \end{center}
 \end{figure}

The \psitwos nuclear modification factors integrated over $y^*$  as a function of \pt are shown in Fig.~\ref{fig:rpa_psi2s_pt}.
The kinematic 
dependence of the prompt \psitwos nuclear modification factor is 
similar to that of the \jpsi meson seen in Ref.~\cite{LHCb-PAPER-2017-014}, 
with additional suppression. The nonprompt 
\psitwos nuclear modification factor is consistent 
with the \jpsi one~\cite{LHCb-PAPER-2017-014} with larger uncertainties 
due to the smaller sample size.

\subsection{\texorpdfstring{Forward-to-backward ratio for $\boldsymbol{\psitwos}$ production}{Forward-to-backward ratio for psi(2S) production}}

\begin{figure}[htpb]
 \begin{center}
 \includegraphics[width=7.85cm]{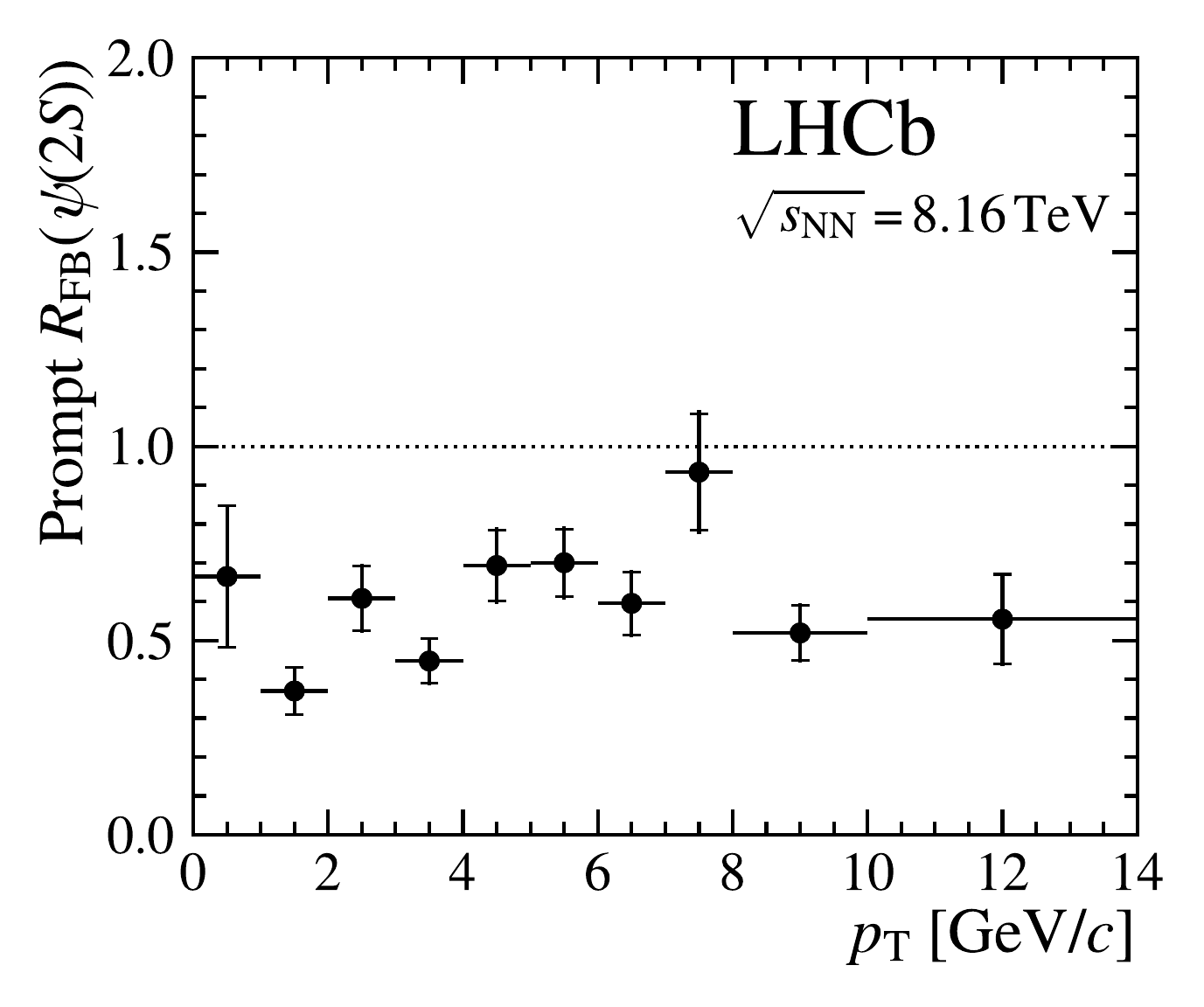}
 \includegraphics[width=7.85cm]{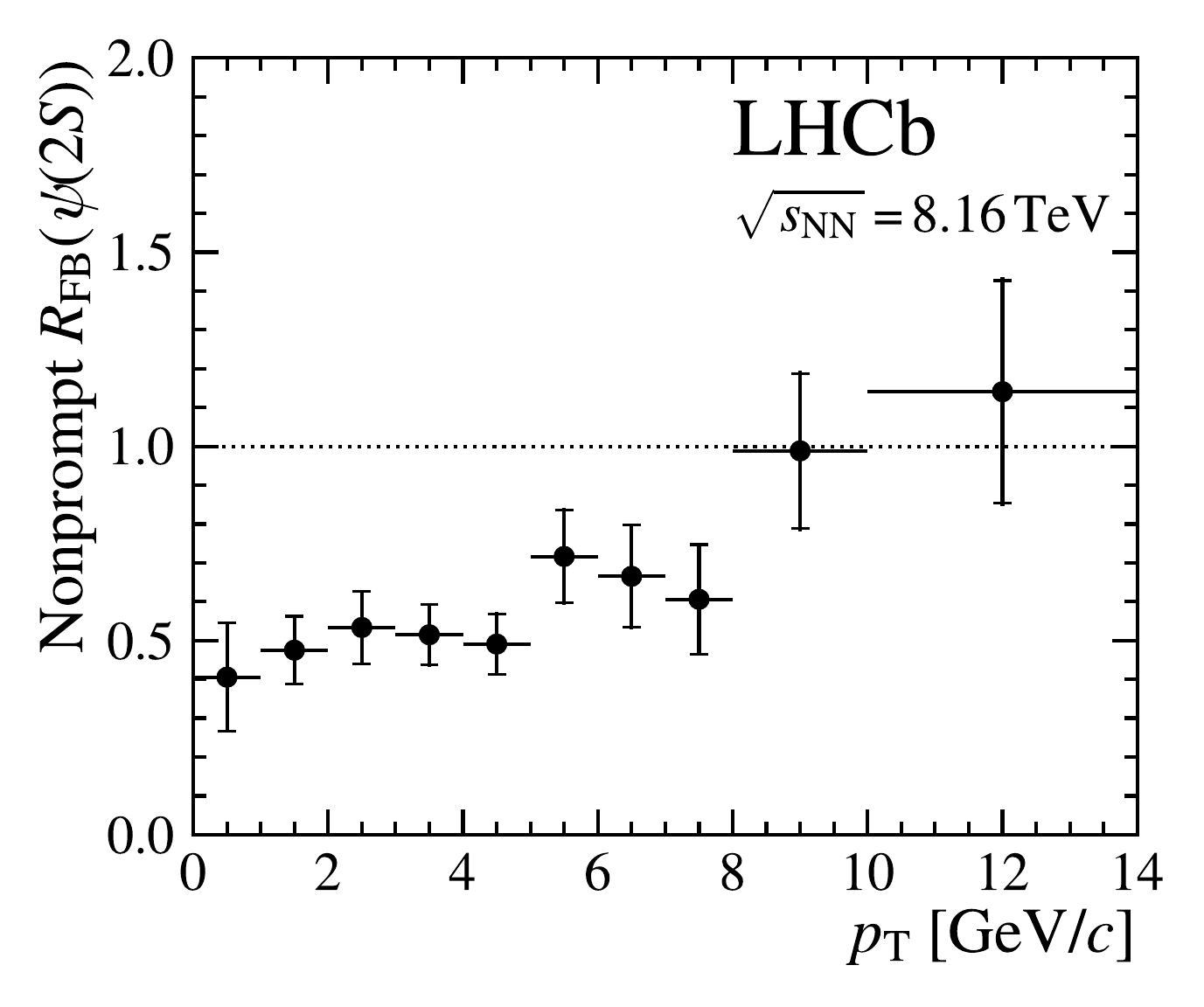}
  \includegraphics[width=7.85cm]{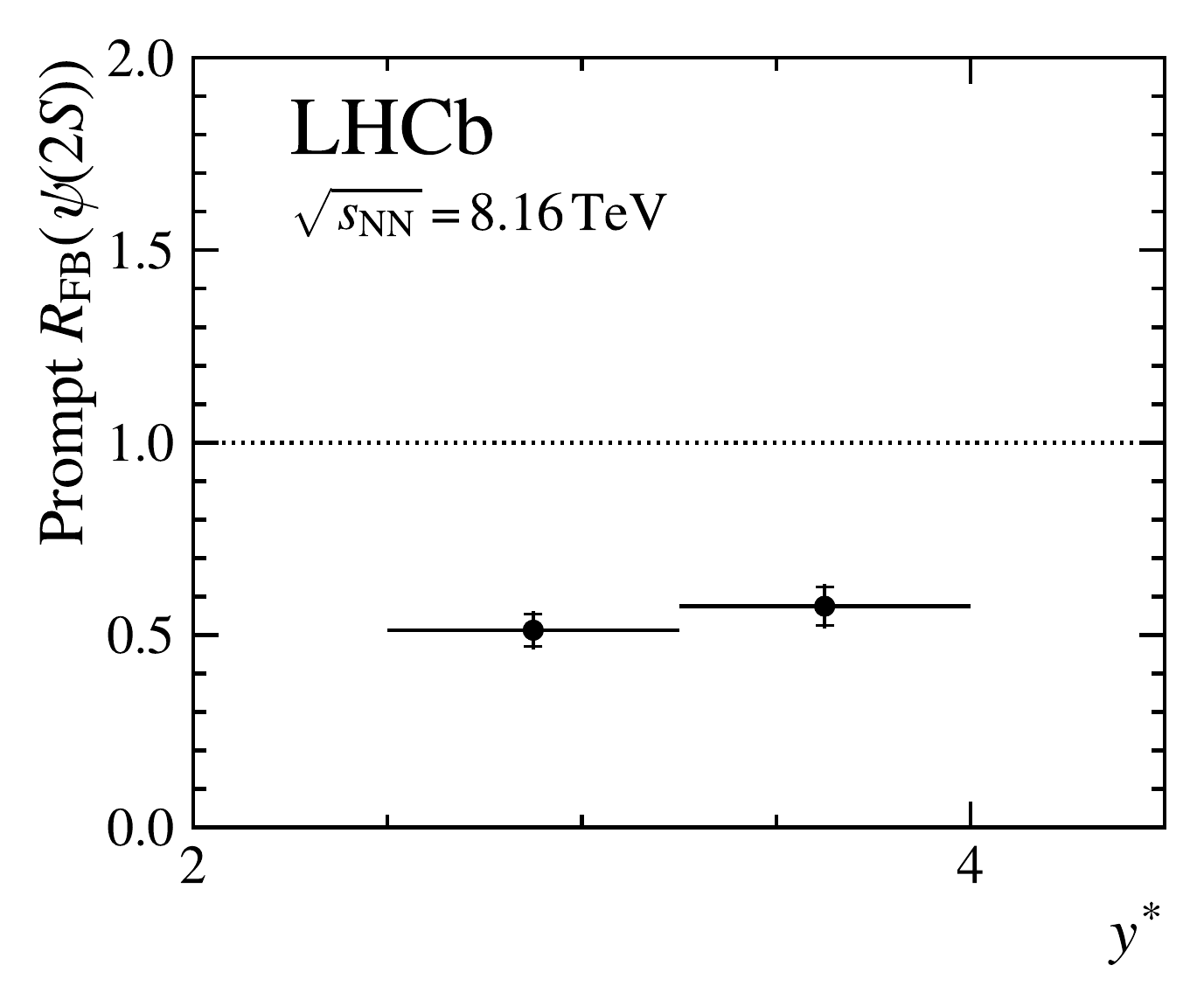}
\includegraphics[width=7.85cm]{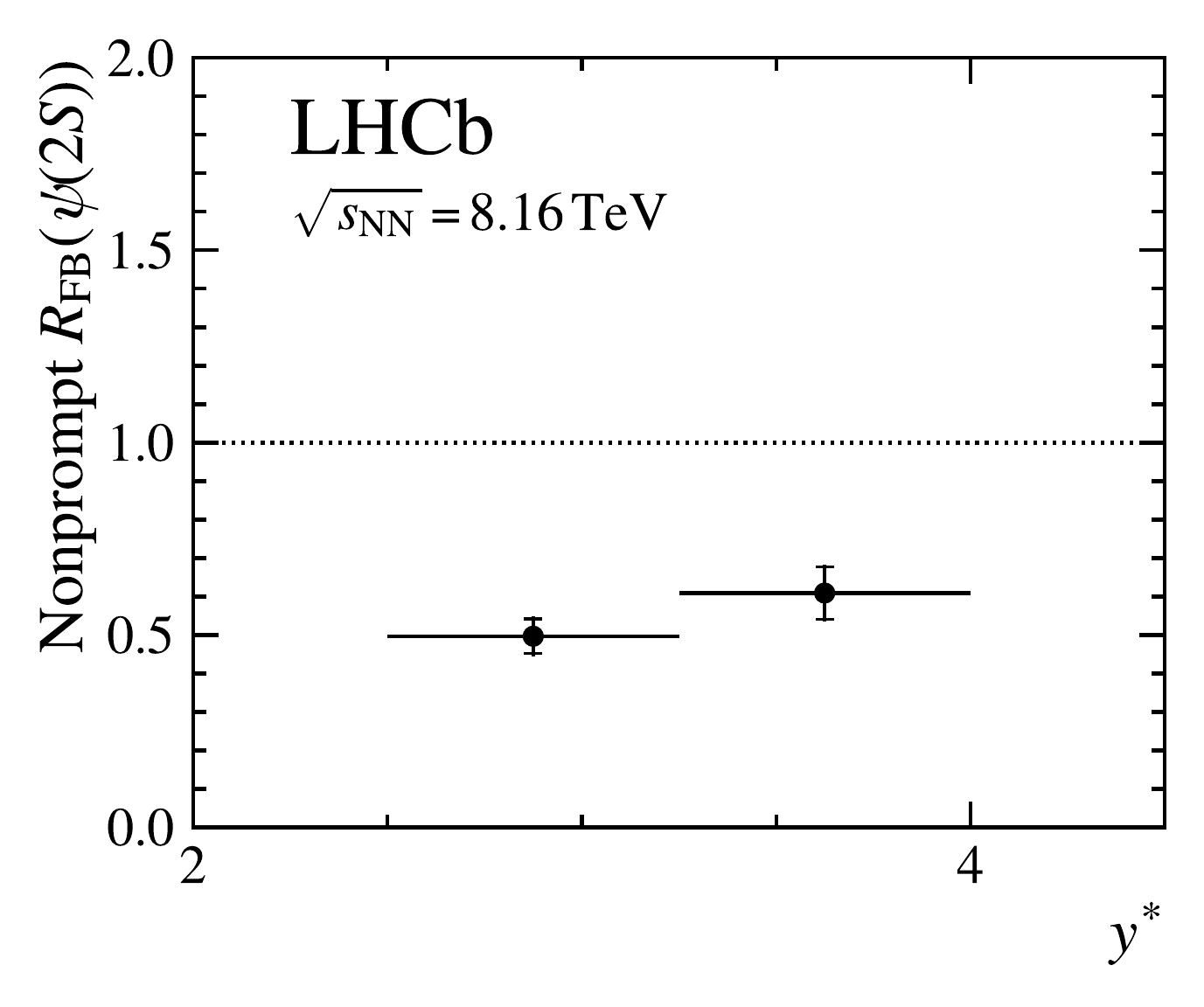}
 \caption{\small Forward-to-backward ratio as a function of \pt
 and integrated over $2.5<|y^{*}|<4.0$ for (top left) prompt and (top right) nonprompt \psitwos;
 as a function of $y^*$ and integrated over \pt  for (bottom left) 
 prompt and (bottom right) nonprompt \psitwos as a function of \pt.  Horizontal error bars are the bin widths,
vertical error bars represent the statistical and total uncertainties.}
    \label{fig:rfb_psi2s}
 \end{center}
 \end{figure}

The \psitwos forward-to-backward ratios
for prompt and nonprompt \psitwos production integrated over $y^*$ as a function of \pt 
and integrated over \pt as a function of $y^*$ are
shown in Fig.~\ref{fig:rfb_psi2s}.  

\clearpage
\subsection{Double cross-section ratios}

In this section, several predictions of phenomenological approaches are compared with the experimental data. 
These comparisons use ratios of nuclear modification factors, ``cross-section double ratios", since many sources of uncertainty, both theoretical and experimental, cancel out to a large extent.
The double ratios are defined as
\begin{equation}
R_{\psitwos/\jpsi}^{p{\rm Pb}} = 
\frac{R_{p{\rm Pb}}(\psitwos)}{R_{p{\rm Pb}}(\jpsi)} =
\frac{\left[\frac{\sigma(\psitwos)}{\sigma(\jpsi)}\right]_{p{\rm Pb}}}{\left[\frac{\sigma(\psitwos)}{\sigma(\jpsi)}\right]_{pp}}.
\end{equation}
The value of this ratio is expected to be equal to one in the case of 
nonprompt production since the modifications of the production due
to the medium affect only the  $b$-hadron production and not the 
final \psitwos or \jpsi states. It should also be equal to one if 
the production of the \psitwos and \jpsi in proton-lead collisions
is only modified by initial-state effects and not final-state effects.

\begin{figure}[htpb]
\begin{center}
\includegraphics[width=7.85cm]{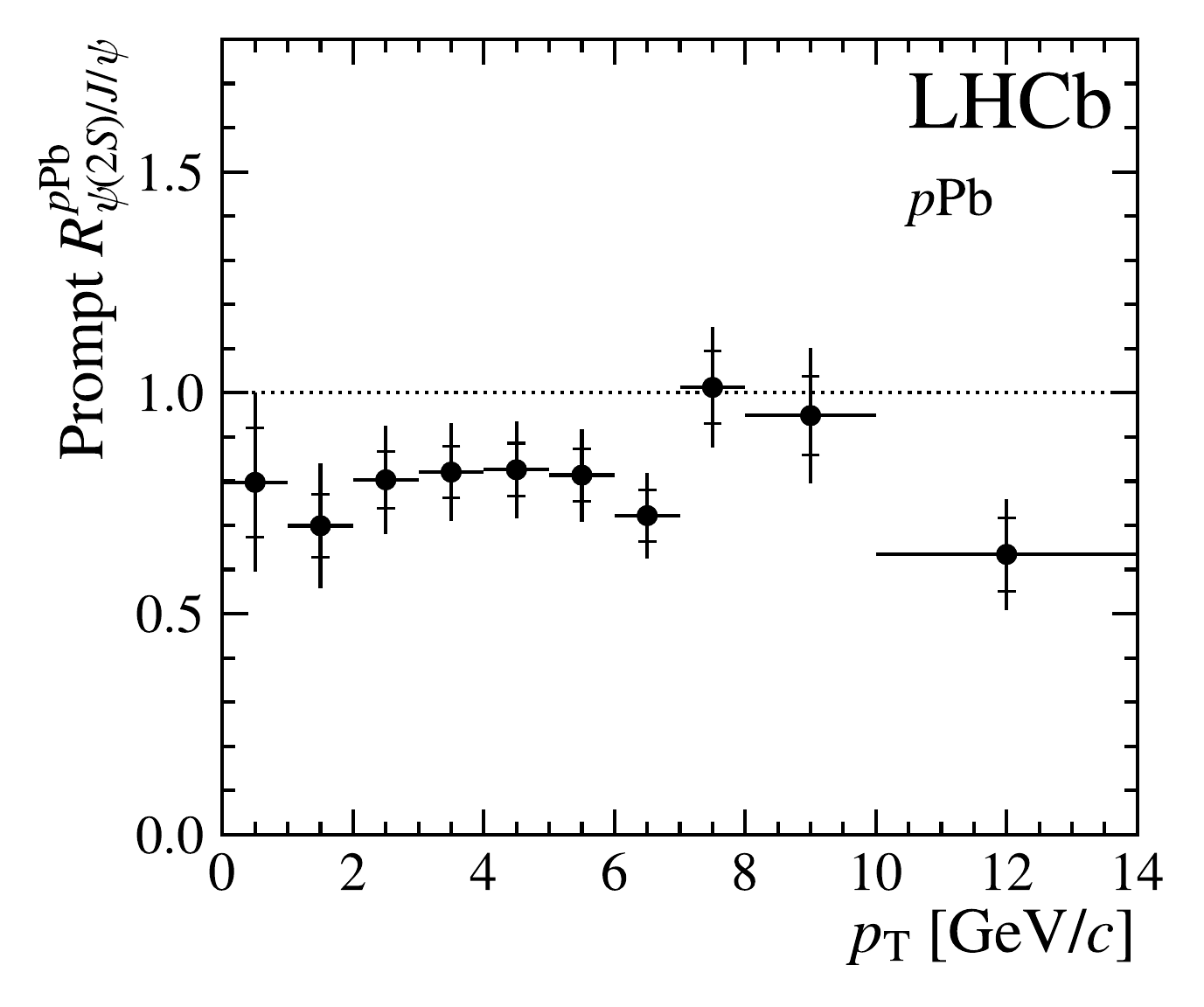}
\includegraphics[width=7.85cm]{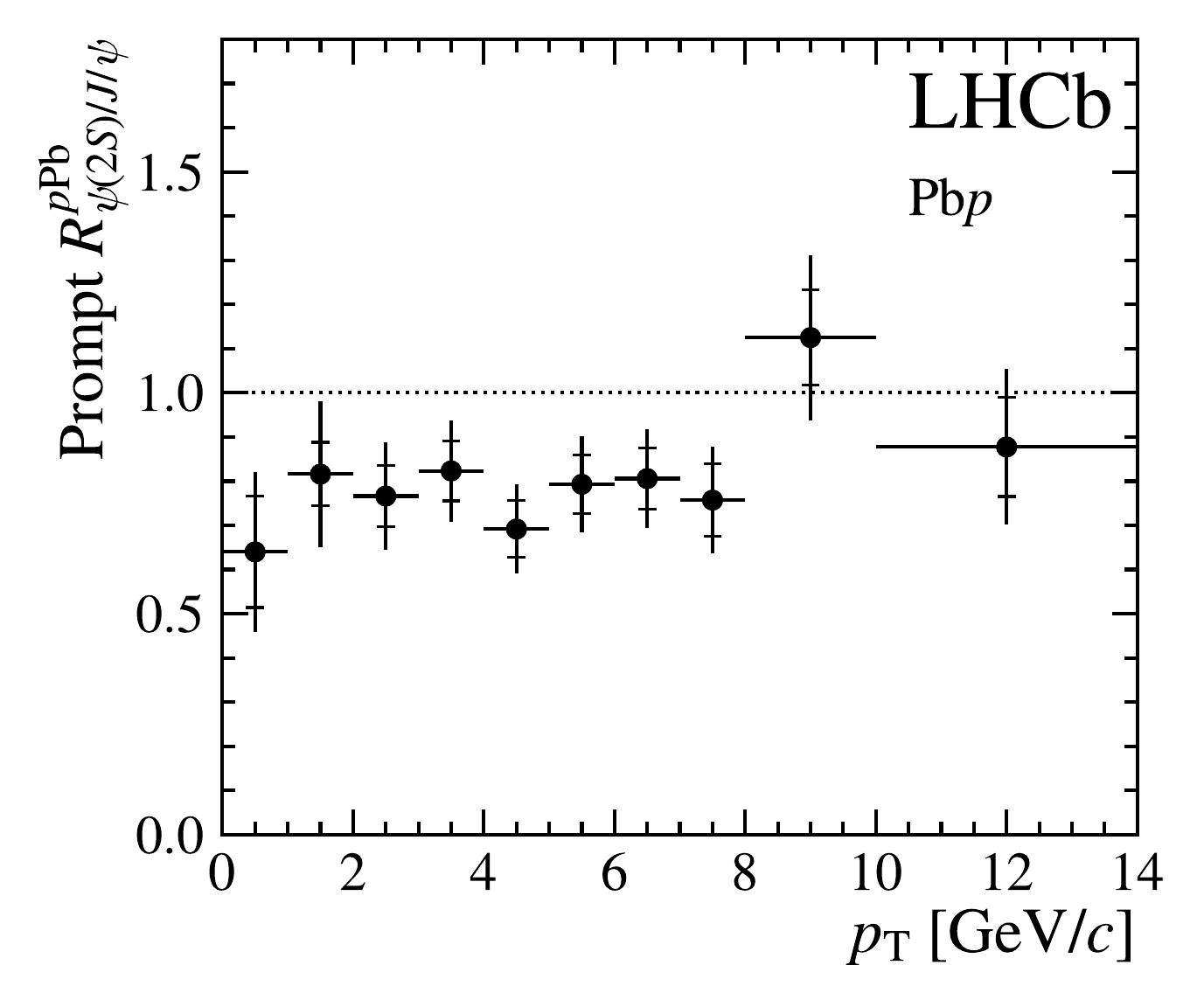}
\includegraphics[width=7.85cm]{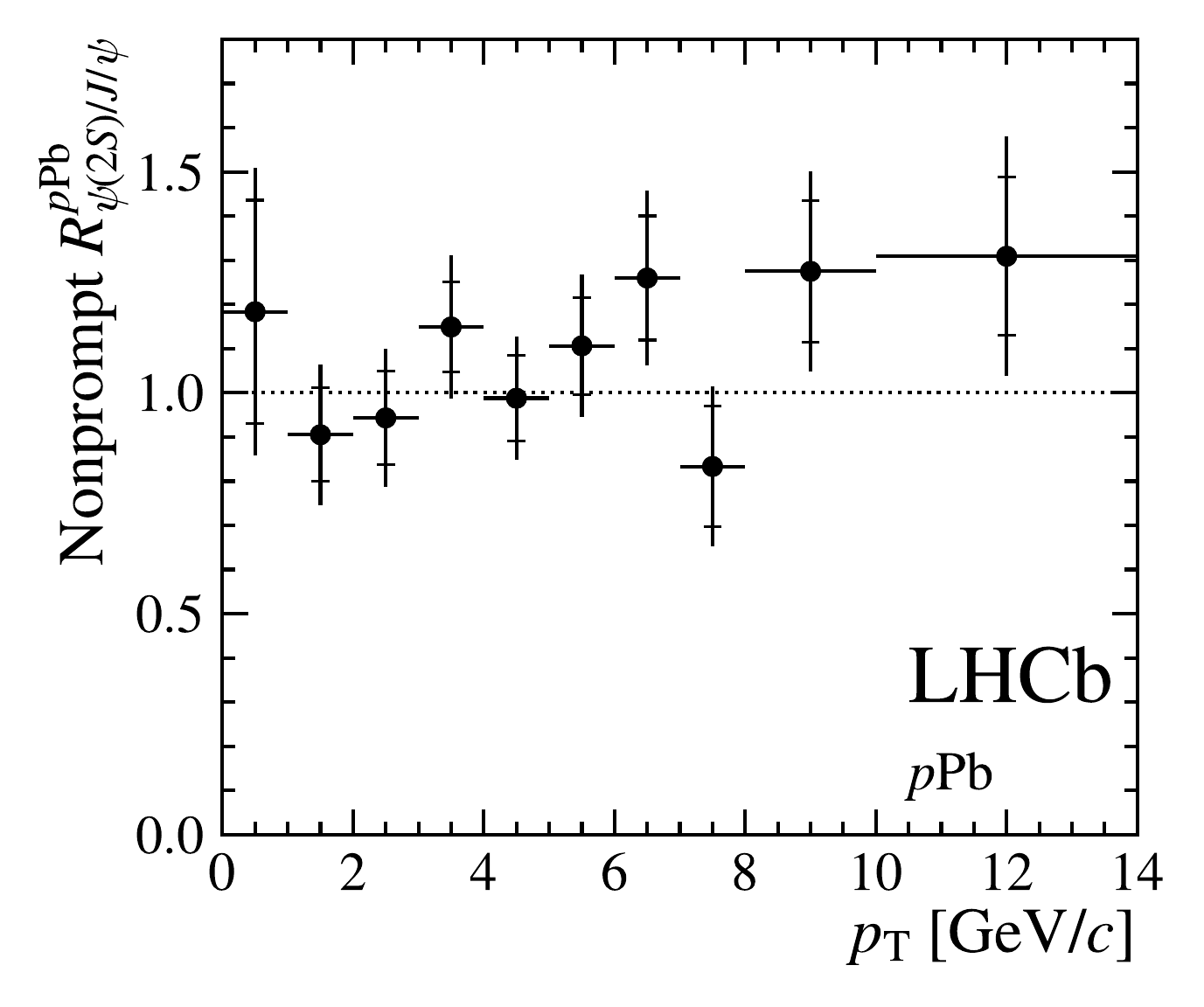}
\includegraphics[width=7.85cm]{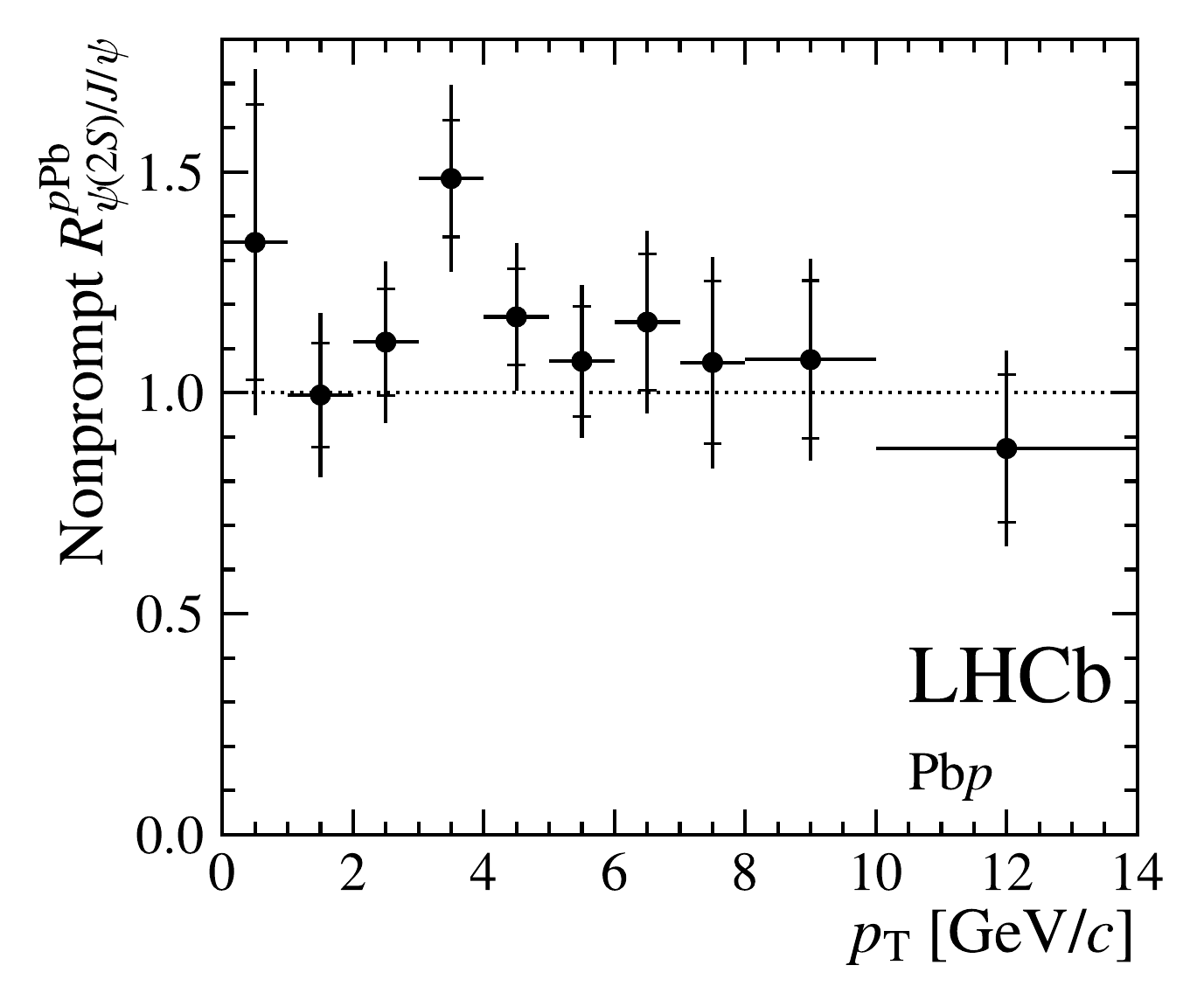}
\caption{\small Cross-section double ratios 
$R_{\psitwos/\jpsi}^{p{\rm Pb}}$
as a function of $\pt$ for (top left) prompt production in $p$Pb collisions, integrated over $1.5<y^*<4.0$, (top right) prompt production in Pb$p$ collisions, 
integrated over $-5.0<y^*<-2.5$,
(bottom left) nonprompt production in $p$Pb collisions, integrated over $1.5<y^*<4.0$, (bottom right) nonprompt production in Pb$p$ collisions, 
integrated over $-5.0<y^*<-2.5$. Horizontal error bars are the bin widths,
vertical error bars represent the statistical and total uncertainties.
}
\label{fig:double_ratio_pt}
\end{center}
\end{figure}

\begin{figure}[htpb]
\begin{center}
\includegraphics[width=7.8cm]{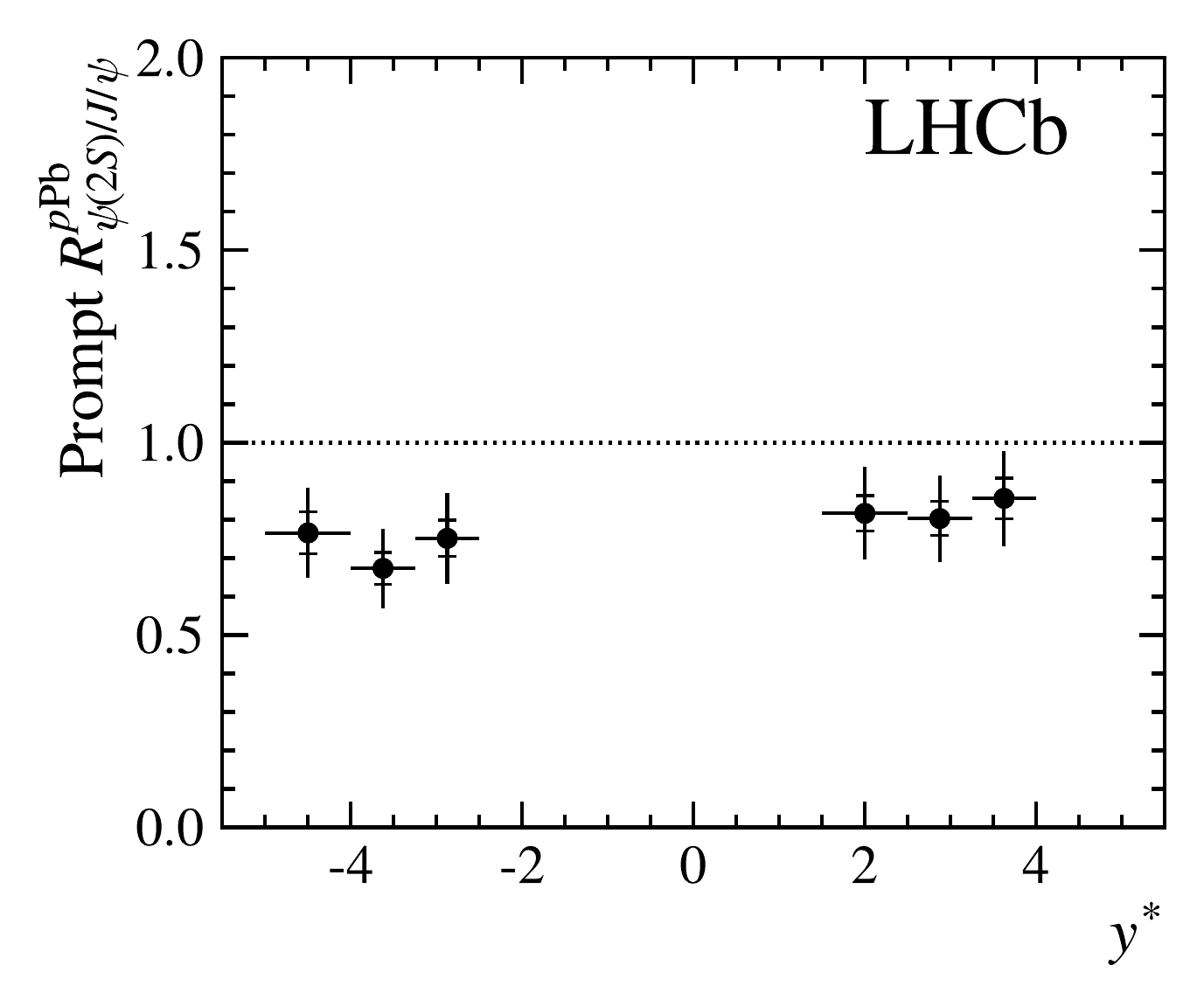}
\includegraphics[width=7.8cm]{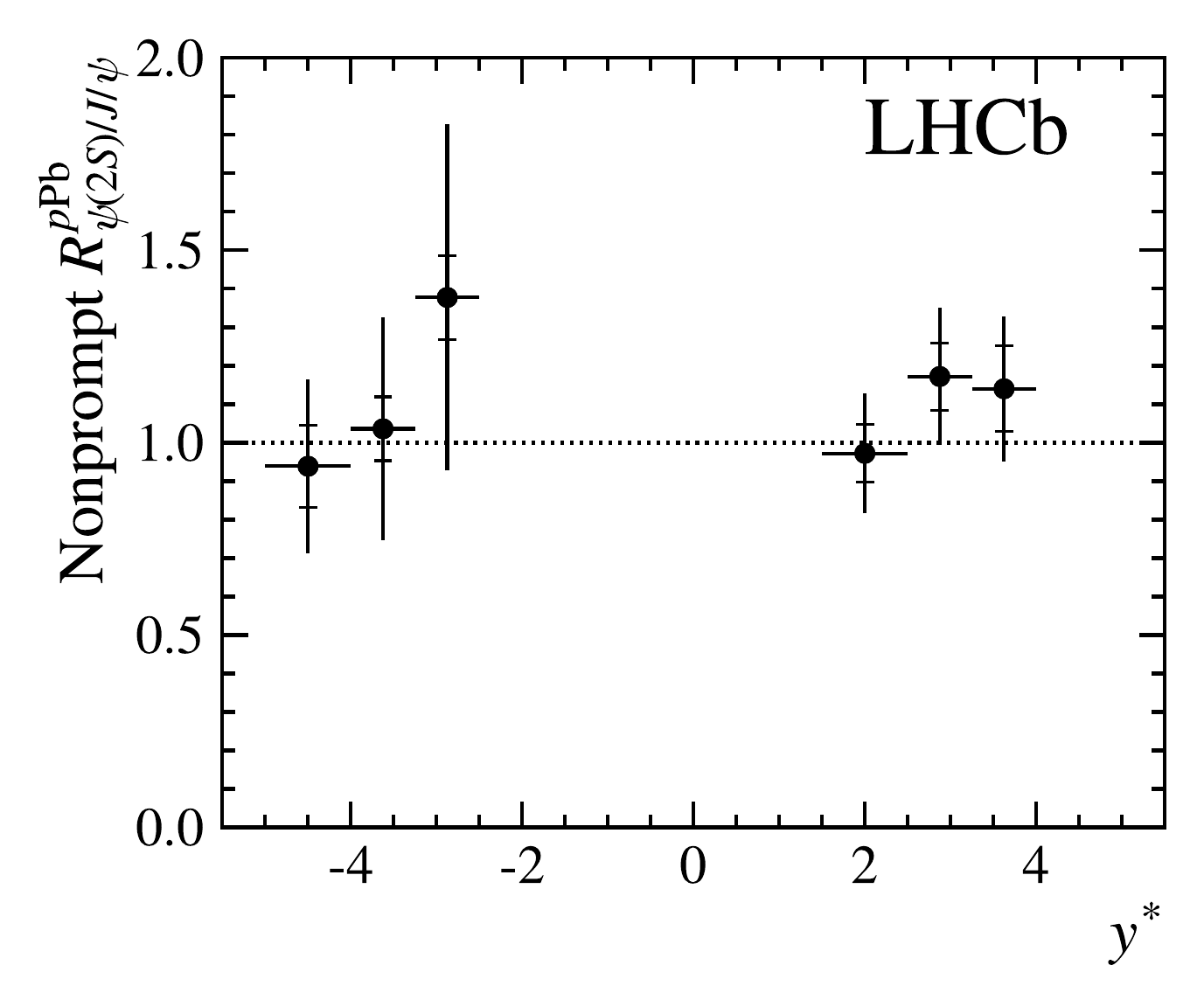}
\caption{\small Cross-section double ratios $R_{\psitwos/\jpsi}^{p{\rm Pb}}$ as 
a function of $y^{*}$, for (left) prompt and (right) nonprompt production. Horizontal error bars are the bin widths,
vertical error bars represent the statistical and total uncertainties.}
 \label{fig:double_ratio_y}
 \end{center}
 \end{figure}

\begin{figure}[htpb]
\begin{center}
\includegraphics[width=7.8cm]{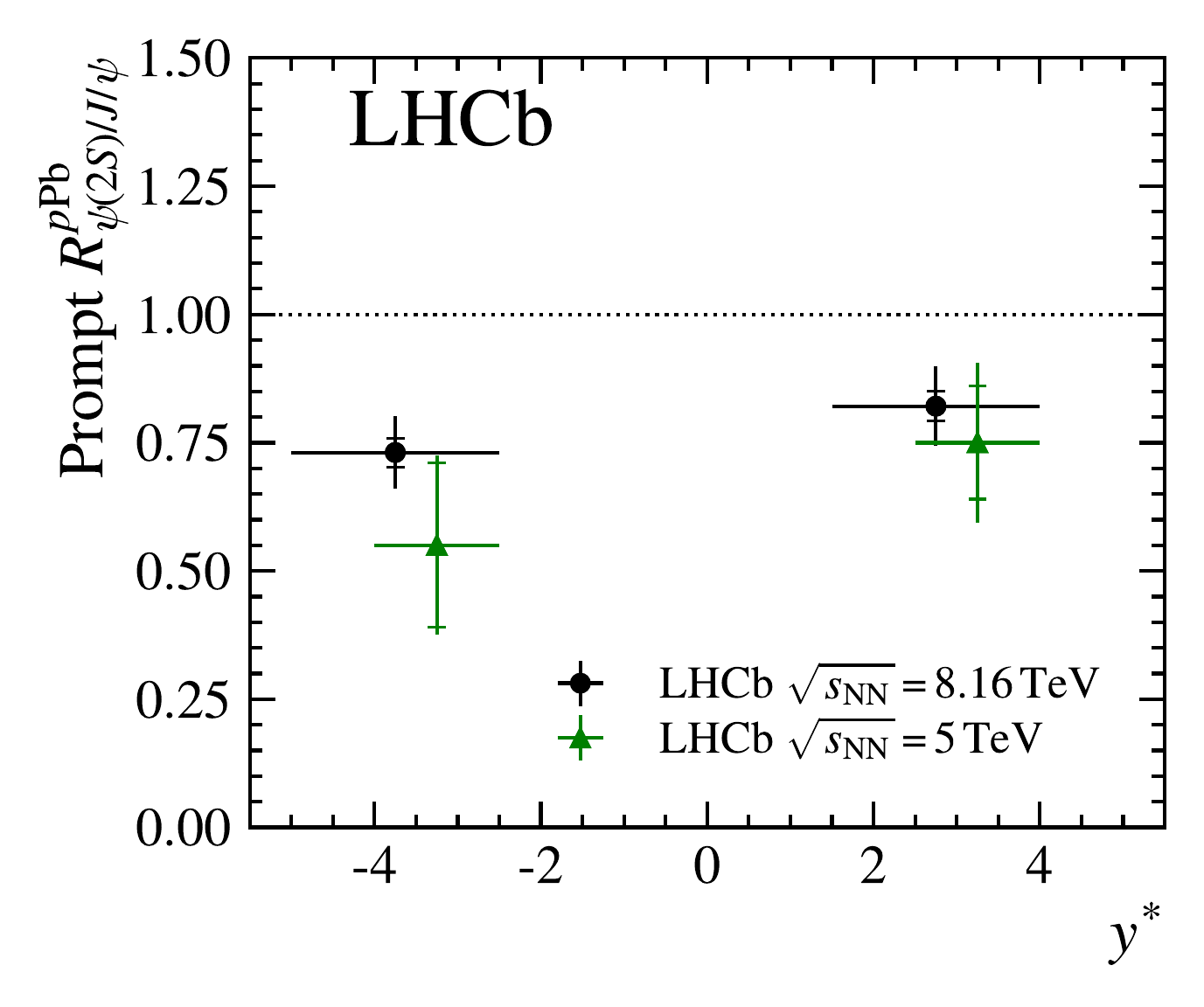}
\includegraphics[width=7.8cm]{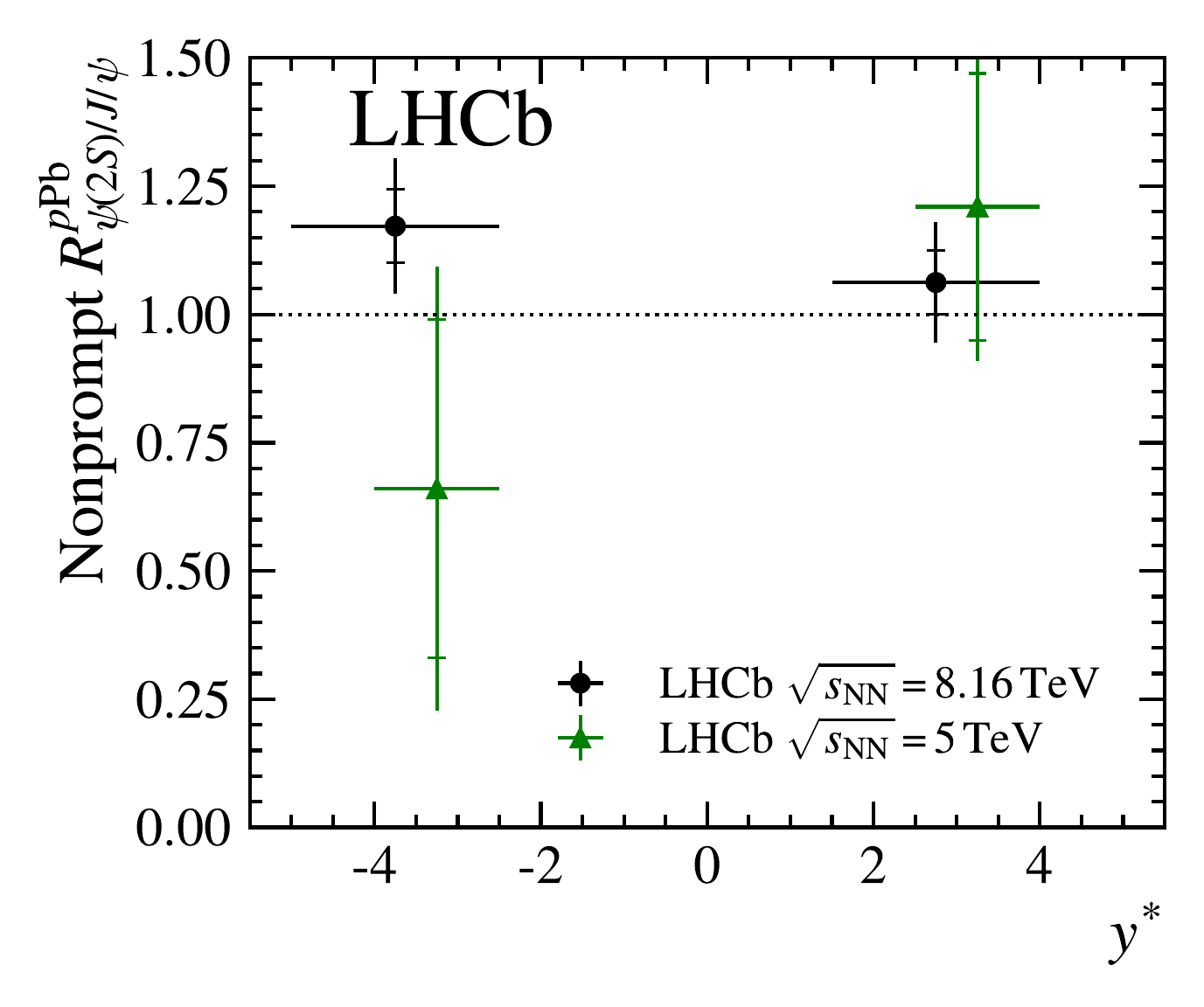}
\caption{\small Cross-section double ratios $R_{\psitwos/\jpsi}^{p{\rm Pb}}$ 
as a function of $y^{*}$, for (left) prompt and (right) 
nonprompt production. The green triangles with error bars correspond 
to the LHCb measurement at 5\tev~\cite{LHCb-PAPER-2015-058}. Horizontal error bars are the bin widths,
vertical error bars represent the statistical and total uncertainties.}
\label{fig:double_ratio_tot}
\end{center}
\end{figure}

Their values as a function of \pt and integrated over $y^{*}$ for prompt and nonprompt production are shown in
Fig.~\ref{fig:double_ratio_pt} and as a function of rapidity and integrated over \pt in
Fig.~\ref{fig:double_ratio_y}. The ratio integrated over \pt is shown in Fig.~\ref{fig:double_ratio_tot}. It is compared with the  measurement 
at $\sqsnn=5\tev$~\cite{LHCb-PAPER-2015-058}.  The results obtained at 8.16\tev 
are more precise than those at 5\tev; they are compatible with each other within uncertainties.

The double-ratio results for nonprompt production are compatible with unity as expected.  
The results for prompt production indicate a larger suppression of the excited \psitwos state compared to the \jpsi state. 
This confirms earlier, similar results obtained at 5\tev by the ALICE~\cite{Abelev:2014zpa}, CMS~\cite{CMS:2018gbb} and LHCb~\cite{LHCb-PAPER-2015-058} collaborations, 
at 8.16\tev by the ALICE collaboration~\cite{ALICE:2020vjy} and at 200\gev 
by the PHENIX collaboration~\cite{Adare:2016psx}. This phenomenon is not explained with the phenomenological models used for the description of \jpsi
production. They assume that the nuclear suppression of the quarkonium state is induced at a timescale shorter than the hadronisation timescale. 
It follows that the mechanisms should apply universally to the \jpsi and \psitwos states. Due to the proximity of the \jpsi and \psitwos masses, 
the nuclear effects are expected to be the same for the two mesons. These considerations apply to all main classes of modification models
usually considered: nPDF modifications~\cite{Lansberg:2016deg,Shao:2015vga,Shao:2012iz}, 
the standard CGC framework~\cite{Ma:2015sia,Ducloue:2015gfa} or coherent energy loss~\cite{Arleo:2012rs}.

A good understanding of the suppression of \psitwos over \jpsi production is crucial for the interpretation of measurements in heavy-ion collisions such as in PbPb 
collisions at the LHC. In PbPb collisions,  the previously mentioned models predict different behaviours of the \psitwos to \jpsi cross-section ratio in the presence of 
a deconfined phase. Therefore, the measurement of this production ratio in $p$Pb or Pb$p$ collisions, where a deconfined system is not expected 
to be created, provides important inputs to the models to describe the charmonium behaviour in heavy-ion collisions.

The  modification of prompt \psitwos production compared to that of the \jpsi in $p$Pb or Pb$p$ collisions can be caused by interactions at late stages of the collision
that do not obey simple QCD factorisation. There are different models exploiting this idea:
\begin{itemize}
\item The Comover model~\cite{Ferreiro:2014bia}: according to the observed final-state particle density, it assumes an interaction cross-section between the 
quarkonium and a ``comoving" medium, composed of particles travelling along with the $c\bar{c}$ 
quark pair. This cross-section depends on the size, hence binding enery, of the quarkonium state, \ie is larger for the \psitwos excited state 
than for the \jpsi ground state.
\item A combined CGC and improved Color Evaporation Model (GCG+ICEM)~\cite{Ma:2017rsu}: the short distance production of the charm and anti-charm pair is described by the standard CGC model,
while the hadronisation into the \jpsi or \psitwos states is computed with the ICEM.
The effects breaking factorisation in the hadronisation are due to additional parton comovers.
They are modelled with a cutoff parameter $\Lambda$ that represents the average gluon 
\pt kick.
\end{itemize}

\begin{figure}[htpb]
\begin{center}
\includegraphics[width=7.85cm]{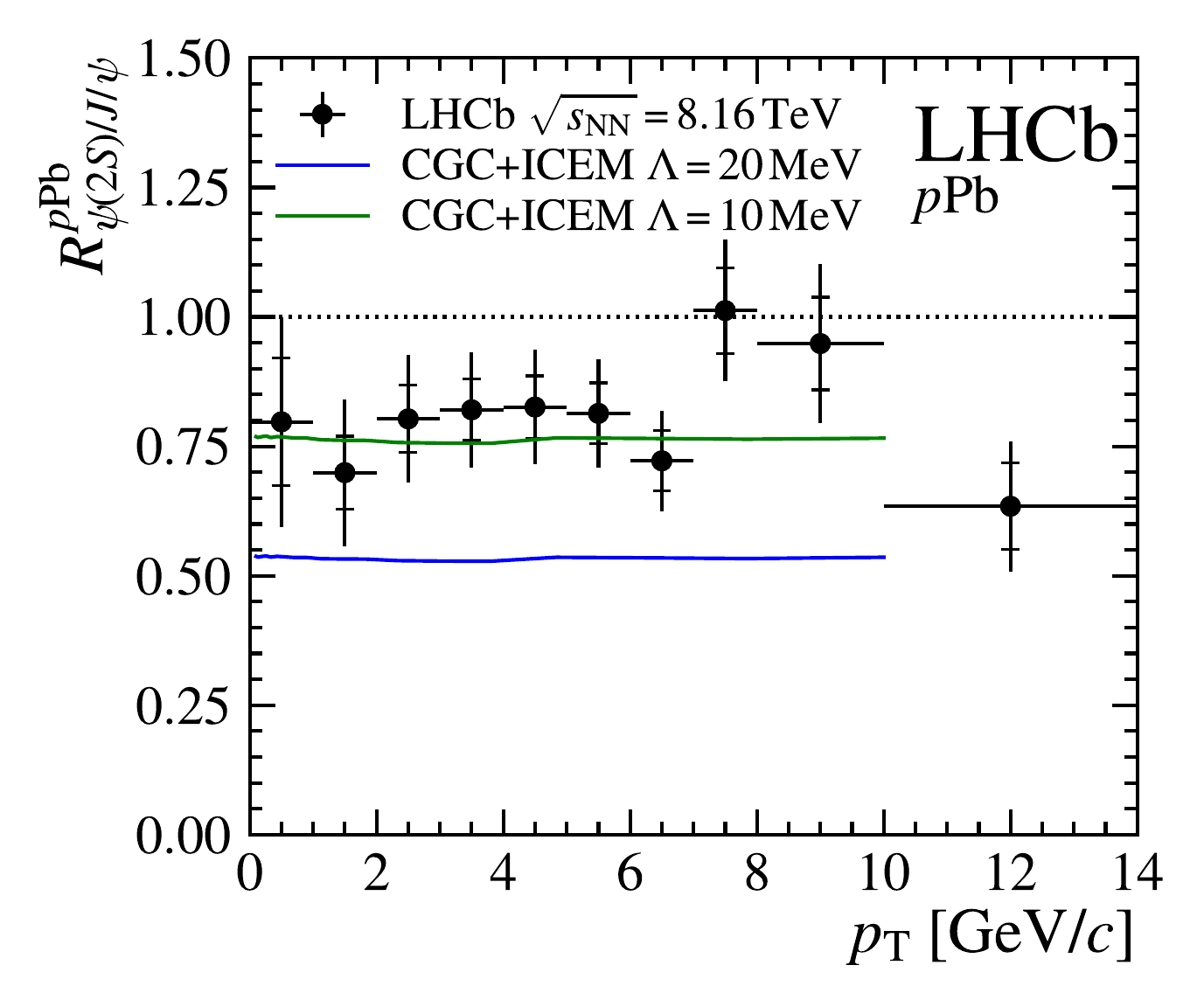}
\includegraphics[width=7.85cm]{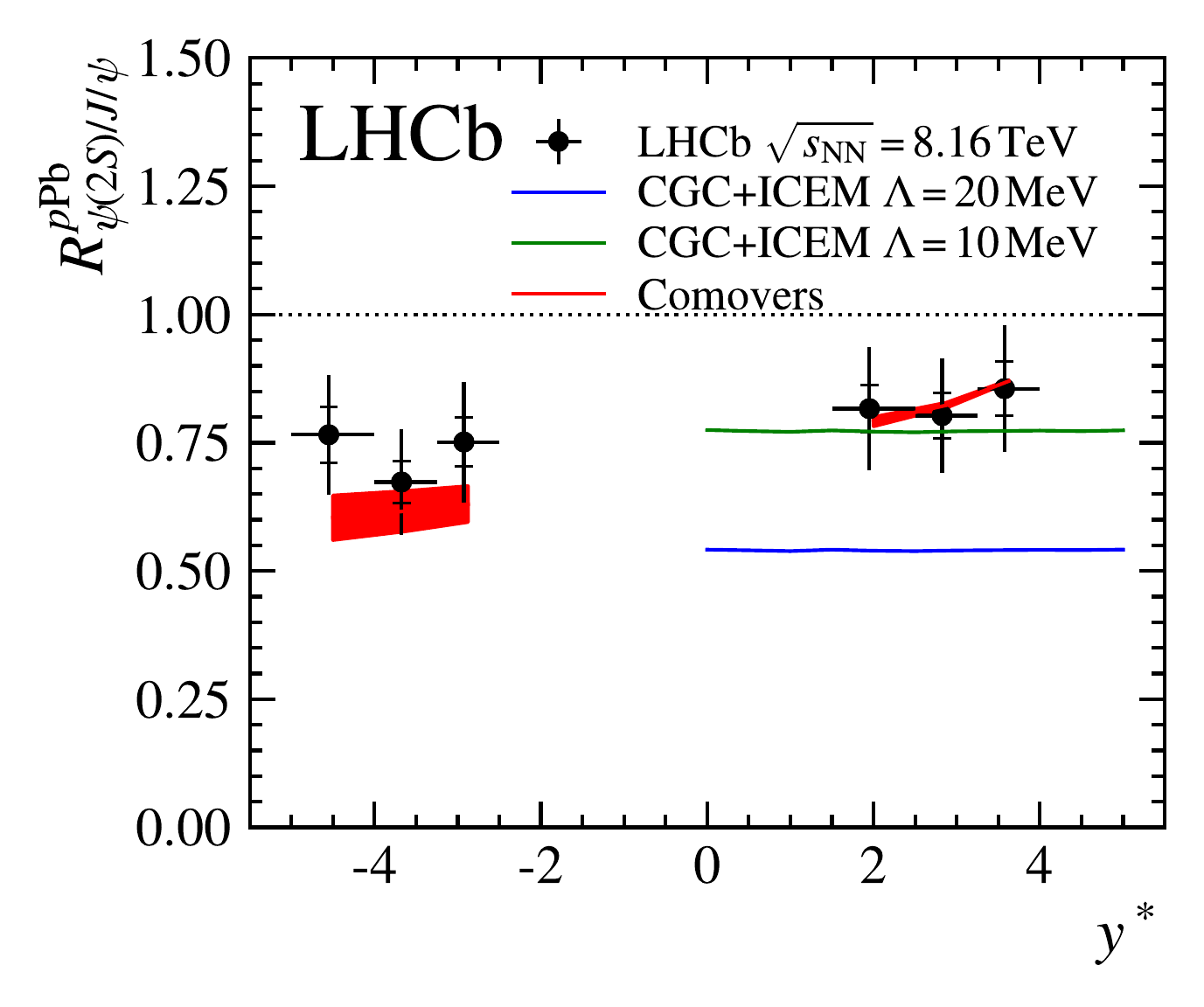}
\caption{\small Prompt cross-section double ratios $R_{\psitwos/\jpsi}^{p{\rm Pb}}$  (left) as a function of \pt and integrated over $1.5<y^*<4.0$, and (right) as a function of $y^{*}$ 
and integrated over \pt, compared with a CGC soft gluon interaction model with a scale (green) $\Lambda=10\mev$ and (blue) $\Lambda=20\mev$ and a (red) comover model (only for the $y^*$ dependence). Horizontal error bars are the bin widths,
vertical error bars the statistical and total uncertainties.}
 \label{fig:double_ratio_pt_models}
   \end{center}
      \end{figure}

Figure~\ref{fig:double_ratio_pt_models} (left) shows the comparison between the data and the CGC+ICEM model~\cite{Ma:2017rsu} with two different values of $\Lambda$, 10\mev and 20\mev, as a function of \pt in the $p$Pb configuration for prompt production. 
Figure~\ref{fig:double_ratio_pt_models} (right) shows, in addition, the comparison with the Comover model~\cite{Ferreiro:2014bia}, as a function of $y^*$. 
The models can describe the data reasonably well with appropriate parameter choices that also describe RHIC data at lower collision energy~\cite{Ferreiro:2014bia,Ma:2017rsu}. 
Due to the feed-down of \psitwos decays contributing to the \jpsi production, the difference in nuclear modification between the two states  
implies an effect on the \jpsi production which is not included in the computations. However, given the small contribution of the \psitwos feed-down decays (10 to
20\%), and the additional observed suppression of 25\%, these effects are below the size of the data uncertainties.

\section{Conclusion}

The measurement of \psitwos production in $p$Pb and Pb$p$ collisions at $\sqsnn=8.16\tev$ is presented and is compared 
with \jpsi production in the same collision system and with the production in $pp$ collisions, at the same center-of-mass energy.
The ratio of prompt \psitwos and \jpsi modification factors, $R_{p{\rm Pb}}(\psitwos)/R_{p{\rm Pb}}(\jpsi)$, is measured 
 to be below unity, indicating factorisation breaking with respect to the final state. The ratio of nonprompt \psitwos and \jpsi modification factors is compatible with one, as expected since the nuclear effects affect $b$-hadron production but 
 not their decays. 
 These findings are supported by the corresponding measurements 
 of the nuclear modification factor. This additional suppression seen 
 in prompt \psitwos production is compatible between the forward 
 ($p$Pb collisions) and 
 backward (Pb$p$ collisions) rapidity ranges and amounts to approximately 25\% in the modification factor ratio integrated over \pt. With increasing \pt, the suppression tends to be smaller
 although large statistical uncertainties at high \pt prevent firm statements. The observed behaviour can be described by models featuring late-stage 
 interactions  breaking preferentially \psitwos mesons compared to \jpsi mesons. These findings are important to constrain
 factorisation breaking with 
 respect to the final state in nuclear collisions in order to interpret quarkonium data in heavy-ion collisions. 

%% file: acknowledgements.tex
\section*{Acknowledgements}
%
%
\noindent We express our gratitude to our colleagues in the CERN
accelerator departments for the excellent performance of the LHC. We
thank the technical and administrative staff at the LHCb
institutes.
We acknowledge support from CERN and from the national agencies:
CAPES, CNPq, FAPERJ and FINEP (Brazil); 
MOST and NSFC (China); 
CNRS/IN2P3 (France); 
BMBF, DFG and MPG (Germany); 
INFN (Italy); 
NWO (Netherlands); 
MNiSW and NCN (Poland); 
MCID/IFA (Romania); 
MICINN (Spain); 
SNSF and SER (Switzerland); 
NASU (Ukraine); 
STFC (United Kingdom); 
DOE NP and NSF (USA).
We acknowledge the computing resources that are provided by CERN, IN2P3
(France), KIT and DESY (Germany), INFN (Italy), SURF (Netherlands),
PIC (Spain), GridPP (United Kingdom), 
CSCS (Switzerland), IFIN-HH (Romania), CBPF (Brazil),
and Polish WLCG (Poland).
We are indebted to the communities behind the multiple open-source
software packages on which we depend.
Individual groups or members have received support from
ARC and ARDC (Australia);
Key Research Program of Frontier Sciences of CAS, CAS PIFI, CAS CCEPP, 
Fundamental Research Funds for the Central Universities, 
and Sci. \& Tech. Program of Guangzhou (China);
Minciencias (Colombia);
EPLANET, Marie Sk\l{}odowska-Curie Actions, ERC and NextGenerationEU (European Union);
A*MIDEX, ANR, IPhU and Labex P2IO, and R\'{e}gion Auvergne-Rh\^{o}ne-Alpes (France);
AvH Foundation (Germany);
ICSC (Italy); 
GVA, XuntaGal, GENCAT, Inditex, InTalent and Prog.~Atracci\'on Talento, CM (Spain);
SRC (Sweden);
the Leverhulme Trust, the Royal Society
 and UKRI (United Kingdom).

%% file: appendix.tex
\clearpage
\section*{Tables with numerical results}

The tables in the following appendices show the numerical values
corresponding to the figures presenting the measurements. 
The cross-section ratios of \psitwos over \jpsi, in bins of \pt and $y^*$, are given in Table~\ref{tab:promptcrosssectionresultspPb_ratio} for prompt production in $p$Pb collisions and Table~\ref{tab:promptcrosssectionresultsPbp_ratio} in Pb$p$ collisions. The same ratios for 
nonprompt production are in Table~\ref{tab:nonpromptcrosssectionresultspPb_ratio} in $p$Pb collisions and in Table~\ref{tab:nonpromptcrosssectionresultsPbp_ratio} in Pb$p$ collisions. 
These ratios integrated over $y^*$, in bins of \pt are shown in Table~\ref{tab:promptcrosssectionresultspPb_ratio_p} for $p$Pb collisions and Table~\ref{tab:promptcrosssectionresultsPbp_ratio_p} for Pb$p$ collisions for prompt production, and in Table~\ref{tab:nonpromptcrosssectionresultspPb_ratio_p}
and Table~\ref{tab:nonpromptcrosssectionresultsPbp_ratio_p} for
Pb$p$ collisions. The ratios integrated over
\pt, in bins of $y^*$ are given in Table~\ref{tab:promptcrosssectionresults_ratio_y} for
prompt production and Table~\ref{tab:nonpromptcrosssectionresults_ratio_y} for nonprompt production. 

The asbolute \psitwos cross-sections in bins of \pt and $y^*$ are available in Table~\ref{tab:promptcrosssectionresultspPb} for $p$Pb
collisions and Table~\ref{tab:promptcrosssectionresultsPbp} 
for Pb$p$ collisions, for prompt production, and in Table~\ref{tab:nonpromptcrosssectionresultspPb} for 
$p$Pb collisions and Table~\ref{tab:nonpromptcrosssectionresultsPbp} for Pb$p$
collisions for nonprompt production. The values
integrated over $y^*$, in bins of \pt, are in Table~\ref{tab:promptcrosssectionresultspPb_p} for prompt
production in $p$Pb collisions and in Table~\ref{tab:promptcrosssectionresultsPbp_p} in Pb$p$ collisions; the values for nonprompt production are in 
Table~\ref{tab:nonpromptcrosssectionresultspPb_p} for
$p$Pb collisions and in Table~\ref{tab:nonpromptcrosssectionresultsPbp_p} Pb$p$ collisions. 

The \psitwos nuclear modification factors in bins of \pt, integrated over $y^*$, are given in Table~\ref{tab:rpAprompt_pt} for prompt production in $p$Pb collisions, in Table~\ref{tab:rApprompt_pt} for prompt
production in Pb$p$ collisions, in Table~\ref{tab:rpAnonprompt_pt} for nonprompt production in 
$p$Pb collisions and Table~\ref{tab:rApnonprompt_pt} for 
nonprompt production in Pb$p$ collisions. 
The \psitwos forward-to-backward ratios are shown, in bins
of \pt integrated over $y^*$, in Table~\ref{tab:rFBprompt_pt} for prompt production and 
in Table~\ref{tab:rFBnonprompt_pt} for nonprompt production. They are also given in bins of $y^*$ integrated over \pt in 
Table~\ref{tab:rFBprompt_y} for prompt production and 
in Table~\ref{tab:rFBnonprompt_y} for nonprompt production.

The \psitwos to \jpsi double cross-section ratios are shown 
in bins of \pt, integrated over $y^*$, in Table~\ref{tab:doublepromptpPb_pt} for prompt production
in $p$Pb collisions, in Table~\ref{tab:doublepromptPbp_pt}
for prompt production in $p$Pb collisions, in Table~\ref{tab:doublenonpromptpPb_pt}
for prompt production in $p$Pb collisions and in 
Table~\ref{tab:doublenonpromptPbp_pt}
 for nonprompt production in Pb$p$ collisions. The values
 in bins of $y^*$, integrated over \pt, are given in 
 Table~\ref{tab:doubleprompt_y} for prompt production 
 and in Table~\ref{tab:doublenonprompt_y} for nonprompt
 production. Finally, the ratios integrated over \pt and 
 $y^*$ are given in Table~\ref{tab:doubleprompt} for prompt
 production and in Table~\ref{tab:doublenonprompt} for 
 nonprompt production. 
 \clearpage
\appendix

\section{Cross-section ratios}

\input{tables/section_prompt_ratio_pPb}
\input{tables/section_prompt_ratio_Pbp}
\input{tables/section_nonprompt_ratio_pPb}
\input{tables/section_nonprompt_ratio_Pbp}

\input{tables/section_prompt_ratio_pPb_p}
\input{tables/section_prompt_ratio_Pbp_p}
\input{tables/section_nonprompt_ratio_pPb_p}
\input{tables/section_nonprompt_ratio_Pbp_p}

\input{tables/section_prompt_ratio_pPb_y}
\input{tables/section_nonprompt_ratio_pPb_y}
\clearpage
\section{\texorpdfstring{Absolute $\boldsymbol{\psitwos}$ cross-sections}{Absolute psi(2S) cross-sections}}
\input{tables/section_prompt_psi2s_pPb}
\input{tables/section_prompt_psi2s_Pbp}
\input{tables/section_nonprompt_psi2s_pPb}
\input{tables/section_nonprompt_psi2s_Pbp}

\input{tables/section_prompt_psi2s_pPb_p}
\input{tables/section_prompt_psi2s_Pbp_p}
\input{tables/section_nonprompt_psi2s_pPb_p}
\input{tables/section_nonprompt_psi2s_Pbp_p}

\clearpage

\section{\texorpdfstring{$\boldsymbol{\psitwos}$ nuclear modification factors}{psi(2S) nuclear modification factors}}

\input{tables/r_pA_psi2s_prompt_pt}
\input{tables/r_Ap_psi2s_prompt_pt}
\input{tables/r_pA_psi2s_nonprompt_pt}
\input{tables/r_Ap_psi2s_nonprompt_pt}

\clearpage

\section{\texorpdfstring{$\boldsymbol{\psitwos}$ forward-to-backward ratios}{psi(2S) forward-to-backward ratios}}

\input{tables/r_FB_psi2s_prompt_pt}
\input{tables/r_FB_psi2s_nonprompt_pt}

\input{tables/r_FB_psi2s_prompt_y}
\input{tables/r_FB_psi2s_nonprompt_y}

\clearpage

\section{\texorpdfstring{$\boldsymbol{\psitwos}$ to $\boldsymbol{\jpsi}$ double cross-section ratios}{psi(2S) to J/psi double cross-section ratios}}

\input{tables/double_ratio_prompt_pPb_pt}
\input{tables/double_ratio_prompt_Pbp_pt}
\input{tables/double_ratio_nonprompt_pPb_pt}
\input{tables/double_ratio_nonprompt_Pbp_pt}

\input{tables/double_ratio_prompt_y}
\input{tables/double_ratio_nonprompt_y}

\input{tables/double_ratio_prompt}
\input{tables/double_ratio_nonprompt}

\clearpage

%% file: tables/section_prompt_ratio_pPb.tex
\renewcommand{\arraystretch}{1.2}
\begin{table}[!htb]
\caption{\small Cross-section ratios of \psitwos over \jpsi prompt 
production in $p$Pb collisions. The first uncertainty is statistical
and the second systematic. 
}
\label{tab:promptcrosssectionresultspPb_ratio}
\centering
\begin{tabular}{@{}cl@{$\,<y^*<\,$}lr@{$\,\pm\,$}r@{$\,\pm\,$}r@{}}
\toprule
 \multicolumn{1}{c}{\pt interval (\!\gevc)} & \multicolumn{2}{c}{$y^*$ interval} & \multicolumn{3}{c}{$ \sigma_{\psi(2S)}/\sigma_{J/\psi}$} \\ 
 \midrule
 \phantom{1}0 $<\pt<$ 1 & 1.50 & 2.50 & $0.08$ & $0.02$ & $0.01$ \\
 \phantom{1}0 $<\pt<$  1 & 2.50 & 3.25 & $0.08$ & $0.02$ & $0.01$ \\
 \phantom{1}0 $<\pt<$  1 & 3.25 & 4.00 & $0.11$ & $0.02$ & $0.01$ \\
 \phantom{1}1 $<\pt<$  2 & 1.50 & 2.50 & $0.10$ & $0.02$ & $0.01$ \\
 \phantom{1}1 $<\pt<$  2 & 2.50 & 3.25 & $0.08$ & $0.01$ & $0.01$ \\
 \phantom{1}1 $<\pt<$  2 & 3.25 & 4.00 & $0.08$ & $0.02$ & $0.01$ \\
 \phantom{1}2 $<\pt<$  3 & 1.50 & 2.50 & $0.10$ & $0.01$ & $0.01$ \\
 \phantom{1}2 $<\pt<$  3 & 2.50 & 3.25 & $0.13$ & $0.02$ & $0.01$ \\
 \phantom{1}2 $<\pt<$  3 & 3.25 & 4.00 & $0.12$ & $0.02$ & $0.01$ \\
 \phantom{1}3 $<\pt<$  4 & 1.50 & 2.50 & $0.16$ & $0.02$ & $0.02$ \\
 \phantom{1}3 $<\pt<$  4 & 2.50 & 3.25 & $0.11$ & $0.02$ & $0.01$ \\
 \phantom{1}3 $<\pt<$  4 & 3.25 & 4.00 & $0.16$ & $0.02$ & $0.01$ \\
 \phantom{1}4 $<\pt<$  5 & 1.50 & 2.50 & $0.15$ & $0.02$ & $0.01$ \\
 \phantom{1}4 $<\pt<$  5 & 2.50 & 3.25 & $0.17$ & $0.02$ & $0.02$ \\
 \phantom{1}4 $<\pt<$  5 & 3.25 & 4.00 & $0.16$ & $0.02$ & $0.01$ \\
 \phantom{1}5 $<\pt<$  6 & 1.50 & 2.50 & $0.15$ & $0.02$ & $0.01$ \\
 \phantom{1}5 $<\pt<$  6 & 2.50 & 3.25 & $0.20$ & $0.02$ & $0.02$ \\
 \phantom{1}5 $<\pt<$  6 & 3.25 & 4.00 & $0.20$ & $0.03$ & $0.02$ \\
 \phantom{1}6 $<\pt<$  7 & 1.50 & 2.50 & $0.17$ & $0.02$ & $0.02$ \\
 \phantom{1}6 $<\pt<$  7 & 2.50 & 3.25 & $0.18$ & $0.02$ & $0.02$ \\
 \phantom{1}6 $<\pt<$  7 & 3.25 & 4.00 & $0.23$ & $0.03$ & $0.02$ \\
 \phantom{1}7 $<\pt<$  8 & 1.50 & 2.50 & $0.24$ & $0.03$ & $0.02$ \\
 \phantom{1}7 $<\pt<$  8 & 2.50 & 3.25 & $0.25$ & $0.03$ & $0.02$ \\
 \phantom{1}7 $<\pt<$  8 & 3.25 & 4.00 & $0.33$ & $0.05$ & $0.03$ \\
 \phantom{1}8 $<\pt<$ 10 & 1.50 & 2.50 & $0.28$ & $0.03$ & $0.03$ \\
 \phantom{1}8 $<\pt<$ 10 & 2.50 & 3.25 & $0.27$ & $0.03$ & $0.02$ \\
 \phantom{1}8 $<\pt<$ 10 & 3.25 & 4.00 & $0.20$ & $0.04$ & $0.02$ \\
10 $<\pt<$ 14 & 1.50 & 2.50 & $0.23$ & $0.04$ & $0.02$ \\
10 $<\pt<$ 14 & 2.50 & 3.25 & $0.18$ & $0.04$ & $0.02$ \\
10 $<\pt<$ 14 & 3.25 & 4.00 & $0.35$ & $0.09$ & $0.04$ \\
\bottomrule
\end{tabular}
\end{table}

%% file: tables/section_prompt_ratio_Pbp.tex
\renewcommand{\arraystretch}{1.2}
\begin{table}[!htb]
\caption{\small Cross-section ratios of \psitwos over \jpsi prompt 
production in Pb$p$ collisions. The first uncertainty is statistical
and the second systematic. 
}
\label{tab:promptcrosssectionresultsPbp_ratio}
\centering
\begin{tabular}{@{}cl@{$\,<y^*<\,$}lr@{$\,\pm\,$}r@{$\,\pm\,$}r@{}}
\toprule
 \multicolumn{1}{c}{\pt interval (\!\gevc)} & \multicolumn{2}{c}{$y^*$ interval} & \multicolumn{3}{c}{$\sigma_{\psi(2S)}/\sigma_{J/\psi}$} \\ 
 \midrule
 \phantom{1}0 $<\pt<$  1 & $-3.25$ & $-2.50$ & $0.07$ & $0.02$ & $0.01$ \\
 \phantom{1}0 $<\pt<$  1 & $-4.00$ & $-3.25$ & $0.07$ & $0.02$ & $0.01$ \\
 \phantom{1}0 $<\pt<$  1 & $-5.00$ & $-4.00$ & $0.06$ & $0.02$ & $0.01$ \\
 \phantom{1}1 $<\pt<$  2 & $-3.25$ & $-2.50$ & $0.11$ & $0.02$ & $0.01$ \\
 \phantom{1}1 $<\pt<$  2 & $-4.00$ & $-3.25$ & $0.09$ & $0.01$ & $0.01$ \\
 \phantom{1}1 $<\pt<$  2 & $-5.00$ & $-4.00$ & $0.12$ & $0.02$ & $0.01$ \\
 \phantom{1}2 $<\pt<$  3 & $-3.25$ & $-2.50$ & $0.10$ & $0.01$ & $0.01$ \\
 \phantom{1}2 $<\pt<$  3 & $-4.00$ & $-3.25$ & $0.09$ & $0.01$ & $0.01$ \\
 \phantom{1}2 $<\pt<$  3 & $-5.00$ & $-4.00$ & $0.11$ & $0.02$ & $0.01$ \\
 \phantom{1}3 $<\pt<$  4 & $-3.25$ & $-2.50$ & $0.13$ & $0.02$ & $0.01$ \\
 \phantom{1}3 $<\pt<$  4 & $-4.00$ & $-3.25$ & $0.15$ & $0.02$ & $0.02$ \\
 \phantom{1}3 $<\pt<$  4 & $-5.00$ & $-4.00$ & $0.10$ & $0.02$ & $0.01$ \\
 \phantom{1}4 $<\pt<$  5 & $-3.25$ & $-2.50$ & $0.12$ & $0.02$ & $0.01$ \\
 \phantom{1}4 $<\pt<$  5 & $-4.00$ & $-3.25$ & $0.12$ & $0.02$ & $0.01$ \\
 \phantom{1}4 $<\pt<$  5 & $-5.00$ & $-4.00$ & $0.13$ & $0.02$ & $0.01$ \\
 \phantom{1}5 $<\pt<$  6 & $-3.25$ & $-2.50$ & $0.15$ & $0.02$ & $0.02$ \\
 \phantom{1}5 $<\pt<$  6 & $-4.00$ & $-3.25$ & $0.15$ & $0.02$ & $0.02$ \\
 \phantom{1}5 $<\pt<$  6 & $-5.00$ & $-4.00$ & $0.18$ & $0.03$ & $0.02$ \\
 \phantom{1}6 $<\pt<$  7 & $-3.25$ & $-2.50$ & $0.18$ & $0.02$ & $0.02$ \\
 \phantom{1}6 $<\pt<$  7 & $-4.00$ & $-3.25$ & $0.17$ & $0.02$ & $0.02$ \\
 \phantom{1}6 $<\pt<$  7 & $-5.00$ & $-4.00$ & $0.21$ & $0.04$ & $0.02$ \\
 \phantom{1}7 $<\pt<$  8 & $-3.25$ & $-2.50$ & $0.15$ & $0.03$ & $0.02$ \\
 \phantom{1}7 $<\pt<$  8 & $-4.00$ & $-3.25$ & $0.16$ & $0.03$ & $0.02$ \\
 \phantom{1}7 $<\pt<$  8 & $-5.00$ & $-4.00$ & $0.28$ & $0.05$ & $0.03$ \\
 \phantom{1}8 $<\pt<$ 10 & $-3.25$ & $-2.50$ & $0.21$ & $0.03$ & $0.02$ \\
 \phantom{1}8 $<\pt<$ 10 & $-4.00$ & $-3.25$ & $0.35$ & $0.03$ & $0.04$ \\
 \phantom{1}8 $<\pt<$ 10 & $-5.00$ & $-4.00$ & $0.33$ & $0.06$ & $0.04$ \\
10 $<\pt<$ 14 & $-3.25$ & $-2.50$ & $0.20$ & $0.04$ & $0.02$ \\
10 $<\pt<$ 14 & $-4.00$ & $-3.25$ & $0.38$ & $0.05$ & $0.04$ \\
10 $<\pt<$ 14 & $-5.00$ & $-4.00$ & $0.41$ & $0.09$ & $0.05$ \\
\bottomrule
\end{tabular}
\end{table}

%% file: tables/section_nonprompt_ratio_pPb.tex
\renewcommand{\arraystretch}{1.2}
\begin{table}[!htb]
\caption{\small Cross-section ratios of \psitwos over \jpsi nonprompt 
production in $p$Pb collisions. The first uncertainty is statistical
and the second systematic. 
}
\label{tab:nonpromptcrosssectionresultspPb_ratio}
\centering
\begin{tabular}{@{}cl@{$\,<y^*<\,$}lr@{$\,\pm\,$}r@{$\,\pm\,$}r@{}}
\toprule
 \multicolumn{1}{c}{\pt interval (\!\gevc)} & \multicolumn{2}{c}{$y^*$ interval} & \multicolumn{3}{c}{$\sigma_{\psi(2S)}/\sigma_{J/\psi}$} \\ 
 \midrule
 \phantom{1}0 $<\pt<$  1 & 1.50 & 2.50 & $0.19$ & $0.06$ & $0.02$ \\
 \phantom{1}0 $<\pt<$  1 & 2.50 & 3.25 & $0.27$ & $0.07$ & $0.03$ \\
 \phantom{1}0 $<\pt<$  1 & 3.25 & 4.00 & $0.07$ & $0.05$ & $0.01$ \\
 \phantom{1}1 $<\pt<$  2 & 1.50 & 2.50 & $0.19$ & $0.04$ & $0.02$ \\
 \phantom{1}1 $<\pt<$  2 & 2.50 & 3.25 & $0.28$ & $0.05$ & $0.03$ \\
 \phantom{1}1 $<\pt<$  2 & 3.25 & 4.00 & $0.18$ & $0.05$ & $0.02$ \\
 \phantom{1}2 $<\pt<$  3 & 1.50 & 2.50 & $0.18$ & $0.03$ & $0.02$ \\
 \phantom{1}2 $<\pt<$  3 & 2.50 & 3.25 & $0.18$ & $0.03$ & $0.02$ \\
 \phantom{1}2 $<\pt<$  3 & 3.25 & 4.00 & $0.29$ & $0.05$ & $0.03$ \\
 \phantom{1}3 $<\pt<$  4 & 1.50 & 2.50 & $0.24$ & $0.03$ & $0.02$ \\
 \phantom{1}3 $<\pt<$  4 & 2.50 & 3.25 & $0.25$ & $0.03$ & $0.02$ \\
 \phantom{1}3 $<\pt<$  4 & 3.25 & 4.00 & $0.33$ & $0.06$ & $0.03$ \\
 \phantom{1}4 $<\pt<$  5 & 1.50 & 2.50 & $0.27$ & $0.04$ & $0.03$ \\
 \phantom{1}4 $<\pt<$  5 & 2.50 & 3.25 & $0.27$ & $0.04$ & $0.02$ \\
 \phantom{1}4 $<\pt<$  5 & 3.25 & 4.00 & $0.33$ & $0.07$ & $0.03$ \\
 \phantom{1}5 $<\pt<$  6 & 1.50 & 2.50 & $0.29$ & $0.05$ & $0.03$ \\
 \phantom{1}5 $<\pt<$  6 & 2.50 & 3.25 & $0.43$ & $0.05$ & $0.04$ \\
 \phantom{1}5 $<\pt<$  6 & 3.25 & 4.00 & $0.26$ & $0.07$ & $0.02$ \\
 \phantom{1}6 $<\pt<$  7 & 1.50 & 2.50 & $0.35$ & $0.06$ & $0.03$ \\
 \phantom{1}6 $<\pt<$  7 & 2.50 & 3.25 & $0.34$ & $0.06$ & $0.03$ \\
 \phantom{1}6 $<\pt<$  7 & 3.25 & 4.00 & $0.36$ & $0.09$ & $0.03$ \\
 \phantom{1}7 $<\pt<$  8 & 1.50 & 2.50 & $0.24$ & $0.05$ & $0.02$ \\
 \phantom{1}7 $<\pt<$  8 & 2.50 & 3.25 & $0.27$ & $0.06$ & $0.02$ \\
 \phantom{1}7 $<\pt<$  8 & 3.25 & 4.00 & $0.30$ & $0.12$ & $0.03$ \\
 \phantom{1}8 $<\pt<$ 10 & 1.50 & 2.50 & $0.28$ & $0.05$ & $0.03$ \\
 \phantom{1}8 $<\pt<$ 10 & 2.50 & 3.25 & $0.50$ & $0.07$ & $0.05$ \\
 \phantom{1}8 $<\pt<$ 10 & 3.25 & 4.00 & $0.49$ & $0.12$ & $0.05$ \\
10 $<\pt<$ 14 & 1.50 & 2.50 & $0.41$ & $0.07$ & $0.04$ \\
10 $<\pt<$ 14 & 2.50 & 3.25 & $0.43$ & $0.08$ & $0.04$ \\
10 $<\pt<$ 14 & 3.25 & 4.00 & $0.51$ & $0.15$ & $0.05$ \\
\bottomrule
\end{tabular}
\end{table}

%% file: tables/section_nonprompt_ratio_Pbp.tex
\renewcommand{\arraystretch}{1.2}
\begin{table}[!htb]
\caption{\small Cross-section ratios of \psitwos over \jpsi nonprompt 
production in Pb$p$ collisions. The first uncertainty is statistical
and the second systematic. 
}
\label{tab:nonpromptcrosssectionresultsPbp_ratio}
\centering
\begin{tabular}{@{}cl@{$\,<y^*<\,$}lr@{$\,\pm\,$}r@{$\,\pm\,$}r@{}}
\toprule
 \multicolumn{1}{c}{\pt interval (\!\gevc)} & \multicolumn{2}{c}{$y^*$ interval} & \multicolumn{3}{c}{$\sigma_{\psi(2S)}/\sigma_{J/\psi}$} \\ 
 \midrule
\phantom{1}0 $<\pt<$  1 & $-3.25$ & $-2.50$ & $0.22$ & $0.08$ & $0.03$ \\
\phantom{1}0 $<\pt<$  1 & $-4.00$ & $-3.25$ & $0.23$ & $0.06$ & $0.03$ \\
 \phantom{1}0 $<\pt<$  1 & $-5.00$ & $-4.00$ & $0.11$ & $0.07$ & $0.01$ \\
 \phantom{1}1 $<\pt<$  2 & $-3.25$ & $-2.50$ & $0.32$ & $0.05$ & $0.04$ \\
 \phantom{1}1 $<\pt<$  2 & $-4.00$ & $-3.25$ & $0.20$ & $0.03$ & $0.02$ \\
 \phantom{1}1 $<\pt<$  2 & $-5.00$ & $-4.00$ & $0.11$ & $0.03$ & $0.01$ \\
 \phantom{1}2 $<\pt<$  3 & $-3.25$ & $-2.50$ & $0.23$ & $0.04$ & $0.03$ \\
 \phantom{1}2 $<\pt<$  3 & $-4.00$ & $-3.25$ & $0.24$ & $0.04$ & $0.03$ \\
 \phantom{1}2 $<\pt<$  3 & $-5.00$ & $-4.00$ & $0.20$ & $0.05$ & $0.02$ \\
 \phantom{1}3 $<\pt<$  4 & $-3.25$ & $-2.50$ & $0.36$ & $0.05$ & $0.04$ \\
 \phantom{1}3 $<\pt<$  4 & $-4.00$ & $-3.25$ & $0.23$ & $0.04$ & $0.02$ \\
 \phantom{1}3 $<\pt<$  4 & $-5.00$ & $-4.00$ & $0.36$ & $0.07$ & $0.04$ \\
 \phantom{1}4 $<\pt<$  5 & $-3.25$ & $-2.50$ & $0.39$ & $0.05$ & $0.04$ \\
 \phantom{1}4 $<\pt<$  5 & $-4.00$ & $-3.25$ & $0.24$ & $0.04$ & $0.02$ \\
 \phantom{1}4 $<\pt<$  5 & $-5.00$ & $-4.00$ & $0.32$ & $0.07$ & $0.03$ \\
 \phantom{1}5 $<\pt<$  6 & $-3.25$ & $-2.50$ & $0.32$ & $0.06$ & $0.03$ \\
 \phantom{1}5 $<\pt<$  6 & $-4.00$ & $-3.25$ & $0.30$ & $0.05$ & $0.03$ \\
 \phantom{1}5 $<\pt<$  6 & $-5.00$ & $-4.00$ & $0.27$ & $0.08$ & $0.03$ \\
 \phantom{1}6 $<\pt<$  7 & $-3.25$ & $-2.50$ & $0.32$ & $0.06$ & $0.03$ \\
 \phantom{1}6 $<\pt<$  7 & $-4.00$ & $-3.25$ & $0.31$ & $0.06$ & $0.03$ \\
 \phantom{1}6 $<\pt<$  7 & $-5.00$ & $-4.00$ & $0.29$ & $0.10$ & $0.03$ \\
 \phantom{1}7 $<\pt<$  8 & $-3.25$ & $-2.50$ & $0.26$ & $0.07$ & $0.03$ \\
 \phantom{1}7 $<\pt<$  8 & $-4.00$ & $-3.25$ & $0.41$ & $0.08$ & $0.04$ \\
 \phantom{1}7 $<\pt<$  8 & $-5.00$ & $-4.00$ & $0.20$ & $0.12$ & $0.02$ \\
 \phantom{1}8 $<\pt<$ 10 & $-3.25$ & $-2.50$ & $0.33$ & $0.06$ & $0.03$ \\
 \phantom{1}8 $<\pt<$ 10 & $-4.00$ & $-3.25$ & $0.27$ & $0.07$ & $0.03$ \\
 \phantom{1}8 $<\pt<$ 10 & $-5.00$ & $-4.00$ & $0.21$ & $0.13$ & $0.03$ \\
10 $<\pt<$ 14 & $-3.25$ & $-2.50$ & $0.21$ & $0.06$ & $0.02$ \\
10 $<\pt<$ 14 & $-4.00$ & $-3.25$ & $0.34$ & $0.09$ & $0.04$ \\
10 $<\pt<$ 14 & $-5.00$ & $-4.00$ & $0.45$ & $0.21$ & $0.06$ \\
\bottomrule
\end{tabular}
\end{table}

%% file: tables/section_prompt_ratio_pPb_p.tex
\renewcommand{\arraystretch}{1.2}
\begin{table}[!htb]
\caption{\small Cross-section ratios of \psitwos over \jpsi prompt 
production in $p$Pb collisions, as a function of \pt, integrated over
$y^*$. The first uncertainty is statistical
and the second systematic. 
}
\label{tab:promptcrosssectionresultspPb_ratio_p}
\centering
\begin{tabular}{@{}cl@{$\,\pm\,$}l@{$\,\pm\,$}l@{}}
\toprule
 \multicolumn{1}{c}{\pt interval (\!\gevc)} & \multicolumn{3}{c}{$\sigma_{\psi(2S)}/\sigma_{J/\psi}$} \\ 
 \midrule
 \phantom{1}0 $<\pt<$  1 & $0.08$ & $0.01$ & $0.01$ \\
 \phantom{1}1 $<\pt<$  2 & $0.09$ & $0.01$ & $0.01$ \\
 \phantom{1}2 $<\pt<$  3 & $0.12$ & $0.01$ & $0.01$ \\
 \phantom{1}3 $<\pt<$  4 & $0.14$ & $0.01$ & $0.01$ \\
 \phantom{1}4 $<\pt<$  5 & $0.16$ & $0.01$ & $0.01$ \\
 \phantom{1}5 $<\pt<$  6 & $0.18$ & $0.01$ & $0.02$ \\
 \phantom{1}6 $<\pt<$  7 & $0.19$ & $0.02$ & $0.02$ \\
 \phantom{1}7 $<\pt<$  8 & $0.26$ & $0.02$ & $0.02$ \\
 \phantom{1}8 $<\pt<$ 10 & $0.27$ & $0.02$ & $0.02$ \\
10 $<\pt<$ 14 & $0.23$ & $0.03$ & $0.02$ \\
\bottomrule
\end{tabular}
\end{table}

%% file: tables/section_prompt_ratio_Pbp_p.tex
\renewcommand{\arraystretch}{1.2}
\begin{table}[!htb]
\caption{\small Cross-section ratios of \psitwos over \jpsi prompt 
production in Pb$p$ collisions, as a function of \pt, integrated over
$y^*$. The first uncertainty is statistical
and the second systematic. 
}
\label{tab:promptcrosssectionresultsPbp_ratio_p}
\centering
\begin{tabular}{@{}cr@{$\,\pm\,$}r@{$\,\pm\,$}r@{}}
\toprule
 \multicolumn{1}{c}{\pt interval (\!\gevc)} & \multicolumn{3}{c}{$\sigma_{\psi(2S)}/\sigma_{J/\psi}$} \\ 
 \midrule
 \phantom{1}0 $<\pt<$  1 & $0.07$ & $0.01$ & $0.01$ \\
 \phantom{1}1 $<\pt<$  2 & $0.11$ & $0.01$ & $0.01$ \\
 \phantom{1}2 $<\pt<$  3 & $0.11$ & $0.01$ & $0.01$ \\
 \phantom{1}3 $<\pt<$  4 & $0.15$ & $0.01$ & $0.01$ \\
 \phantom{1}4 $<\pt<$  5 & $0.14$ & $0.01$ & $0.01$ \\
 \phantom{1}5 $<\pt<$  6 & $0.17$ & $0.01$ & $0.02$ \\
 \phantom{1}6 $<\pt<$  7 & $0.21$ & $0.02$ & $0.02$ \\
 \phantom{1}7 $<\pt<$  8 & $0.20$ & $0.02$ & $0.02$ \\
 \phantom{1}8 $<\pt<$ 10 & $0.32$ & $0.02$ & $0.03$ \\
10 $<\pt<$ 14 & $0.32$ & $0.04$ & $0.03$ \\
\bottomrule
\end{tabular}
\end{table}

%% file: tables/section_nonprompt_ratio_pPb_p.tex
\renewcommand{\arraystretch}{1.2}
\begin{table}[!htb]
\caption{\small Cross-section ratios of \psitwos over \jpsi nonprompt 
production in $p$Pb collisions, as a function of \pt, integrated over
$y^*$. The first uncertainty is statistical
and the second systematic. 
}
\label{tab:nonpromptcrosssectionresultspPb_ratio_p}
\centering
\begin{tabular}{@{}cr@{$\,\pm\,$}r@{$\,\pm\,$}r@{}}
\toprule
 \multicolumn{1}{c}{\pt interval (\!\gevc)} & \multicolumn{3}{c}{$\sigma_{\psi(2S)}/\sigma_{J/\psi}$} \\ 
 \midrule
 \phantom{1}0 $<\pt<$  1 & $0.18$ & $0.04$ & $0.02$ \\
 \phantom{1}1 $<\pt<$  2 & $0.21$ & $0.02$ & $0.02$ \\
 \phantom{1}2 $<\pt<$  3 & $0.20$ & $0.02$ & $0.02$ \\
 \phantom{1}3 $<\pt<$  4 & $0.27$ & $0.02$ & $0.03$ \\
 \phantom{1}4 $<\pt<$  5 & $0.29$ & $0.03$ & $0.03$ \\
 \phantom{1}5 $<\pt<$  6 & $0.33$ & $0.03$ & $0.03$ \\
 \phantom{1}6 $<\pt<$  7 & $0.35$ & $0.04$ & $0.03$ \\
 \phantom{1}7 $<\pt<$  8 & $0.24$ & $0.04$ & $0.02$ \\
 \phantom{1}8 $<\pt<$ 10 & $0.37$ & $0.04$ & $0.03$ \\
10 $<\pt<$ 14 & $0.43$ & $0.05$ & $0.04$ \\
\bottomrule
\end{tabular}
\end{table}

%% file: tables/section_nonprompt_ratio_Pbp_p.tex
\renewcommand{\arraystretch}{1.2}
\begin{table}[!htb]
\caption{\small Cross-section ratios of \psitwos over \jpsi nonprompt 
production in Pb$p$ collisions, as a function of \pt, integrated over
$y^*$. The first uncertainty is statistical
and the second systematic. 
}
\label{tab:nonpromptcrosssectionresultsPbp_ratio_p}
\centering
\begin{tabular}{@{}cr@{$\,\pm\,$}r@{$\,\pm\,$}r@{}}
\toprule
 \multicolumn{1}{c}{\pt interval (\!\gevc)} & \multicolumn{3}{c}{$\sigma_{\psi(2S)}/\sigma_{J/\psi}$} \\ 
 \midrule
    \phantom{1}0 $<\pt<$  1 & $0.20$ & $0.04$ & $0.02$ \\
 \phantom{1}1 $<\pt<$  2 & $0.23$ & $0.03$ & $0.03$ \\
 \phantom{1}2 $<\pt<$  3 & $0.24$ & $0.03$ & $0.03$ \\
 \phantom{1}3 $<\pt<$  4 & $0.35$ & $0.03$ & $0.03$ \\
 \phantom{1}4 $<\pt<$  5 & $0.34$ & $0.03$ & $0.03$ \\
 \phantom{1}5 $<\pt<$  6 & $0.32$ & $0.04$ & $0.03$ \\
 \phantom{1}6 $<\pt<$  7 & $0.32$ & $0.04$ & $0.03$ \\
 \phantom{1}7 $<\pt<$  8 & $0.31$ & $0.05$ & $0.03$ \\
 \phantom{1}8 $<\pt<$ 10 & $0.32$ & $0.05$ & $0.03$ \\
10 $<\pt<$ 14 & $0.29$ & $0.05$ & $0.03$ \\
\bottomrule
\end{tabular}
\end{table}

%% file: tables/section_prompt_ratio_pPb_y.tex
\renewcommand{\arraystretch}{1.2}
\begin{table}[!htb]
\caption{\small Cross-section ratios of \psitwos over \jpsi prompt 
production, as a function of $y^*$, integrated over
\pt. The first uncertainty is statistical
and the second systematic. 
}
\label{tab:promptcrosssectionresults_ratio_y}
\centering
\begin{tabular}{@{}l@{$\,<y^{*}<\,$}lr@{$\,\pm\,$}r@{$\,\pm\,$}r@{}}
\toprule
 \multicolumn{2}{c}{$y^*$ interval} & \multicolumn{3}{c}{$\sigma_{\psi(2S)}/\sigma_{J/\psi}$} \\ 
 \midrule
$-3.25$ & $-2.50$ & $0.11$ & $0.01$ & $0.01$ \\
$-4.00$ & $-3.25$ & $0.10$ & $0.01$ & $0.01$ \\
$-5.00$ & $-4.00$ & $0.12$ & $0.01$ & $0.01$ \\
$\phantom{-}1.50$ & $\phantom{-}2.50$ & $0.12$ & $0.01$ & $0.01$ \\
$\phantom{-}2.50$ & $\phantom{-}3.25$ & $0.12$ & $0.01$ & $0.01$ \\
$\phantom{-}3.25$ & $\phantom{-}4.00$ & $0.13$ & $0.01$ & $0.01$ \\
\bottomrule
\end{tabular}
\end{table}

%% file: tables/section_nonprompt_ratio_pPb_y.tex
\renewcommand{\arraystretch}{1.2}
\begin{table}[!htb]
\caption{\small Cross-section ratios of \psitwos over \jpsi nonprompt 
production, as a function of $y^*$, integrated over
\pt. The first uncertainty is statistical
and the second systematic. 
}
\label{tab:nonpromptcrosssectionresults_ratio_y}
\centering
\begin{tabular}{@{}l@{$\,<y^{*}<\,$}lr@{$\,\pm\,$}r@{$\,\pm\,$}r@{}}
\toprule
 \multicolumn{2}{c}{$y^*$ interval} & \multicolumn{3}{c}{$\sigma_{\psi(2S)}/\sigma_{J/\psi}$} \\ 
 \midrule
$-3.25$ & $-2.50$ & $0.32$ & $0.02$ & $0.03$ \\
$-4.00$ & $-3.25$ & $0.24$ & $0.02$ & $0.03$ \\
$-5.00$ & $-4.00$ & $0.22$ & $0.02$ & $0.02$ \\
$\phantom{-}1.50$ & $\phantom{-}2.50$ & $0.23$ & $0.01$ & $0.02$ \\
$\phantom{-}2.50$ & $\phantom{-}3.25$ & $0.27$ & $0.02$ & $0.03$ \\
$\phantom{-}3.25$ & $\phantom{-}4.00$ & $0.26$ & $0.02$ & $0.02$ \\
\bottomrule
\end{tabular}
\end{table}

%% file: tables/section_prompt_psi2s_pPb.tex
\renewcommand{\arraystretch}{1.2}
\begin{table}[!htb]
\caption{\small Absolute \psitwos prompt 
production cross-section in $p$Pb collisions. The first uncertainty is statistical
and the second systematic. 
}
\label{tab:promptcrosssectionresultspPb}
\centering
\begin{tabular}{@{}l@{$\,<\pt<\,$}ll@{$\,<y^*<\,$}lr@{$\,\pm\,$}r@{$\,\pm\,$}r@{}}
\toprule
 \multicolumn{2}{c}{\pt interval (\!\gevc)} & \multicolumn{2}{c}{$y^*$ interval} & \multicolumn{3}{c}{${\rm d}^2 \sigma/({\rm d} y^{*} {\rm d} \pt)$ [nb/(\!\gevc)]} \\ 
 \midrule
 \phantom{1111}0 &  1 & 1.50 & 2.50 &  $\phantom{00000000}7583$ & $2196$ & $1094\phantom{000}$ \\
 \phantom{1111}0 &  1 & 2.50 & 3.25 &  $6311$ & $1628$ &  $674\phantom{000}$ \\
 \phantom{1111}0 &  1 & 3.25 & 4.00 &  $7599$ & $1529$ &  $934\phantom{000}$ \\
 \phantom{1111}1 &  2 & 1.50 & 2.50 & $20666$ & $3108$ & $2673\phantom{000}$ \\
 \phantom{1111}1 &  2 & 2.50 & 3.25 & $12453$ & $2175$ & $1351\phantom{000}$ \\
 \phantom{1111}1 &  2 & 3.25 & 4.00 & $10852$ & $2057$ & $1222\phantom{000}$ \\
 \phantom{1111}2 &  3 & 1.50 & 2.50 & $19380$ & $2772$ & $2430\phantom{000}$ \\
 \phantom{1111}2 &  3 & 2.50 & 3.25 & $20220$ & $2386$ & $2178\phantom{000}$ \\
 \phantom{1111}2 &  3 & 3.25 & 4.00 & $14116$ & $1991$ & $1584\phantom{000}$ \\
 \phantom{1111}3 &  4 & 1.50 & 2.50 & $20058$ & $2118$ & $2294\phantom{000}$ \\
 \phantom{1111}3 &  4 & 2.50 & 3.25 & $11529$ & $1567$ & $1200\phantom{000}$ \\
 \phantom{1111}3 &  4 & 3.25 & 4.00 & $11643$ & $1525$ & $1232\phantom{000}$ \\
 \phantom{1111}4 &  5 & 1.50 & 2.50 & $11912$ & $1455$ & $1292\phantom{000}$ \\
 \phantom{1111}4 &  5 & 2.50 & 3.25 & $10620$ & $1043$ & $1069\phantom{000}$ \\
 \phantom{1111}4 &  5 & 3.25 & 4.00 &  $6561$ &  $945$ &  $665\phantom{000}$ \\
 \phantom{1111}5 &  6 & 1.50 & 2.50 &  $6438$ &  $814$ &  $682\phantom{000}$ \\
 \phantom{1111}5 &  6 & 2.50 & 3.25 &  $6758$ &  $683$ &  $677\phantom{000}$ \\
 \phantom{1111}5 &  6 & 3.25 & 4.00 &  $4930$ &  $687$ &  $493\phantom{000}$ \\
 \phantom{1111}6 &  7 & 1.50 & 2.50 &  $4281$ &  $553$ &  $453\phantom{000}$ \\
 \phantom{1111}6 &  7 & 2.50 & 3.25 &  $3205$ &  $407$ &  $323\phantom{000}$ \\
 \phantom{1111}6 &  7 & 3.25 & 4.00 &  $2972$ &  $437$ &  $310\phantom{000}$ \\
 \phantom{1111}7 &  8 & 1.50 & 2.50 &  $3330$ &  $412$ &  $348\phantom{000}$ \\
 \phantom{1111}7 &  8 & 2.50 & 3.25 &  $2399$ &  $310$ &  $245\phantom{000}$ \\
 \phantom{1111}7 &  8 & 3.25 & 4.00 &  $2217$ &  $357$ &  $241\phantom{000}$ \\
 \phantom{1111}8 & 10 & 1.50 & 2.50 &  $1758$ &  $187$ &  $184\phantom{000}$ \\
 \phantom{1111}8 & 10 & 2.50 & 3.25 &  $1199$ &  $145$ &  $122\phantom{000}$ \\
 \phantom{1111}8 & 10 & 3.25 & 4.00 &   $605$ &  $124$ &   $70\phantom{000}$ \\
\phantom{111}10 & 14 & 1.50 & 2.50 &   $351$ &   $55$ &   $36\phantom{000}$ \\
\phantom{111}10 & 14 & 2.50 & 3.25 &   $194$ &   $47$ &   $20\phantom{000}$ \\
\phantom{111}10 & 14 & 3.25 & 4.00 &   $229$ &   $55$ &   $28\phantom{000}$ \\
\bottomrule
\end{tabular}
\end{table}

%% file: tables/section_prompt_psi2s_Pbp.tex
\renewcommand{\arraystretch}{1.2}
\begin{table}[!htb]
\caption{\small Absolute \psitwos prompt 
production cross-section in Pb$p$ collisions. The first uncertainty is statistical
and the second systematic. 
}
\label{tab:promptcrosssectionresultsPbp}
\centering
\begin{tabular}{@{}l@{$\,<\pt<\,$}ll@{$\,<y^*<\,$}lr@{$\,\pm\,$}r@{$\,\pm\,$}r@{}}
\toprule
 \multicolumn{2}{c}{\pt interval (\!\gevc)} & \multicolumn{2}{c}{$y^*$ interval} & \multicolumn{3}{c}{${\rm d}^2 \sigma/({\rm d} y^{*} {\rm d}\pt$) [nb/(\gevc)]} \\ 
 \midrule
 \phantom{1111}0 &  1 & $-3.25$ & $-2.50$ & $\phantom{00000000}8376$ & $2712$ & $1346\phantom{000}$ \\
 \phantom{1111}0 &  1 & $-4.00$ & $-3.25$ &  $6593$ & $1894$ &  $886\phantom{000}$ \\
 \phantom{1111}0 &  1 & $-5.00$ & $-4.00$ &  $4716$ & $1638$ &  $608\phantom{000}$ \\
 \phantom{1111}1 &  2 & $-3.25$ & $-2.50$ & $27207$ & $3822$ & $4113\phantom{000}$ \\
 \phantom{1111}1 &  2 & $-4.00$ & $-3.25$ & $12760$ & $1979$ & $1670\phantom{000}$ \\
 \phantom{1111}1 &  2 & $-5.00$ & $-4.00$ & $18484$ & $2313$ & $2285\phantom{000}$ \\
 \phantom{1111}2 &  3 & $-3.25$ & $-2.50$ & $23515$ & $3270$ & $3167\phantom{000}$ \\
 \phantom{1111}2 &  3 & $-4.00$ & $-3.25$ & $15894$ & $2480$ & $2017\phantom{000}$ \\
 \phantom{1111}2 &  3 & $-5.00$ & $-4.00$ & $12462$ & $1988$ & $1543\phantom{000}$ \\
 \phantom{1111}3 &  4 & $-3.25$ & $-2.50$ & $19105$ & $2442$ & $2376\phantom{000}$ \\
 \phantom{1111}3 &  4 & $-4.00$ & $-3.25$ & $16264$ & $1872$ & $1941\phantom{000}$ \\
 \phantom{1111}3 &  4 & $-5.00$ & $-4.00$ &  $6856$ & $1316$ &  $828\phantom{000}$ \\
 \phantom{1111}4 &  5 & $-3.25$ & $-2.50$ &  $9423$ & $1413$ & $1092\phantom{000}$ \\
 \phantom{1111}4 &  5 & $-4.00$ & $-3.25$ &  $7576$ & $1030$ &  $897\phantom{000}$ \\
 \phantom{1111}4 &  5 & $-5.00$ & $-4.00$ &  $4467$ &  $773$ &  $536\phantom{000}$ \\
 \phantom{1111}5 &  6 & $-3.25$ & $-2.50$ &  $6512$ &  $847$ &  $755\phantom{000}$ \\
 \phantom{1111}5 &  6 & $-4.00$ & $-3.25$ &  $4432$ &  $549$ &  $528\phantom{000}$ \\
 \phantom{1111}5 &  6 & $-5.00$ & $-4.00$ &  $3061$ &  $450$ &  $368\phantom{000}$ \\
 \phantom{1111}6 &  7 & $-3.25$ & $-2.50$ &  $4063$ &  $552$ &  $490\phantom{000}$ \\
 \phantom{1111}6 &  7 & $-4.00$ & $-3.25$ &  $2875$ &  $347$ &  $335\phantom{000}$ \\
 \phantom{1111}6 &  7 & $-5.00$ & $-4.00$ &  $1769$ &  $297$ &  $217\phantom{000}$ \\
 \phantom{1111}7 &  8 & $-3.25$ & $-2.50$ &  $1837$ &  $327$ &  $222\phantom{000}$ \\
 \phantom{1111}7 &  8 & $-4.00$ & $-3.25$ &  $1317$ &  $219$ &  $161\phantom{000}$ \\
 \phantom{1111}7 &  8 & $-5.00$ & $-4.00$ &  $1080$ &  $179$ &  $145\phantom{000}$ \\
 \phantom{1111}8 & 10 & $-3.25$ & $-2.50$ &  $1105$ &  $151$ &  $133\phantom{000}$ \\
 \phantom{1111}8 & 10 & $-4.00$ & $-3.25$ &  $1200$ &  $115$ &  $144\phantom{000}$ \\
 \phantom{1111}8 & 10 & $-5.00$ & $-4.00$ &   $449$ &   $74$ &   $65\phantom{000}$ \\
\phantom{111}10 & 14 & $-3.25$ & $-2.50$ &   $240$ &   $47$ &   $31\phantom{000}$ \\
\phantom{111}10 & 14 & $-4.00$ & $-3.25$ &   $262$ &   $36$ &   $34\phantom{000}$ \\
\phantom{111}10 & 14 & $-5.00$ & $-4.00$ &    $95$ &   $21$ &   $14\phantom{000}$ \\
\bottomrule
\end{tabular}
\end{table}

%% file: tables/section_nonprompt_psi2s_pPb.tex
\renewcommand{\arraystretch}{1.2}
\begin{table}[!htb]
\caption{\small Absolute \psitwos nonprompt 
production cross-section in $p$Pb collisions. The first uncertainty is statistical
and the second systematic. 
}
\label{tab:nonpromptcrosssectionresultspPb}
\centering
\begin{tabular}{@{}l@{$\,<\pt<$}ll@{$\,<y^*<\,$}lr@{$\,\pm\,$}r@{$\,\pm\,$}r@{}}
\toprule
 \multicolumn{2}{c}{\pt interval (\!\gevc)} & \multicolumn{2}{c}{$y^*$ interval} & \multicolumn{3}{c}{${\rm d}^2 \sigma/({\rm d} y^{*} {\rm d} \pt)$ [nb/(\!\gevc)]} \\ 
 \midrule
 \phantom{1111}0 &  1 & 1.50 & 2.50 & $\phantom{00000000}2692$ &  $798$ & $389\phantom{000}$ \\
 \phantom{1111}0 &  1 & 2.50 & 3.25 & $2666$ &  $688$ & $285\phantom{000}$ \\
 \phantom{1111}0 &  1 & 3.25 & 4.00 &  $592$ &  $445$ &  $73\phantom{000}$ \\
 \phantom{1111}1 &  2 & 1.50 & 2.50 & $5911$ & $1111$ & $764\phantom{000}$ \\
 \phantom{1111}1 &  2 & 2.50 & 3.25 & $6634$ & $1097$ & $720\phantom{000}$ \\
 \phantom{1111}1 &  2 & 3.25 & 4.00 & $3446$ &  $891$ & $388\phantom{000}$ \\
 \phantom{1111}2 &  3 & 1.50 & 2.50 & $5705$ & $1084$ & $715\phantom{000}$ \\
 \phantom{1111}2 &  3 & 2.50 & 3.25 & $4484$ &  $820$ & $483\phantom{000}$ \\
 \phantom{1111}2 &  3 & 3.25 & 4.00 & $4965$ &  $878$ & $557\phantom{000}$ \\
 \phantom{1111}3 &  4 & 1.50 & 2.50 & $6062$ &  $789$ & $693\phantom{000}$ \\
 \phantom{1111}3 &  4 & 2.50 & 3.25 & $4521$ &  $613$ & $471\phantom{000}$ \\
 \phantom{1111}3 &  4 & 3.25 & 4.00 & $4220$ &  $807$ & $446\phantom{000}$ \\
 \phantom{1111}4 &  5 & 1.50 & 2.50 & $4419$ &  $630$ & $479\phantom{000}$ \\
 \phantom{1111}4 &  5 & 2.50 & 3.25 & $3105$ &  $442$ & $312\phantom{000}$ \\
 \phantom{1111}4 &  5 & 3.25 & 4.00 & $2535$ &  $563$ & $257\phantom{000}$ \\
 \phantom{1111}5 &  6 & 1.50 & 2.50 & $2841$ &  $449$ & $301\phantom{000}$ \\
 \phantom{1111}5 &  6 & 2.50 & 3.25 & $3092$ &  $384$ & $310\phantom{000}$ \\
 \phantom{1111}5 &  6 & 3.25 & 4.00 & $1156$ &  $289$ & $116\phantom{000}$ \\
 \phantom{1111}6 &  7 & 1.50 & 2.50 & $2107$ &  $336$ & $223\phantom{000}$ \\
 \phantom{1111}6 &  7 & 2.50 & 3.25 & $1521$ &  $251$ & $153\phantom{000}$ \\
 \phantom{1111}6 &  7 & 3.25 & 4.00 &  $924$ &  $233$ &  $96\phantom{000}$ \\
 \phantom{1111}7 &  8 & 1.50 & 2.50 &  $784$ &  $175$ &  $82\phantom{000}$ \\
 \phantom{1111}7 &  8 & 2.50 & 3.25 &  $786$ &  $166$ &  $80\phantom{000}$ \\
 \phantom{1111}7 &  8 & 3.25 & 4.00 &  $496$ &  $191$ &  $54\phantom{000}$ \\
 \phantom{1111}8 & 10 & 1.50 & 2.50 &  $573$ &  $102$ &  $60\phantom{000}$ \\
 \phantom{1111}8 & 10 & 2.50 & 3.25 &  $675$ &   $95$ &  $69\phantom{000}$ \\
 \phantom{1111}8 & 10 & 3.25 & 4.00 &  $404$ &   $96$ &  $47\phantom{000}$ \\
\phantom{111}10 & 14 & 1.50 & 2.50 &  $263$ &   $43$ &  $27\phantom{000}$ \\
\phantom{111}10 & 14 & 2.50 & 3.25 &  $186$ &   $36$ &  $19\phantom{000}$ \\
\phantom{111}10 & 14 & 3.25 & 4.00 &  $118$ &   $33$ &  $14\phantom{000}$ \\
\bottomrule
\end{tabular}
\end{table}

%% file: tables/section_nonprompt_psi2s_Pbp.tex
\renewcommand{\arraystretch}{1.2}
\begin{table}[!htb]
\caption{\small Absolute \psitwos nonprompt 
production cross-section in Pb$p$ collisions. The first uncertainty is 
statistical and the second systematic. 
}
\label{tab:nonpromptcrosssectionresultsPbp}
\centering
\begin{tabular}{@{}l@{$\,<\pt<$}ll@{$\,<y^*<\,$}lr@{$\,\pm\,$}r@{$\,\pm\,$}r@{}}
\toprule
 \multicolumn{2}{c}{\pt interval (\!\gevc)} & \multicolumn{2}{c}{$y^*$ interval} & \multicolumn{3}{c}{${\rm d}^2 \sigma/({\rm d} y^{*} {\rm d} \pt)$ [nb/(\!\gevc)]} \\ 
 \midrule
 \phantom{1111}0 &  1 & $-3.25$ & $-2.50$ &  $\phantom{000000}3257$ & $118$ &  $523\phantom{000}$ \\
 \phantom{1111}0 &  1 & $-4.00$ & $-3.25$ &  $2564$ &  $625$ &  $345\phantom{000}$ \\
 \phantom{1111}0 &  1 & $-5.00$ & $-4.00$ &   $753$ &  $469$ &   $97\phantom{000}$ \\
 \phantom{1111}1 &  2 & $-3.25$ & $-2.50$ & $10845$ & $1677$ & $1639\phantom{000}$ \\
 \phantom{1111}1 &  2 & $-4.00$ & $-3.25$ &  $3441$ &  $589$ &  $450\phantom{000}$ \\
 \phantom{1111}1 &  2 & $-5.00$ & $-4.00$ &  $1504$ &  $441$ &  $186\phantom{000}$ \\
 \phantom{1111}2 &  3 & $-3.25$ & $-2.50$ &  $7165$ & $1217$ &  $965\phantom{000}$ \\
\phantom{1111}2 &  3 & $-4.00$ & $-3.25$ &  $5337$ &  $873$ &  $677\phantom{000}$ \\
 \phantom{1111}2 &  3 & $-5.00$ & $-4.00$ &  $2438$ &  $562$ &  $302\phantom{000}$ \\
 \phantom{1111}3 &  4 & $-3.25$ & $-2.50$ &  $8195$ & $1017$ & $1019\phantom{000}$ \\
 \phantom{1111}3 &  4 & $-4.00$ & $-3.25$ &  $3767$ &  $567$ &  $450\phantom{000}$ \\
 \phantom{1111}3 &  4 & $-5.00$ & $-4.00$ &  $2765$ &  $522$ &  $334\phantom{000}$ \\
 \phantom{1111}4 &  5 & $-3.25$ & $-2.50$ &  $5532$ &  $660$ &  $641\phantom{000}$ \\
 \phantom{1111}4 &  5 & $-4.00$ & $-3.25$ &  $2142$ &  $382$ &  $254\phantom{000}$ \\
 \phantom{1111}4 &  5 & $-5.00$ & $-4.00$ &  $1376$ &  $294$ &  $165\phantom{000}$ \\
 \phantom{1111}5 &  6 & $-3.25$ & $-2.50$ &  $2565$ &  $446$ &  $297\phantom{000}$ \\
 \phantom{1111}5 &  6 & $-4.00$ & $-3.25$ &  $1563$ &  $242$ &  $186\phantom{000}$ \\
 \phantom{1111}5 &  6 & $-5.00$ & $-4.00$ &   $598$ &  $170$ &   $72\phantom{000}$ \\
 \phantom{1111}6 &  7 & $-3.25$ & $-2.50$ &  $1482$ &  $291$ &  $179\phantom{000}$ \\
 \phantom{1111}6 &  7 & $-4.00$ & $-3.25$ &   $926$ &  $166$ &  $108\phantom{000}$ \\
 \phantom{1111}6 &  7 & $-5.00$ & $-4.00$ &   $352$ &  $120$ &   $43\phantom{000}$ \\
 \phantom{1111}7 &  8 & $-3.25$ & $-2.50$ &   $707$ &  $182$ &   $85\phantom{000}$ \\
 \phantom{1111}7 &  8 & $-4.00$ & $-3.25$ &   $718$ &  $139$ &   $88\phantom{000}$ \\
 \phantom{1111}7 &  8 & $-5.00$ & $-4.00$ &   $117$ &   $67$ &   $16\phantom{000}$ \\
 \phantom{1111}8 & 10 & $-3.25$ & $-2.50$ &   $519$ &   $99$ &   $63\phantom{000}$ \\
 \phantom{1111}8 & 10 & $-4.00$ & $-3.25$ &   $215$ &   $55$ &   $26\phantom{000}$ \\
 \phantom{1111}8 & 10 & $-5.00$ & $-4.00$ &    $56$ &   $33$ &    $8\phantom{000}$ \\
\phantom{111}10 & 14 & $-3.25$ & $-2.50$ &    $98$ &   $27$ &   $13\phantom{000}$ \\
\phantom{111}10 & 14 & $-4.00$ & $-3.25$ &    $76$ &   $20$ &    $10\phantom{000}$ \\
\phantom{111}10 & 14 & $-5.00$ & $-4.00$ &    $27$ &   $12$ &    $4\phantom{000}$ \\
\bottomrule
\end{tabular}
\end{table}

%% file: tables/section_prompt_psi2s_pPb_p.tex
\renewcommand{\arraystretch}{1.2}
\begin{table}[!htb]
\caption{\small Absolute \psitwos prompt 
production cross-section in $p$Pb collisions, as a function of \pt, 
integrated over $y^*$. The first uncertainty is 
statistical and the second systematic. 
}
\label{tab:promptcrosssectionresultspPb_p}
\centering
\begin{tabular}{@{}l@{$\,<\pt<\,$}lr@{$\,\pm\,$}r@{$\,\pm\,$}r@{}}
\toprule
 \multicolumn{2}{c}{\pt interval (\!\gevc)} & \multicolumn{3}{c}{${\rm d} \sigma/{\rm d}\pt$ [nb/(\gevc)]} \\ 
 \midrule
\phantom{111}0 & 1 & $\phantom{00000}18015$ & $2762$ &  $656\phantom{000}$ \\
\phantom{111}1 & 2 & $38145$ & $3834$ & $1595\phantom{000}$ \\
\phantom{111}2 & 3 & $45132$ & $3621$ & $1495\phantom{000}$ \\
\phantom{111}3 & 4 & $37437$ & $2679$ & $1380\phantom{000}$ \\
\phantom{111}4 & 5 & $24798$ & $1797$ &  $905\phantom{000}$ \\
\phantom{111}5 & 6 & $15204$ & $1091$ &  $618\phantom{000}$ \\
\phantom{111}6 & 7 &  $8914$ &  $712$ &  $408\phantom{000}$ \\
\phantom{111}7 & 8 &  $6792$ &  $544$ &  $180\phantom{000}$ \\
\phantom{111}8 & 10 &  $3111$ &  $236$ &  $145\phantom{000}$ \\
\phantom{11}10 & 14 &   $669$ &   $78$ &   $51\phantom{000}$ \\
\bottomrule
\end{tabular}
\end{table}

%% file: tables/section_prompt_psi2s_Pbp_p.tex
\renewcommand{\arraystretch}{1.2}
\begin{table}[!htb]
\caption{\small Absolute \psitwos prompt 
production cross-section in Pb$p$ collisions, as a function of \pt, 
integrated over $y^*$. The first uncertainty is 
statistical and the second systematic. 
}
\label{tab:promptcrosssectionresultsPbp_p}
\centering
\begin{tabular}{@{}l@{$\,<\pt<\,$}lr@{$\,\pm\,$}r@{$\,\pm\,$}r@{}}
\toprule
 \multicolumn{2}{c}{\pt interval (\!\gevc)} & \multicolumn{3}{c}{${\rm d} \sigma/ {\rm d} \pt$ [nb/(\!\gevc)]} \\ 
 \midrule
 \phantom{111}0 &  1 & $\phantom{00000}16857$ & $3298$ & $4383\phantom{000}$ \\
 \phantom{111}1 &  2 & $50640$ & $4452$ & $11141\phantom{000}$ \\
 \phantom{111}2 &  3 & $44782$ & $4047$ & $5822\phantom{000}$ \\
 \phantom{111}3 &  4 & $36445$ & $2985$ & $6560\phantom{000}$ \\
 \phantom{111}4 &  5 & $18454$ & $1711$ & $4245\phantom{000}$ \\
 \phantom{111}5 &  6 & $12132$ & $1001$ & $2548\phantom{000}$ \\
 \phantom{111}6 &  7 &  $7547$ &  $649$ & $1660\phantom{000}$ \\
 \phantom{111}7 &  8 &  $3635$ &  $390$ & $836\phantom{000}$ \\
 \phantom{111}8 & 10 &  $2343$ &  $183$ & $422\phantom{000}$ \\
\phantom{11}10 & 14 &   $507$ &   $57$ & $76\phantom{000}$ \\
\bottomrule
\end{tabular}
\end{table}

%% file: tables/section_nonprompt_psi2s_pPb_p.tex
\renewcommand{\arraystretch}{1.2}
\begin{table}[!htb]
\caption{\small Absolute \psitwos nonprompt 
production cross-section in $p$Pb collisions, as a function of \pt, 
integrated over $y^*$. The first uncertainty is 
statistical and the second systematic. 
}
\label{tab:nonpromptcrosssectionresultspPb_p}
\centering
\begin{tabular}{@{}l@{$\,<\pt<\,$}lr@{$\,\pm\,$}r@{$\,\pm\,$}r@{}}
\toprule
 \multicolumn{2}{c}{\pt interval (\!\gevc)} & \multicolumn{3}{c}{${\rm d} \sigma/ {\rm d} \pt$ [nb/(\!\gevc)]} \\ 
 \midrule
 \phantom{111}0 &  1 &  $\phantom{000}5136$ & $1007$ &  $656\phantom{000}$ \\
 \phantom{111}1 &  2 & $13471$ & $1535$ & $1595\phantom{000}$ \\
 \phantom{111}2 &  3 & $12791$ & $1409$ & $1495\phantom{000}$ \\
 \phantom{111}3 &  4 & $12617$ & $1096$ & $1380\phantom{000}$ \\
 \phantom{111}4 &  5 &  $8648$ &  $828$ &  $905\phantom{000}$ \\
 \phantom{111}5 &  6 &  $6027$ &  $575$ &  $618\phantom{000}$ \\
 \phantom{111}6 &  7 &  $3940$ &  $423$ &  $408\phantom{000}$ \\
 \phantom{111}7 &  8 &  $1745$ &  $258$ &  $180\phantom{000}$ \\
 \phantom{111}8 & 10 &  $1383$ &  $144$ &  $145\phantom{000}$ \\
\phantom{11}10 & 14 &   $491$ &   $57$ &   $51\phantom{000}$ \\
\bottomrule
\end{tabular}
\end{table}

%% file: tables/section_nonprompt_psi2s_Pbp_p.tex
\renewcommand{\arraystretch}{1.2}
\begin{table}[!htb]
\caption{\small Absolute \psitwos nonprompt 
production cross-section in Pb$p$ collisions, as a function of \pt, 
integrated over $y^*$. The first uncertainty is 
statistical and the second systematic. 
}
\label{tab:nonpromptcrosssectionresultsPbp_p}
\centering
\begin{tabular}{@{}l@{$\,<\pt<\,$}lr@{$\,\pm\,$}r@{$\,\pm\,$}r@{}}
\toprule
 \multicolumn{2}{c}{\pt interval (\!\gevc)} & \multicolumn{3}{c}{${\rm d} \sigma/ {\rm d} \pt$ [nb/(\!\gevc)]} \\ 
 \midrule
 \phantom{111}0 &  1 &  $\phantom{000}5119$ & $1109$ & $1329\phantom{000}$ \\
 \phantom{111}1 &  2 & $12219$ & $1404$ & $2706\phantom{000}$ \\
 \phantom{111}2 &  3 & $11814$ & $1256$ & $1574\phantom{000}$ \\
 \phantom{111}3 &  4 & $11737$ & $1017$ & $2145\phantom{000}$ \\
 \phantom{111}4 &  5 &  $7132$ &  $643$ & $1645\phantom{000}$ \\
 \phantom{111}5 &  6 &  $3694$ &  $417$ &  $804\phantom{000}$ \\
 \phantom{111}6 &  7 &  $2157$ &  $279$ &  $493\phantom{000}$ \\
 \phantom{111}7 &  8 &  $1186$ &  $184$ &  $275\phantom{000}$ \\
 \phantom{111}8 & 10 &   $606$ &   $91$ &  $114\phantom{000}$ \\
\phantom{11}10 & 14 &   $158$ &   $28$ &   $24\phantom{000}$ \\
\bottomrule
\end{tabular}
\end{table}

%% file: tables/r_pA_psi2s_prompt_pt.tex
\renewcommand{\arraystretch}{1.2}
\begin{table}[!htb]
\caption{\small Prompt \psitwos modification factor $R_{p{\rm Pb}}$ 
in $p$Pb collisions, as a function of \pt, 
integrated over $y^*$. The first uncertainty is 
statistical and the second systematic. 
}
\label{tab:rpAprompt_pt}
\centering
\begin{tabular}{@{}l@{$\,<\pt<\,$}lr@{$\,\pm\,$}r@{$\,\pm\,$}r@{}}
\toprule
\multicolumn{2}{c}{\pt interval (\!\gevc)} & \multicolumn{3}{c}{$R_{p{\rm Pb}}$} \\ 
\midrule
\phantom{111}0 & 1 & 0.42 & 0.07 & 0.07 \\
\phantom{111}1 & 2 & 0.39 & 0.04 & 0.07 \\
\phantom{111}2 & 3 & 0.52 & 0.04 & 0.08 \\
\phantom{111}3 & 4 & 0.59 & 0.04 & 0.09 \\
\phantom{111}4 & 5 & 0.63 & 0.05 & 0.10 \\
\phantom{111}5 & 6 & 0.66 & 0.05 & 0.10 \\
\phantom{111}6 & 7 & 0.62 & 0.05 & 0.11 \\
\phantom{111}7 & 8 & 0.88 & 0.07 & 0.12 \\
\phantom{111}8 & 10 & 0.85 & 0.08 & 0.12 \\
\phantom{11}10 & 14 & 0.58 & 0.08 & 0.13 \\
\bottomrule
\end{tabular}
\end{table}

%% file: tables/r_Ap_psi2s_prompt_pt.tex
\renewcommand{\arraystretch}{1.2}
\begin{table}[!htb]
\caption{\small Prompt \psitwos modification factor $R_{p{\rm Pb}}$ 
in Pb$p$ collisions, as a function of \pt, 
integrated over $y^*$. The first uncertainty is 
statistical and the second systematic. 
}
\label{tab:rApprompt_pt}
\centering
\begin{tabular}{@{}l@{$\,<\pt<\,$}lr@{$\,\pm\,$}r@{$\,\pm\,$}r@{}}
\toprule
\multicolumn{2}{c}{\pt interval (\!\gevc)} & \multicolumn{3}{c}{$R_{p{\rm Pb}}$} \\ 
\midrule
\phantom{111}0 & 1 & 0.48 & 0.09 & 0.12 \\
\phantom{111}1 & 2 & 0.66 & 0.06 & 0.12 \\
\phantom{111}2 & 3 & 0.71 & 0.06 & 0.14 \\
\phantom{111}3 & 4 & 0.82 & 0.07 & 0.16 \\
\phantom{111}4 & 5 & 0.71 & 0.07 & 0.17 \\
\phantom{111}5 & 6 & 0.84 & 0.07 & 0.18 \\
\phantom{111}6 & 7 & 0.87 & 0.08 & 0.19 \\
\phantom{111}7 & 8 & 0.81 & 0.09 & 0.20 \\
\phantom{111}8 & 10 & 1.19 & 0.12 & 0.21 \\
\phantom{11}10 & 14 & 0.88 & 0.11 & 0.22 \\
\bottomrule
\end{tabular}
\end{table}

%% file: tables/r_pA_psi2s_nonprompt_pt.tex
\renewcommand{\arraystretch}{1.2}
\begin{table}[!htb]
\caption{\small Nonprompt \psitwos modification factor $R_{p{\rm Pb}}$ 
in $p$Pb collisions, as a function of \pt, 
integrated over $y^*$. The first uncertainty is 
statistical and the second systematic. 
}
\label{tab:rpAnonprompt_pt}
\centering
\begin{tabular}{@{}l@{$\,<\pt<\,$}lr@{$\,\pm\,$}r@{$\,\pm\,$}r@{}}
\toprule
\multicolumn{2}{c}{\pt interval (\!\gevc)} & \multicolumn{3}{c}{$R_{p{\rm Pb}}$} \\ 
\midrule
\phantom{111}0 & 1 & 0.88 & 0.19 & 0.13 \\
\phantom{111}1 & 2 & 0.72 & 0.08 & 0.10 \\
\phantom{111}2 & 3 & 0.78 & 0.09 & 0.11 \\
\phantom{111}3 & 4 & 0.98 & 0.09 & 0.12 \\
\phantom{111}4 & 5 & 0.86 & 0.09 & 0.12 \\
\phantom{111}5 & 6 & 1.00 & 0.10 & 0.13 \\
\phantom{111}6 & 7 & 1.15 & 0.13 & 0.14 \\
\phantom{111}7 & 8 & 0.82 & 0.14 & 0.15 \\
\phantom{111}8 & 10 & 1.20 & 0.15 & 0.15 \\
\phantom{11}10 & 14 & 1.18 & 0.16 & 0.16 \\
\bottomrule
\end{tabular}
\end{table}

%% file: tables/r_Ap_psi2s_nonprompt_pt.tex
\renewcommand{\arraystretch}{1.2}
\begin{table}[!htb]
\caption{\small Nonprompt \psitwos modification factor $R_{p{\rm Pb}}$ 
in Pb$p$ collisions, as a function of \pt, 
integrated over $y^*$. The first uncertainty is 
statistical and the second systematic. 
}
\label{tab:rApnonprompt_pt}
\centering
\begin{tabular}{@{}l@{$\,<\pt<\,$}lr@{$\,\pm\,$}r@{$\,\pm\,$}r@{}}
\toprule
\multicolumn{2}{c}{\pt interval (\!\gevc)} & \multicolumn{3}{c}{$R_{p{\rm Pb}}$} \\ 
\midrule
\phantom{111}0 & 1 & 1.41 & 0.33 & 0.21 \\
\phantom{111}1 & 2 & 1.04 & 0.12 & 0.18 \\
\phantom{111}2 & 3 & 1.19 & 0.13 & 0.18 \\
\phantom{111}3 & 4 & 1.62 & 0.15 & 0.20 \\
\phantom{111}4 & 5 & 1.31 & 0.12 & 0.21 \\
\phantom{111}5 & 6 & 1.13 & 0.13 & 0.21 \\
\phantom{111}6 & 7 & 1.18 & 0.16 & 0.21 \\
\phantom{111}7 & 8 & 1.06 & 0.18 & 0.21 \\
\phantom{111}8 & 10 & 1.10 & 0.18 & 0.23 \\
\phantom{11}10 & 14 & 0.86 & 0.17 & 0.26 \\
\bottomrule
\end{tabular}
\end{table}

%% file: tables/r_FB_psi2s_prompt_pt.tex
\renewcommand{\arraystretch}{1.2}
\begin{table}[!htb]
\caption{\small Prompt \psitwos forward to backward ratio $R_{\rm FB}$, as a function of \pt, 
integrated over $y^*$. The first uncertainty is 
statistical and the second systematic. 
}
\label{tab:rFBprompt_pt}
\centering
\begin{tabular}{@{}l@{$\,<\pt<\,$}lr@{$\,\pm\,$}r@{$\,\pm\,$}r@{}}
\toprule
\multicolumn{2}{c}{\pt interval (\!\gevc)} & \multicolumn{3}{c}{$R_{\rm FB}$} \\ 
\midrule
\phantom{111}0 & 1 & 0.66 & 0.18 & 0.04 \\
\phantom{111}1 & 2 & 0.37 & 0.06 & 0.02 \\
\phantom{111}2 & 3 & 0.61 & 0.08 & 0.03 \\
\phantom{111}3 & 4 & 0.45 & 0.06 & 0.02 \\
\phantom{111}4 & 5 & 0.69 & 0.09 & 0.04 \\
\phantom{111}5 & 6 & 0.70 & 0.09 & 0.04 \\
\phantom{111}6 & 7 & 0.60 & 0.08 & 0.03 \\
\phantom{111}7 & 8 & 0.93 & 0.15 & 0.05 \\
\phantom{111}8 & 10 & 0.52 & 0.07 & 0.03 \\
\phantom{11}10 & 14 & 0.55 & 0.12 & 0.03 \\
\bottomrule
\end{tabular}
\end{table}

%% file: tables/r_FB_psi2s_nonprompt_pt.tex
\renewcommand{\arraystretch}{1.2}
\begin{table}[!htb]
\caption{\small Nonprompt \psitwos forward to backward ratio $R_{\rm FB}$, as a function of \pt, 
integrated over $y^*$. The first uncertainty is 
statistical and the second systematic. 
}
\label{tab:rFBnonprompt_pt}
\centering
\begin{tabular}{@{}l@{$\,<\pt<\,$}lr@{$\,\pm\,$}r@{$\,\pm\,$}r@{}}
\toprule
\multicolumn{2}{c}{\pt interval (\!\gevc)} & \multicolumn{3}{c}{$R_{\rm FB}$} \\ 
\midrule
\phantom{111}0 & 1 & 0.41 & 0.14 & 0.02 \\
\phantom{111}1 & 2 & 0.48 & 0.09 & 0.03 \\
\phantom{111}2 & 3 & 0.53 & 0.09 & 0.03 \\
\phantom{111}3 & 4 & 0.51 & 0.08 & 0.03 \\
\phantom{111}4 & 5 & 0.49 & 0.08 & 0.03 \\
\phantom{111}5 & 6 & 0.72 & 0.12 & 0.04 \\
\phantom{111}6 & 7 & 0.67 & 0.13 & 0.04 \\
\phantom{111}7 & 8 & 0.61 & 0.14 & 0.03 \\
\phantom{111}8 & 10 & 0.99 & 0.20 & 0.06 \\
\phantom{11}10 & 14 & 1.14 & 0.29 & 0.07 \\
\bottomrule
\end{tabular}
\end{table}

%% file: tables/r_FB_psi2s_prompt_y.tex
\renewcommand{\arraystretch}{1.2}
\begin{table}[!htb]
\caption{\small Prompt \psitwos forward to backward ratio $R_{\rm FB}$, as a function of $y^*$, 
integrated over \pt. The first uncertainty is 
statistical and the second systematic. 
}
\label{tab:rFBprompt_y}
\centering
\begin{tabular}{@{}l@{$\,<y^*<\,$}lr@{$\,\pm\,$}r@{$\,\pm\,$}r@{}}
\toprule
\multicolumn{2}{c}{$y^*$ interval} & \multicolumn{3}{c}{$R_{\rm FB}$} \\ 
\midrule
2.50 & 3.25 & 0.51 & 0.04 & 0.03 \\
3.25 & 4.00 & 0.57 & 0.05 & 0.03 \\
\bottomrule
\end{tabular}
\end{table}

%% file: tables/r_FB_psi2s_nonprompt_y.tex
\renewcommand{\arraystretch}{1.2}
\begin{table}[!htb]
\caption{\small Nonprompt \psitwos forward to backward ratio $R_{\rm FB}$, as a function of $y^*$, 
integrated over \pt. The first uncertainty is 
statistical and the second systematic. 
}
\label{tab:rFBnonprompt_y}
\centering
\begin{tabular}{@{}l@{$\,<y^*<\,$}lr@{$\,\pm\,$}r@{$\,\pm\,$}r@{}}
\toprule
\multicolumn{2}{c}{$y^*$ interval} & \multicolumn{3}{c}{$R_{\rm FB}$} \\ 
\midrule
2.50 & 3.25 & 0.50 & 0.05 & 0.03 \\
3.25 & 4.00 & 0.61 & 0.07 & 0.03 \\
\bottomrule
\end{tabular}
\end{table}

%% file: tables/double_ratio_prompt_pPb_pt.tex
\renewcommand{\arraystretch}{1.2}
\begin{table}[!htb]
\caption{\small Double ratio $R_{\psitwos/\jpsi}^{p{\rm Pb}}$ 
in $p$Pb collisions for prompt production, as a function of \pt, 
integrated over $y^*$. The first uncertainty is 
statistical and the second systematic. 
}
\label{tab:doublepromptpPb_pt}
\centering
\begin{tabular}{@{}l@{$\,<\pt<\,$}lr@{$\,\pm\,$}r@{$\,\pm\,$}r@{}}
\toprule
\multicolumn{2}{c}{\pt interval (\!\gevc)} & \multicolumn{3}{c}{$R_{\psitwos/\jpsi}^{p{\rm Pb}}$} \\ 
\midrule
\phantom{111}0 & 1 & 0.80 & 0.12 & 0.16 \\
\phantom{111}1 & 2 & 0.70 & 0.07 & 0.12 \\
\phantom{111}2 & 3 & 0.80 & 0.06 & 0.10 \\
\phantom{111}3 & 4 & 0.82 & 0.06 & 0.09 \\
\phantom{111}4 & 5 & 0.83 & 0.06 & 0.09 \\
\phantom{111}5 & 6 & 0.81 & 0.06 & 0.09 \\
\phantom{111}6 & 7 & 0.72 & 0.06 & 0.08 \\
\phantom{111}7 & 8 & 1.01 & 0.08 & 0.11 \\
\phantom{111}8 & 10 & 0.95 & 0.09 & 0.13 \\
\phantom{11}10 & 14 & 0.63 & 0.08 & 0.09 \\
\bottomrule
\end{tabular}
\end{table}

%% file: tables/double_ratio_prompt_Pbp_pt.tex
\renewcommand{\arraystretch}{1.2}
\begin{table}[!htb]
\caption{\small Double ratio $R_{\psitwos/\jpsi}^{p{\rm Pb}}$ 
in Pb$p$ collisions for prompt production, as a function of \pt, 
integrated over $y^*$. The first uncertainty is 
statistical and the second systematic. 
}
\label{tab:doublepromptPbp_pt}
\centering
\begin{tabular}{@{}l@{$\,<\pt<\,$}lr@{$\,\pm\,$}r@{$\,\pm\,$}r@{}}
\toprule
\multicolumn{2}{c}{\pt interval (\!\gevc)} & \multicolumn{3}{c}{$R_{\psitwos/\jpsi}^{p{\rm Pb}}$} \\ 
\midrule
\phantom{111}0 & 1 & 0.64 & 0.13 & 0.27 \\
\phantom{111}1 & 2 & 0.82 & 0.07 & 0.18 \\
\phantom{111}2 & 3 & 0.77 & 0.07 & 0.14 \\
\phantom{111}3 & 4 & 0.82 & 0.07 & 0.17 \\
\phantom{111}4 & 5 & 0.69 & 0.06 & 0.13 \\
\phantom{111}5 & 6 & 0.79 & 0.07 & 0.12 \\
\phantom{111}6 & 7 & 0.81 & 0.07 & 0.13 \\
\phantom{111}7 & 8 & 0.76 & 0.08 & 0.12 \\
\phantom{111}8 & 10 & 1.13 & 0.11 & 0.15 \\
\phantom{11}10 & 14 & 0.88 & 0.11 & 0.14 \\
\bottomrule
\end{tabular}
\end{table}

%% file: tables/double_ratio_nonprompt_pPb_pt.tex
\renewcommand{\arraystretch}{1.2}
\begin{table}[!htb]
\caption{\small Double ratio $R_{\psitwos/\jpsi}^{p{\rm Pb}}$ 
in $p$Pb collisions for nonprompt production, as a function of \pt, 
integrated over $y^*$. The first uncertainty is 
statistical and the second systematic. 
}
\label{tab:doublenonpromptpPb_pt}
\centering
\begin{tabular}{@{}l@{$\,<\pt<\,$}lr@{$\,\pm\,$}r@{$\,\pm\,$}r@{}}
\toprule
\multicolumn{2}{c}{\pt interval (\!\gevc)} & \multicolumn{3}{c}{$R_{\psitwos/\jpsi}^{p{\rm Pb}}$} \\ 
\midrule
\phantom{111}0 & 1 & 1.18 & 0.25 & 0.21 \\
\phantom{111}1 & 2 & 0.91 & 0.11 & 0.12 \\
\phantom{111}2 & 3 & 0.94 & 0.11 & 0.11 \\
\phantom{111}3 & 4 & 1.15 & 0.10 & 0.12 \\
\phantom{111}4 & 5 & 0.99 & 0.10 & 0.10 \\
\phantom{111}5 & 6 & 1.11 & 0.11 & 0.12 \\
\phantom{111}6 & 7 & 1.26 & 0.14 & 0.14 \\
\phantom{111}7 & 8 & 0.83 & 0.14 & 0.12 \\
\phantom{111}8 & 10 & 1.27 & 0.16 & 0.16 \\
\phantom{11}10 & 14 & 1.31 & 0.18 & 0.20 \\
\bottomrule
\end{tabular}
\end{table}

%% file: tables/double_ratio_nonprompt_Pbp_pt.tex
\renewcommand{\arraystretch}{1.2}
\begin{table}[!htb]
\caption{\small Double ratio $R_{\psitwos/\jpsi}^{p{\rm Pb}}$ 
in Pb$p$ collisions for nonprompt production, as a function of \pt, 
integrated over $y^*$. The first uncertainty is 
statistical and the second systematic. 
}
\label{tab:doublenonpromptPbp_pt}
\centering
\begin{tabular}{@{}l@{$\,<\pt<\,$}lr@{$\,\pm\,$}r@{$\,\pm\,$}r@{}}
\toprule
\multicolumn{2}{c}{\pt interval (\!\gevc)} & \multicolumn{3}{c}{$R_{\psitwos/\jpsi}^{p{\rm Pb}}$} \\ 
\midrule
\phantom{111}0 & 1 & 1.34 & 0.31 & 0.24 \\
\phantom{111}1 & 2 & 0.99 & 0.12 & 0.14 \\
\phantom{111}2 & 3 & 1.11 & 0.12 & 0.14 \\
\phantom{111}3 & 4 & 1.49 & 0.13 & 0.17 \\
\phantom{111}4 & 5 & 1.17 & 0.11 & 0.13 \\
\phantom{111}5 & 6 & 1.07 & 0.12 & 0.12 \\
\phantom{111}6 & 7 & 1.16 & 0.15 & 0.14 \\
\phantom{111}7 & 8 & 1.07 & 0.18 & 0.15 \\
\phantom{111}8 & 10 & 1.08 & 0.18 & 0.14 \\
\phantom{11}10 & 14 & 0.87 & 0.17 & 0.15 \\
\bottomrule
\end{tabular}
\end{table}

%% file: tables/double_ratio_prompt_y.tex
\renewcommand{\arraystretch}{1.2}
\begin{table}[!htb]
\caption{\small Double ratio $R_{\psitwos/\jpsi}^{p{\rm Pb}}$ 
for prompt production, as a function of $y^*$, 
integrated over \pt. The first uncertainty is 
statistical and the second systematic. 
}
\label{tab:doubleprompt_y}
\centering
\begin{tabular}{@{}l@{$\,<y^*<\,$}lr@{$\,\pm\,$}r@{$\,\pm\,$}r@{}}
\toprule
\multicolumn{2}{c}{$y^*$ interval} & \multicolumn{3}{c}{$R_{\psitwos/\jpsi}^{p{\rm Pb}}$} \\
\midrule
$-5.00$ & $-4.00$ &0.77 & 0.05 & 0.10 \\
$-4.00$ & $-3.25$ & 0.67 & 0.04 & 0.10 \\
$-3.25$ & $-2.50$  & 0.75 & 0.05 & 0.11 \\
$\phantom{-}1.50$ & $\phantom{-}2.50$ & 0.82 & 0.05 & 0.11 \\
$\phantom{-}2.50$ & $\phantom{-}3.25$ & 0.80 & 0.04 & 0.10 \\
$\phantom{-}3.25$ & $\phantom{-}4.00$ & 0.86 & 0.05 & 0.11 \\
\bottomrule
\end{tabular}
\end{table}

%% file: tables/double_ratio_nonprompt_y.tex
\renewcommand{\arraystretch}{1.2}
\begin{table}[!htb]
\caption{\small Double ratio $R_{\psitwos/\jpsi}^{p{\rm Pb}}$ 
for nonprompt production, as a function of $y^*$, 
integrated over \pt. The first uncertainty is 
statistical and the second systematic. 
}
\label{tab:doublenonprompt_y}
\centering
\begin{tabular}{@{}l@{$\,<y^*<\,$}lr@{$\,\pm\,$}r@{$\,\pm\,$}r@{}}
\toprule
\multicolumn{2}{c}{$y^*$ interval} & \multicolumn{3}{c}{$R_{\psitwos/\jpsi}^{p{\rm Pb}}$} \\
\midrule
$-5.00$ & $-4.00$ & 0.94 & 0.11 & 0.20 \\
$-4.00$ & $-3.25$ & 1.04 & 0.08 & 0.28 \\
$-3.25$ & $-2.50$ & 1.38 & 0.11 & 0.44 \\
$\phantom{-}1.50$ & $\phantom{-}2.50$ & 0.97 & 0.07 & 0.14 \\
$\phantom{-}2.50$ & $\phantom{-}3.25$ & 1.17 & 0.09 & 0.16 \\
$\phantom{-}3.25$ & $\phantom{-}4.00$ & 1.14 & 0.11 & 0.15 \\
\bottomrule
\end{tabular}
\end{table}

%% file: tables/double_ratio_prompt.tex
\renewcommand{\arraystretch}{1.2}
\begin{table}[!htb]
\caption{\small Double ratio $R_{\psitwos/\jpsi}^{p{\rm Pb}}$ 
for prompt production, 
integrated over \pt and $y^*$. The first uncertainty is 
statistical and the second systematic. 
}
\label{tab:doubleprompt}
\centering
\begin{tabular}{@{}l@{$\,<y^*<\,$}lr@{$\,\pm\,$}r@{$\,\pm\,$}r@{}}
\toprule
\multicolumn{2}{c}{$y^*$ interval} & \multicolumn{3}{c}{$R_{\psitwos/\jpsi}^{p{\rm Pb}}$} \\
\midrule
$-5.00$ & $-2.50$ & 0.73 & 0.03 & 0.06 \\
$\phantom{-}1.50$ & $\phantom{-}4.00$ & 0.82 & 0.03 & 0.07 \\
\bottomrule
\end{tabular}
\end{table}

%% file: tables/double_ratio_nonprompt.tex
\renewcommand{\arraystretch}{1.2}
\begin{table}[!htb]
\caption{\small Double ratio $R_{\psitwos/\jpsi}^{p{\rm Pb}}$ 
for nonprompt production, 
integrated over \pt and $y^*$. The first uncertainty is 
statistical and the second systematic. 
}
\label{tab:doublenonprompt}
\centering
\begin{tabular}{@{}l@{$\,<y^*<\,$}lr@{$\,\pm\,$}r@{$\,\pm\,$}r@{}}
\toprule
\multicolumn{2}{c}{$y^*$ interval} & \multicolumn{3}{c}{$R_{\psitwos/\jpsi}^{p{\rm Pb}}$} \\
\midrule
$-5.00$ & $-2.50$ & 1.17 & 0.07 & 0.11 \\
$\phantom{-}1.50$ & $\phantom{-}4.00$ & 1.06 & 0.06 & 0.10 \\
\bottomrule
\end{tabular}
\end{table}

%% file: Authorship_LHCb-PAPER-2023-024.tex
\centerline
{\large\bf LHCb collaboration}
\begin
{flushleft}
\small
R.~Aaij$^{36}$\lhcborcid{0000-0003-0533-1952},
A.S.W.~Abdelmotteleb$^{55}$\lhcborcid{0000-0001-7905-0542},
C.~Abellan~Beteta$^{49}$,
F.~Abudin{\'e}n$^{55}$\lhcborcid{0000-0002-6737-3528},
T.~Ackernley$^{59}$\lhcborcid{0000-0002-5951-3498},
B.~Adeva$^{45}$\lhcborcid{0000-0001-9756-3712},
M.~Adinolfi$^{53}$\lhcborcid{0000-0002-1326-1264},
P.~Adlarson$^{79}$\lhcborcid{0000-0001-6280-3851},
H.~Afsharnia$^{11}$,
C.~Agapopoulou$^{47}$\lhcborcid{0000-0002-2368-0147}
% ...

{\footnotesize \it

% ...
$^{11}$Universit{\'e} Clermont Auvergne, CNRS/IN2P3, LPC, Clermont-Ferrand, France\\
% ...
$^{36}$Nikhef National Institute for Subatomic Physics, Amsterdam, Netherlands\\
% ...
$^{45}$Instituto Galego de F{\'\i}sica de Altas Enerx{\'\i}as (IGFAE), Universidade de Santiago de Compostela, Santiago de Compostela, Spain\\
% ...
$^{47}$European Organization for Nuclear Research (CERN), Geneva, Switzerland\\
% ...
$^{49}$Physik-Institut, Universit{\"a}t Z{\"u}rich, Z{\"u}rich, Switzerland\\
% ...
$^{53}$H.H. Wills Physics Laboratory, University of Bristol, Bristol, United Kingdom\\
% ...
$^{55}$Department of Physics, University of Warwick, Coventry, United Kingdom\\
% ...
$^{58}$School of Physics and Astronomy, University of Glasgow, Glasgow, United Kingdom\\
$^{59}$Oliver Lodge Laboratory, University of Liverpool, Liverpool, United Kingdom\\
% ...
$^{79}$Department of Physics and Astronomy, Uppsala University, Uppsala, Sweden, associated to $^{58}$\\
% ...
}
\end{flushleft}